\documentclass[a4paper,fleqn,11pt]{article}
\pdfoutput=1

\usepackage[T1]{fontenc}
\usepackage[utf8]{inputenc}
\usepackage[english]{babel}
\usepackage{cite}
\usepackage{hyperref}

 \usepackage{mathtools}
\usepackage{xcolor}
\usepackage{color}
\usepackage{slashed}

\usepackage[absolute]{textpos}
\usepackage{multirow}

\usepackage{relsize}
\usepackage{braket}
\usepackage{xspace}
\usepackage{listings}
\usepackage{amssymb}
\usepackage{ulem} 

\usepackage{amsmath}
\usepackage{amssymb}
\usepackage{array}
\usepackage{calc}
\usepackage{longtable}
\usepackage{multirow}
\usepackage{slashed}
\usepackage{booktabs}
\usepackage{pstricks}
\usepackage{graphicx}
\usepackage{xspace}
\usepackage{units}
\usepackage{textcomp}

\numberwithin{equation}{section}
\usepackage[format=hang,labelfont=bf,hypcap=true]{caption}
\usepackage{subcaption}
\usepackage{sectsty}
\usepackage{enumitem}
\allsectionsfont{\sffamily}
\subsubsectionfont{\mdseries\itshape\large}
\makeatletter
\DeclareRobustCommand*{\bfseries}{%
  \not@math@alphabet\bfseries\mathbf
  \fontseries\bfdefault\selectfont
  \boldmath
}
\makeatletter
\DeclareRobustCommand*{\bfseries}{%
  \not@math@alphabet\bfseries\mathbf
  \fontseries\bfdefault\selectfont
  \boldmath
}
\makeatother
\let\spreprint\empty
\newcommand{\preprint}[1]{\def\spreprint{\protect#1}}
\let\sinstitute\empty
\newcommand{\institute}[1]{\def\sinstitute{\protect#1}}
\makeatletter
\renewcommand{\maketitle}{\begingroup
  \null\thispagestyle{empty}%
    \ifx\spreprint\empty
      \vskip 5ex
    \else
      \flushright\large\spreprint\vskip 10ex
    \fi
    \vskip 5ex
    \flushleft
      {\sffamily\bfseries\huge\@title}\vskip 6ex
      \@author\vskip 2ex
      \ifx\sinstitute\empty
      \else
        {\small\sinstitute}
      \fi
    \vskip 5ex
  \endgroup
}
\makeatother
\renewenvironment{abstract}{\begin{center}
  {\large\sffamily\bfseries Abstract: }
  \begin{minipage}[t]{0.75\textwidth}
}{\end{minipage}\end{center}\vskip 10ex}

\numberwithin{equation}{section}
\allowdisplaybreaks[2]

\newcommand{\Collier}{{\rmfamily\scshape Collier}\xspace}
\newcommand{\Coli}{{\rmfamily\scshape Coli}\xspace}
\newcommand{\DD}{{\rmfamily\scshape DD}\xspace}
\newcommand{\OneLoop}{{\rmfamily\scshape OneLOop}\xspace}
\newcommand{\Cuttools}{{\rmfamily\scshape CutTools}\xspace}
\newcommand{\Matrix}{{\rmfamily \scshape Matrix}\xspace}
\newcommand{\Munich}{{\rmfamily \scshape Munich}\xspace}
\newcommand{\Sherpa}{{\rmfamily\scshape Sherpa}\xspace}
\newcommand{\Rambo}{{\rmfamily\scshape Rambo}\xspace}
\newcommand{\OpenLoops}{{\rmfamily\scshape OpenLoops}\xspace}
\newcommand{\Recola}{{\rmfamily\scshape Recola}\xspace}
\newcommand{\OL}{\OpenLoops}
\newcommand{\OLone}{\mbox{\OpenLoops\hspace{-0.5mm}1}\xspace}
\newcommand{\OLL}{\mbox{\OpenLoops\hspace{-0.5mm}2}\xspace}
\newcommand{\OLtwo}{\OLL}

\newcommand{\SherpaOpenLoops}{{\rmfamily \scshape Sherpa+OpenLoops}\xspace}

\newcommand{\Herwig}{\plusplus{{\rmfamily\scshape Herwig}}\xspace}
\newcommand{\Powheg}{{\rmfamily\scshape Powheg-Box}\xspace}
\newcommand{\Whizard}{{\rmfamily\scshape Whizard}\xspace}
\newcommand{\Geneva}{{\rmfamily\scshape Geneva}\xspace}
\newcommand{\Mathematica}{{\rmfamily\scshape Mathematica}\xspace}
\newcommand{\Python}{{\rmfamily\scshape Python}\xspace}
\newcommand{\Fortran}{{\rmfamily\scshape Fortran}\xspace}
\newcommand{\plusplus}[1]{{#1\nolinebreak[4]\hspace{-.0em}\raisebox{.4ex}{\tiny\bf ++}}}
\def\cpp{\plusplus{C}\xspace}

\renewcommand{\refeq}[1]{\mbox{\eqref{#1}}}

\newcommand{\reffi}[1]{\mbox{Fig.~\ref{#1}}}

\newcommand{\refta}[1]{\mbox{Tab.~\ref{#1}}}

\newcommand{\refse}[1]{\mbox{Section~\ref{#1}}}
\newcommand{\refses}[2]{\mbox{Sections~\ref{#1}--\ref{#2}}}
\newcommand{\refapp}[1]{\mbox{Appendix~\ref{#1}}}

\newcommand{\ie}{i.e.\ }

\def\Djslash{\hspace{.7ex}{\big\slash}{\mbox{\hspace{-1.9ex}$\bar D_j$}}}

\newcommand{\f}[2]{\frac{#1}{#2}}

\newcommand{\ssst}[1]{\scriptscriptstyle{\text{#1}}}
\newcommand{\nosss}[1]{#1}

\newcommand{\bit}{\begin{itemize}}
\newcommand{\eit}{\end{itemize}}
\newcommand{\bce}{\begin{center}}
\newcommand{\ece}{\end{center}}
\newcommand{\bea}{\begin{eqnarray}}
\newcommand{\eea}{\end{eqnarray}}
\newcommand{\be}{\begin{equation}}
\newcommand{\ee}{\end{equation}}
\newcommand{\ba}{\begin{align}}
\newcommand{\ea}{\end{align}}
\newcommand{\beqar}{\begin{eqnarray}}
\newcommand{\eeqar}{\end{eqnarray}}
\newcommand{\beq}{\begin{equation}}
\newcommand{\eeq}{\end{equation}}
\newcommand{\beas}{\begin{eqnarray*}}
\newcommand{\eeas}{\end{eqnarray*}}
\newcommand{\bes}{\begin{equation*}}
\newcommand{\ees}{\end{equation*}}
\newcommand{\bas}{\begin{align*}}
\newcommand{\eas}{\end{align*}}
\newcommand{\Li}{{\rm Li}}

\newcommand{\eps}{{\varepsilon}}
\newcommand{\lb}{\left(}
\newcommand{\rb}{\right)}

\DeclareMathAlphabet{\mathbbold}{U}{bbold}{m}{n}
\newcommand{\idop}{\mathbbold{1}}

\newcommand{\ceps}{C_{\eps}}
\newcommand{\Dbar}[1]{\bar{D}_{\nosss{#1}}}

\newcommand{\momk}[1]{k_{\nosss{#1}}}
\newcommand{\momp}[1]{p_{#1}}
\newcommand{\mass}[1]{m_{\nosss{#1}}}
\newcommand{\heli}{h}

\newcommand{\helicheck}{\check h}

\newcommand{\helihat}{\hat h}

\newcommand{\momq}{\bar{q}}
\newcommand{\tilq}{\tilde{q}}
\newcommand{\npart}{N_{\mathrm{p}}}

\newcommand{\calA}{\mathcal{A}}
\newcommand{\calC}{\mathcal{C}}
\newcommand{\calI}{\mathcal{I}}

\newcommand{\calK}{\mathcal{K}}
\newcommand{\calM}{\mathcal{M}}
\newcommand{\calN}{\mathcal{N}}

\newcommand{\calU}{\mathcal{U}}
\newcommand{\calS}{\mathcal{S}}
\newcommand{\calV}{\mathcal{V}}

\newcommand{\calW}{\mathcal{W}}

\newcommand{\seg}{S}

\newcommand{\col}{\mathrm{col}}
\newcommand{\hel}{\mathrm{hel}}
\newcommand{\re}{\mathrm{Re}}
\newcommand{\Tr}{\mathrm{Tr}}

\newcommand{\smax}{s_{\mathrm{max}}}

\newcommand{\rY}{\mathrm Y}
\newcommand{\rR}{\mathrm R}
\newcommand{\rL}{\mathrm L}
\newcommand{\rS}{\mathrm s}
\newcommand{\rs}{\mathrm s}
\newcommand{\rT}{\mathrm T}
\newcommand{\ri}{\mathrm i}
\newcommand{\rd}{\mathrm d}
\newcommand{\ord}{\mathcal O}

\newcommand{\LO}{\mathrm{LO}}
\newcommand{\NLO}{\mathrm{NLO}}

\definecolor{bluemar}{rgb}{0,0,.5}
\definecolor{redmar}{rgb}{.8,0,0}
\definecolor{greenmar}{rgb}{0,.5,0}

\newcommand{\tree}{00}
\newcommand{\onel}{01}
\newcommand{\onelsq}{11}

\newcommand{\treeiop}{\tree, \mbox{\scriptsize I-op}}
\newcommand{\onelsqiop}{\onelsq,\mbox{\scriptsize I-op}}

\newlength{\defheight}
\newcommand{\heightB}{1.3\defheight}

\newcommand{\alphaS}{\alpha_{\rs}}
\newcommand{\gmu}{G_{\mu}}
\newcommand{\GF}{\gmu}

\newcommand{\MZ}{M_\mathrm{Z}}

\newcommand{\mur}{\mu_{\rm R}}
\newcommand{\muf}{\mu_{\rm F}}
\newcommand{\mudim}{\mu_{D}}

\def\ct{\mathrm{CT}}
\def\ir{\mathrm{IR}}
\def\uv{\mathrm{UV}}
\def\ssLO{{\scriptscriptstyle \LO}}
\def\ssNLO{{\scriptscriptstyle \NLO}}
\def\ssQCD{{\scriptscriptstyle \QCD}}
\def\ssQED{{\scriptscriptstyle \QED}}

\newcommand{\colcorr}[4]{\calC^{(#1|#2)}_{#3,#4}}
\newcommand{\spincorr}[4]{\calB^{(#1|#2|#3)}_{#4}}
\newcommand{\spincorrB}[3]{\calB^{(#1|#2)}_{#3}}
\def\kperp{k_\perp}

\def\nqq{n_{q\bar q}}
\def\smin{s_{\mathrm{min}}}
\def\tmin{t_{\mathrm{min}}}
\def\tmax{t_{\mathrm{max}}}
\def\ntildeqq{\tilde{n}_{q\bar q}}

\def\corrop{\delta}
\def\EW{\mathrm{EW}}
\def\QCD{\mathrm{QCD}}
\def\QED{\mathrm{QED}}

\def\msbar{\overline{\text{MS}}}

\def\ceps{C_\epsilon}
\def\epsuv{\eps_{\mathrm{UV}}}
\def\epsir{\eps_{\mathrm{IR}}}
\def\rE{\mathrm{E}}

\newcommand{\paramyuk}[2]{{#1}_{#2,\mathrm{Y}}}
\newcommand\muyuk[1]{\paramyuk{\mu}{#1}}

\newcommand\massyuk[1]{\paramyuk{M}{#1}}
\newcommand\widthyuk[1]{\paramyuk{\Gamma}{#1}}
\def\mudimuv{\mu_{\mathrm{UV}}}
\def\mudimir{\mu_{\mathrm{IR}}}

\hypersetup{
  pdfauthor={Federico Buccioni, Jean-Nicolas Lang, Jonas M. Lindert, Philipp Maierh\"ofer,  Stefano Pozzorini, Hantian Zhang, Max F. Zoller},
  pdftitle={OpenLoops\;2}
}

\preprint{IPPP/19/62\\FR-PHENO-2019-12\\PSI-PR-19-15\\ZU-TH 37/19\\
}

\title{OpenLoops\;2}

\author{Federico Buccioni$^1$, Jean-Nicolas Lang$^1$, Jonas M. Lindert$^2$, Philipp Maierh\"ofer$^3$, \\  
Stefano Pozzorini$^1$, Hantian Zhang$^1$, Max F. Zoller$^4$}

\institute{
$^1$Physik-Institut, Universit\"at Z\"urich, Winterthurerstrasse 190,  CH-8057 Z\"urich, Switzerland\\
$^2$Institute for Particle Physics Phenomenology, University of Durham, Durham,~DH1~3LE, UK\\
$^3$Physikalisches Institut, Albert-Ludwigs-Universit\"at Freiburg, 79104 Freiburg, Germany\\
$^4$Paul Scherrer Institut, CH-5232 Villigen PSI, Switzerland
}

\begin{document}
\vspace*{10mm}
\maketitle
\vspace*{20mm}

\begin{abstract}
We present the new version of \OpenLoops, an automated generator of tree and
one-loop scattering amplitudes based on the open-loop recursion.  
One main
novelty of \OLtwo is the extension of the original algorithm 
from NLO QCD to the full Standard Model, including 
electroweak (EW) corrections from gauge, Higgs and Yukawa interactions.
In this context, among several new features, 
we discuss  the systematic bookkeeping of QCD--EW interferences,
a flexible implementation of the complex-mass scheme
for processes with on-shell and off-shell unstable particles,
a special treatment of on-shell and off-shell 
external photons, and efficient scale variations.
The other main novelty is the implementation of the recently proposed
on-the-fly reduction algorithm, which supersedes the usage of external
reduction libraries for the calculation of tree--loop interferences. 
This new algorithm is equipped with an automated system that avoids
Gram-determinant instabilities through analytic methods in combination with
a new hybrid-precision approach based on a highly targeted usage of
quadruple precision with minimal CPU overhead.  The resulting 
significant speed and
stability improvements are especially relevant for challenging NLO multi-leg
calculations and for NNLO applications.

\end{abstract}

\newpage
\tableofcontents

\section{Introduction}
\label{sec:intro}

Scattering amplitudes at one loop are a mandatory ingredient 
for any precision calculation at high-energy colliders. At next-to-leading order
(NLO),
the calculation of hard cross sections 
requires one-loop matrix elements with hard kinematics,
while next-to-next-to leading order (NNLO) predictions 
require one-loop amplitudes with one additional unresolved particle.
Nowadays, thanks to a variety of modern techniques~\cite{Britto:2004nc,
delAguila:2004nf,
Bern:2005cq,
Denner:2005nn,
Ossola:2006us,
Forde:2007mi,Giele:2008ve,
vanHameren:2009vq,
Cascioli:2011va},
one-loop calculations can be carried out with 
a number of automated and widely applicable programs~\cite{Ossola:2007ax,
Berger:2008sj,
vanHameren:2009dr,
Hirschi:2011pa,
Mastrolia:2010nb,
Badger:2012pg,
hepforge,
Cullen:2014yla,
Peraro:2014cba,
Denner:2016kdg,
Actis:2016mpe} 
that have strongly boosted the field of precision phenomenology.
Most notably, such tools have extended the reach of NLO calculations 
to highly non-trivial multi-particle processes~\cite{Bern:2013gka,
Badger:2013yda,
Bevilacqua:2015qha,
Hoche:2016elu,
Denner:2016wet}
and have opened the door to the automation of
multi-purpose Monte Carlo generators at NLO~\cite{Gleisberg:2008ta,
Alioli:2010xd,
Bevilacqua:2011xh,
Alwall:2014hca,
Sjostrand:2014zea,
Weiss:2015npa,
Bellm:2015jjp}.

In this paper we present the new version of \OpenLoops,%
\footnote{The original version of the algorithm was presented in a
letter~\cite{Cascioli:2011va}, and its public implementation was only
documented online~\cite{hepforge} so far.  Thus this paper provides the first
thorough description of the \OL program.}
an automated tool for the calculation of tree and one-loop scattering
amplitudes within the Standard Model (SM).
The \OpenLoops algorithm is based on a numerical recursion%
\footnote{This type of recursion was first
proposed in the context of off-shell recurrence relations for colour-ordered
gluon-scattering amplitudes~\cite{vanHameren:2009vq}.}
that generates loop amplitudes in terms of cut-open loop
diagrams~\cite{Cascioli:2011va,Buccioni:2017yxi}.
Such objects, called open loops, are characterised by a 
tree topology but depend on the loop momentum.

In the original version of the algorithm~\cite{Cascioli:2011va},
implemented in \OLone~\cite{hepforge}, 
loop amplitudes are built in two phases.
In the first phase, Feynman diagrams are constructed in terms of 
tensor integrals using the open-loop recursion, 
while in the second phase, loop amplitudes are
reduced to scalar integrals using 
external libraries such as \Collier~\cite{Denner:2016kdg} 
or \Cuttools~\cite{Ossola:2007ax}.
The main strengths of this approach are the high speed of the open-loop
recursion and the possibility of curing numerical instabilities 
through the tensor-reduction techniques~\cite{Denner:2002ii,Denner:2005nn} implemented
in~\Collier~\cite{Denner:2016kdg}.

In the original open-loop algorithm~\cite{Cascioli:2011va}, the rank of open
loops increases at each step of the recursion.  As a consequence, 
the CPU time required for their processing, the memory footprint, and also 
numerical instabilities, tend to grow rather fast with the number of scattering
particles.
%
%
For these reasons, in \OLtwo the construction of loop amplitudes and their reduction
have been unified in a single recursive algorithm~\cite{Buccioni:2017yxi}
that makes it possible to avoid high-rank objects at all stages of the amplitude calculations.
This is achieved by interleaving single steps of the construction of
open loops with reduction operations at the integrand
level~\cite{delAguila:2004nf}.
The implementation of this method,
called on-the-fly reduction, 
is one of the main novelties of \OLtwo.
So far it is restricted to tree--loop interferences at NLO, while squared
loop amplitudes are still processed in the same way as in \OLone.

The on-the-fly reduction algorithm in \OLtwo is equipped with an automated system that
avoids numerical instabilities in a highly efficient way.
This stability system makes use of analytic techniques that have been
introduced in~\cite{Buccioni:2017yxi} and have meanwhile been extended in
various directions, and supplemented by a novel hybrid-precision system.
The latter monitors the level of stability by
exploiting information on the analytic structure of the reduction
identities, and residual instabilities
are stabilised on-the-fly through quadruple precision (qp).
This system is implemented at the level of individual operations. In this 
way, the usage of qp is restricted to a minimal part of the
calculations, which results in a huge speed-up as compared to complete qp
re-evaluations.
Thanks to these features, the on-the-fly reduction method makes it possible
to achieve an unprecedented level of numerical stability, both for multi-leg
NLO calculations with hard kinematics and for NNLO applications with
unresolved partons.

The structure of the open-loop
recursion~\cite{Cascioli:2011va,Buccioni:2017yxi} is model independent, and
the explicit form of its kernels depends only on the Lagrangian of the model
at hand.  The original implementation~\cite{hepforge} was applicable to any
SM process at NLO QCD, and the other major novelty of \OLtwo 
is the extension of NLO automation to the full
SM~\cite{Kallweit:2014xda,Kallweit:2017khh}, including any 
correction effect of $\ord(\alphaS)$ and $\ord(\alpha)$.\footnote{In the
following, by $\ord(\alpha)$ or EW corrections we mean the full
set of NLO corrections in the EW, Higgs and Yukawa couplings.}
In this respect, in this paper we present a detailed discussion of the interplay of QCD and
EW effects in scattering amplitudes with more than one quark chain, which
are relevant for LHC processes with two or more light jets.
In that case, Born
amplitudes consist of towers of terms of order 
$\alphaS^p\alpha^q$ 
with fixed total power $p+q$ but variable powers in the QCD and EW couplings.
In such cases, as is well known, QCD and EW interactions mix through
interference effects and, in general, NLO terms of fixed order
$\alphaS^P\alpha^Q$ 
involve correction effects of QCD {\it and} EW kind.
However, as we will point out, each NLO term of order
$\alphaS^P\alpha^Q$
is always dominated either by QCD
corrections to Born terms of 
order $\alphaS^{P-1}\alpha^Q$ {\it or} by EW corrections to
Born terms of 
order $\alphaS^{P}\alpha^{Q-1}$.

In this paper the renormalisation of the SM and its implementation in 
\OL are discussed in detail. 
In the QCD sector, quark masses and Yukawa couplings can be renormalised in
the on-shell and $\msbar$ schemes, and the $\alphaS$ counterterm can be
flexibly adapted to any flavour-number scheme.
The renormalisation of masses and couplings at 
$\ord(\alpha)$ is based on the on-shell scheme~\cite{Denner:1991kt}
and its extension to complex masses~\cite{Denner:2005fg} for 
off-shell unstable particles.
More precisely, in 
\OLtwo these two approaches are unified in a generic scheme that 
can address processes with combinations of on-shell and off-shell
unstable particles, such as for 
$pp\to t \bar t \ell^+ \ell^-$,
where $Z$-bosons occur as internal resonances, while
top quarks are on-shell external states. 
Besides UV counterterms, \OLtwo implements
also Catani--Seymour's $\mathbf{I}$-operator 
for the subtraction of infrared (IR)
singularities at $\ord(\alphaS)$~\cite{Catani:1996vz,Catani:2002hc}  
and $\ord(\alpha)$~\cite{Dittmaier:1999mb,Dittmaier:2008md,Gehrmann:2010ry,Kallweit:2017khh,Schonherr:2017qcj}.

%
For the definition of EW couplings, three different schemes based on the the
input parameters $\alpha(0)$, $\alpha(M_Z^2)$ and $G_\mu$ are supported. 
Moreover, \OLtwo implements an automated system for the optimal choice of
the coupling of on-shell and off-shell external photons.
Concerning the choice of $\alphaS$ and the renormalisation scale 
$\mur$, a new automated scale-variation mechanism
makes it possible to re-evaluate scattering amplitudes for 
multiple values of $\alphaS$ and $\mur$ with minimal CPU cost.

The \OLtwo program can be combined with any other
code by means of its native \Fortran and C/\cpp interfaces, which allow
one to exploit its functionalities in a flexible way.  Besides the choice of
processes and parameters, the interfaces support the calculation of LO, NLO,
and loop-induced matrix elements and building blocks thereof, as well as
various colour and spin correlators relevant for the subtraction of IR
singularities at NLO and NNLO.  Additional interface functions give access
to the SU(3) colour basis and the colour flow of tree amplitudes.
Besides its native interfaces, \OL offers also a standard interface in the
BLHA format~\cite{Binoth:2010xt,Alioli:2013nda}.

The \OL program can be used as a plug-in by the Monte Carlo programs \Sherpa{}~\cite{Gleisberg:2008ta,Bothmann:2019yzt},
\Powheg{}~\cite{Alioli:2010xd}, \Herwig{}~\cite{Bellm:2015jjp},
\Geneva~\cite{Alioli:2012fc}, and \Whizard{}~\cite{Kilian:2007gr}, which
possess built-in interfaces that control all relevant \OpenLoops
functionalities in a largely automated way, requiring only little user
intervention.
Moreover, \OL is used as a building block of \Matrix{}~\cite{Grazzini:2017mhc}
for the calculation of  NNLO QCD observables. In this context, 
the automation of EW corrections in \OLtwo 
opens the door to ubiquitous NLO\,QCD+NLO\,EW simulations in
\Sherpa{}~\cite{Kallweit:2015dum,Gutschow:2018tuk} and NNLO\,QCD+NLO\,EW calculations in
\Matrix{}~\cite{dibosonnnloew}.

The \OLtwo code is publicly available on the {\sc Hepforge} webpage\\
\rule{0pt}{3ex}    
\centerline{\url{https://openloops.hepforge.org}}
\rule{0pt}{3ex}and via the Git repository 
\url{https://gitlab.com/openloops/OpenLoops}.
%
It consists of a process-independent base code and a process library that
covers several hundred partonic processes, including essentially all
 relevant processes at the LHC.  The desired processes can be easily accessed
through an automated download mechanism.  The set of available processes is
continuously extended, and possible missing processes can be promptly generated
 by the authors upon request.

The paper is organised as follows.  \refse{sec:OLalg} presents  the
structure of the original open-loop recursion and the new
on-the-fly reduction algorithm. Numerical instabilities and
the new hybrid-precision system are discussed in detail.
\refse{se:automation} deals with general aspects of NLO calculations and
their automation in \OL. This includes 
the bookkeeping of towers of terms of variable order $\alphaS^p\alpha^q$,
the treatment of input parameters, optimal couplings for external photons,
the renormalisation of the SM at $\ord(\alphaS)$ and $\ord(\alpha)$, 
the on-shell and complex-mass schemes, and the $\mathbf{I}$-operator.
\refse{sec:program} provides instructions on how to use the program,
starting from installation and process selection, and including the various
interfaces for the calculation of matrix elements, colour/spin
correlators, and tree amplitudes in colour space.
Technical benchmarks concerning the speed and numerical stability of 
\OLtwo are presented in \refse{se:benchmarks}.
A detailed description of the syntax and usage of the \OL interfaces can
be found in the appendices.

While the paper as a whole serves as a detailed documentation of the algorithms implemented
in \OLtwo, Section~\ref{sec:program} together with Appendix~\ref{app:native}  
can be used alone as a manual.

\section{The \OpenLoops{} algorithm} \label{sec:OLalg}
\def\nhcs{N_{\mathrm{hcs}}}

The calculation of loop amplitudes in \OL proceeds through the
recursive construction of open loops and their reduction to 
master integrals.
In this section we outline two variants of this procedure: the original
open-loop method~\cite{Cascioli:2011va}, which was used throughout in
\OLone and is still used  for loop-induced
processes in \OLtwo, and the new on-the-fly method~\cite{Buccioni:2017yxi} used for
tree--loop interferences in \OLtwo.

\subsection{Scattering amplitudes and probability densities}

The main task carried out by \OpenLoops{} is the computation of
the colour and helicity-summed
scattering probability densities
\bea
\label{eq:Wtree}
\calW_{\tree}
&=&
\langle \calM_0|\calM_0\big\rangle \,=\,
\frac{1}{\nhcs}\sum_{\hel}\sum_{\col}|\calM_{0}|^2,\\
\qquad
\label{eq:Wloop}
\calW_{\onel}
&=&
2\,\re\, \langle \calM_0|\calM_1\big\rangle \,=\,
\frac{1}{\nhcs}\sum_{\hel}\sum_{\col} 2\,\re \Big[\calM_{0}^*\calM_{1}\Big],
\qquad \\
\label{eq:Wloop2}
\calW_{\onelsq}&=&
\langle \calM_1|\calM_1\big\rangle \,=\, 
\frac{1}{\nhcs}\sum_{\hel}\sum_{\col}|\calM_1|^2,
\eea
which consist of the various interference terms that involve the Born
amplitude $\calM_0$ and the one-loop amplitude $\calM_1$ for a certain
process selected by the user.
The usual summations and averaging over external helicities%
\footnote{In \OpenLoops it is also possible to 
select polarisations of external particles in \refeq{eq:Wtree}--\refeq{eq:Wloop2}, \ie 
to perform a sum only over a subset of the helicity configurations.}  
and colours, as
well as symmetry factors for identical particles, are included throughout
and implicitly understood in the bra--ket notation in \refeq{eq:Wtree}-\refeq{eq:Wloop2}.
The relevant average factors are encoded in 
\bea
\nhcs&=&
\left(\prod_{p\in\mathcal{P}_{\mathrm{out}}} 
n_p!\right)
\left(\prod_{i\in\calS_{\mathrm{in}}} 
N_{\mathrm{hel},i}
N_{\mathrm{col},i}\right)\,,
\eea
where $\calS_{\mathrm{in}}$ denotes the set of initial-state particles,
while
$N_{\mathrm{hel},i}$ and  $N_{\mathrm{col},i}$ are the number of helicity and
colour states of particle $i$.  The symmetry factors 
depend on the number $n_p$ of identical final-state particles. They 
are applied to all 
types  of final-state particles,
$p\in \mathcal{P}_{\mathrm{out}}$,
treating particles and anti-particles as
different types.

For standard processes with $\calM_{0}\neq 0$, 
leading-order (LO) 
cross sections involve only squared tree contributions $\calW_{\tree}$, while at 
next-to-leading order (NLO) 
virtual one-loop contributions $\calW_{\onel}$
and real-emission contributions of type $\calW_{\tree}$ with 
one additional parton are needed.
The squared one-loop probability density
$\calW_{\onelsq}$ is the main LO building block
for loop-induced processes,
\ie processes with $\calM_{0}=0$. For the calculation of such processes
at NLO also $\calW_{\onelsq}$-type densities with 
one additional parton are needed.
  Otherwise $\calW_{\onelsq}$  is
relevant as ingredient of next-to-next-to-leading order
(NNLO) calculations.

In \OpenLoops,  $L$-loop matrix elements $\calM_L$ are computed 
in terms of Feynman diagrams, whose structures are generated with {\sc Feynarts}~\cite{Hahn:2000kx}. 
The Feynman diagrams are expressed as helicity amplitudes,
\bea
\label{eq:amplitudes}
\calM_{L}(\heli)
&=&
\sum_{\calI\in\Omega_{L}} 
\calM_{L}(\calI,\heli)
\,=\,
\sum_{\calI\in\Omega_{L}} 
\calC(\calI)\,\calA_L(\calI,\heli)\,,
\qquad (L=0,1)
\eea
where $\Omega_{L}$ is the set of all $L$-loop Feynman diagrams,
$\heli$ describes a specific helicity configuration of the external
particles, and each  diagram $\calI$ is factorised into a colour factor $\calC(\calI)$ and 
a colour-stripped diagram amplitude%
\footnote{Quartic gluon couplings involving three different colour structures are split into colour-factorised contributions which are treated as separate diagrams.}
$\calA_L(\calI,\heli)$.
The colour structures $\calC(\calI)$  are algebraically reduced 
to a standard colour basis $\{\calC_i\}$ (see \refse{sec:colourbasis}), 
\be
\calC(\calI)=\sum\limits_i a_i(\calI)\,\calC_i,
\label{eq:colfactdec}
\ee
where scattering amplitudes take the form
\bea
\calM_{L}(\heli) &=&
\sum_{i} \calC_i\, \calA_{L}^{(i)}(\heli)\,,
\label{eq:colourvector}
\eea
and colour-summed interferences in 
\refeq{eq:Wtree}-\refeq{eq:Wloop2}
are built 
by means of the colour-interference matrix
\bea
\calK_{ij} &=&
\sum_{\col}\,\calC_i^\dagger\,\calC_j\,.
\label{eq:colintB}
\eea

In the following we focus on the construction of the colour-stripped amplitudes $\calA_L(\calI,\heli)$.

\subsection{Tree amplitudes} \label{sec:treeamp}

At tree level, each colour-stripped Feynman diagram is built by
contracting two subtrees that are connected through a certain cut
propagator,%
\footnote{The Feynman diagrams in this paper are drawn with {\sc Axodraw}~\cite{Vermaseren:1994je}.}
\be
\calA_0(\calI,h)
\;=\;\;
\vcenter{\hbox{\scalebox{1.}{\includegraphics[width=35mm]{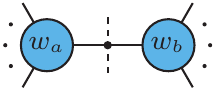}}}}\;\;
\,=\,
w^{\sigma_a}_a(k_a,h_a)
\,\delta_{\sigma_a\sigma_b}
\widetilde{w}^{\sigma_b}_b(k_b,h_b)\,.
\label{eq:cuttreeA}
\ee
Here $\momk{a}=-\momk{b}$ and $\sigma_a,\sigma_b$ 
are the momenta and spinor/Lorentz indices of the subtrees, while
$h_a,h_b$ denote the helicity configurations of the external 
particles connected to the respective subtrees.\footnote{See \cite{Buccioni:2017yxi} for more details.}
The tilde in $\widetilde{w}_b$ marks the absence of the cut propagator, 
which is included in $w_a$.
All relevant subtrees are generated 
through a numerical recursion 
that starts from the external wave functions and 
connects an increasing number of external particles
through operations of the form
\be
w^{\sigma_a}_a(k_a,h_a)
\;=\;\;
\vcenter{\hbox{\scalebox{1.}{\includegraphics[width=60mm]{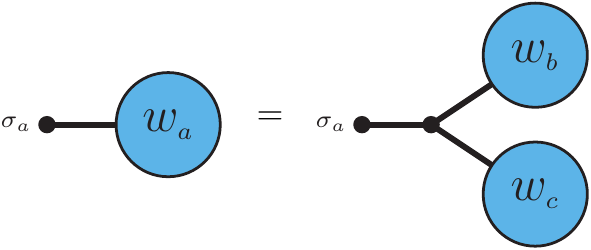}}}}
\;=\,
\f{X_{\sigma_b\sigma_c}^{\sigma_a}(\momk{b},\momk{c})}{\momk{a}^2-\mass{a}^2}\;
w^{\sigma_b}_b(\momk{b},\heli_{b})\;
w^{\sigma_c}_c(\momk{c},\heli_{c}){}\,.
\label{eq:treerecursionA}
\ee
The tensor $X_{\sigma_b\sigma_c}^{\sigma_a}$ corresponds to the 
triple vertex that connects  $w_a, w_b, w_c$, 
combined with the numerator of the propagator attached to $w_a$. 
For quartic vertices an analogous relation is used.
Each step needs to be carried out 
for all independent helicity configurations $\heli_{b}, \heli_{c}$.
The resulting tree recursion is implemented in an efficient way
by caching the amplitudes of subtrees that contribute to 
multiple Feynman diagrams.

\subsection{One-loop amplitudes} \label{sec:loopamp}
\def\fourdim{4\mathrm{D}}

Renormalised one-loop amplitudes are split into three building blocks,
\be
\calM_{1}(h)=\calM_{1,\fourdim}(h)+\calM_{1,R_2}{(h)}+\calM_{1,\text{CT}}{(h)},
\label{eq:diaA1dbar}
\ee
where $\calM_{1,\text{CT}}$ denotes UV counter-terms, while 
bare one-loop amplitudes are 
decomposed into a contribution 
that is computed with four-dimensional 
loop numerator ($\calM_{1,\fourdim}$)
plus a so-called $R_2$ rational term ($\calM_{1,R_2}$)
stemming form the $(D-4)$-dimensional part of loop numerators.
The latter is reconstructed via $R_2$
counter-terms~\cite{Ossola:2008xq,
Draggiotis:2009yb,
Garzelli:2009is,
Garzelli:2010qm, 
Garzelli:2010fq, 
Shao:2011tg, 
Pittau:2011qp,
Page:2013xla},
and $\calM_{1,R_2}+\calM_{1,\text{CT}}$ are generated in a
similar way as tree amplitudes.

The remaining part, $\calM_{1,\fourdim}$,
is constructed in terms of colour-stripped loop 
diagrams,
\be 
{\mathcal{A}}_{1}(\calI_N,\heli) \;=\;
\int\!\rd^D\!\momq\, \f{\Tr\Big[{\calN}(\mathcal{I}_{N},q,\heli)\Big]}{\Dbar{0} \Dbar{1}\cdots \Dbar{N-1}}
\quad=\;
\vcenter{\hbox{\scalebox{1.}{\includegraphics[height=33mm]{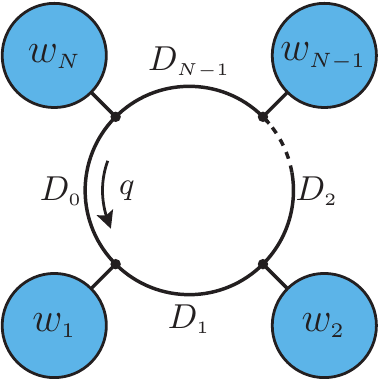}}}}\quad{},
\label{eq:OLfulldia}
\ee
with four-dimensional numerators
${\calN}(\mathcal{I}_{N},q,\heli)$
and denominators $\Dbar{i}(\momq)=(\momq +
\momp{i})^2-\mass{i}^2$,
where the bar is used for quantities in 
$D$ dimensions, and the 
$(D-4)$-dimensional part of the loop momentum is denoted
$\tilq=\momq-q$. 
The trace represents the contraction of spinor/Lorentz 
indices along the loop, and the index $\calI_N$ stands for the $N$-point topology at
hand. 

The numerator is computed by cut-opening the loop 
at a certain propagator, which results into a tree-like structure 
consisting of a product of loop segments,
\be
\Big[\calN(q,\heli)\Big]_{\beta_0}^{\beta_N}= \vcenter{\hbox{\includegraphics[height=33mm]{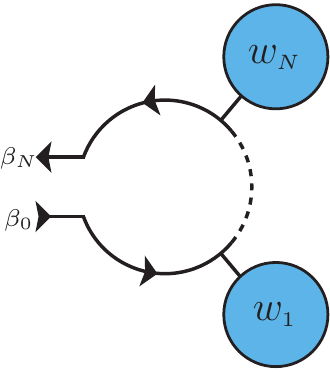}}}=
\Big[\seg_1(q,\heli_1)\Big]_{\beta_0}^{\beta_1}\,
\Big[\seg_2(q,\heli_2)\Big]_{\beta_1}^{\beta_2}\cdots
\Big[\seg_{N}(q,\heli_{N})\Big]_{\beta_{N-1}}^{\beta_N}, \label{eq:fac}
\ee
where $\beta_{0},\beta_{N}$ are the spinor/Lorentz indices of the cut propagator.
Loop segments that are connected to the loop via triple
vertices have the form
\bea
& &\Big[\seg_{i}(q,\heli_i)\Big]_{\beta_{i-1}}^{\beta_{i}}
\,=\quad
\raisebox{3mm}{\parbox{1.45\defheight}{
\includegraphics[height=\heightB]{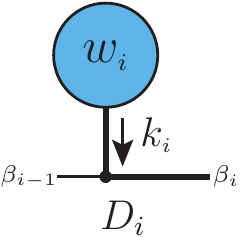}}}= \label{eq:seg3point}
\,\Bigg\{\Big[Y_{\sigma_i}^{i}\Big]_{\beta_{i-1}}^{\beta_{i}}
+ \Big[Z_{\nu;\sigma_i}^{i}\Big]_{\beta_{i-1}}^{\beta_{i}}\,q^\nu
\Bigg\}\, w^{\sigma_i}_i(\momk{i},\heli_i)\,, 
\eea
where an external subtree $w_i$
is connected to a loop vertex and 
to the adjacent loop propagator.
The latter correspond to a 
rank-one polynomial in the loop momentum
with coefficients $Y$ and $Z$.
A similar relation is used for quartic vertices.

The loop numerator is constructed by 
attaching the various segments to each other
through recursive {\it dressing} steps, 
\be
\calN_k(q,\helihat_k)=\calN_{k-1}(q,\helihat_{k-1})\seg_k(q,\heli_k), 
\qquad\mbox{for}\quad
 k=1,\dots, N{},
\label{eq:OLrec}
\ee
starting from the initial condition $\calN_{0}=\idop$.
The labels $\heli_k$ and $\helihat_k$
correspond, respectively,  to the helicity configuration of the external legs entering the
$k^{\mathrm{th}}$ segments and the first $k$ segments.
The partially dressed numerator \refeq{eq:OLrec} is called an open loop.
Schematically it can be represented as
\bea \calN_{k}(q,\helihat_k) &=&
\;\prod\limits_{i=1}^{k}\seg_i(q,\heli_i)
\;=\;
\raisebox{3mm}{\parbox{75mm}{
\includegraphics[height=18mm]{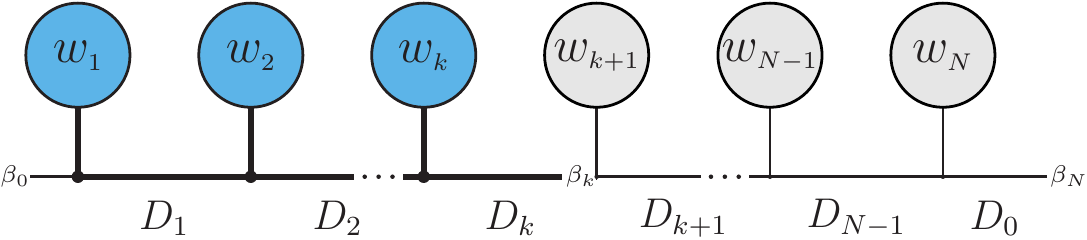}}} 
\eea 

where blue and grey blobs correspond, respectively, 
to those loop segments that are already dressed 
and remain to be dressed.
Each open loop is a polynomial in $q$,
\bea \calN_{k}(q,\helihat_k) &=&
\sum\limits_{r=0}^R
\calN^{(k)}_{\mu_1\dots\mu_r}(\helihat_k)
\,q^{\mu_1}\cdots q^{\mu_r}\,, 
\label{eq:partnum}
\eea 
and all dressing steps are implemented at the 
level of the open-loop tensor coefficients $\calN^{(k)}_{\mu_1\dots\mu_r}$.
%

\subsection{Reduction to master integrals} \label{sec:tensorreduction}

\begin{figure}[t!]\begin{center} \begin{tabular}{ccc}
\includegraphics[height=50mm]{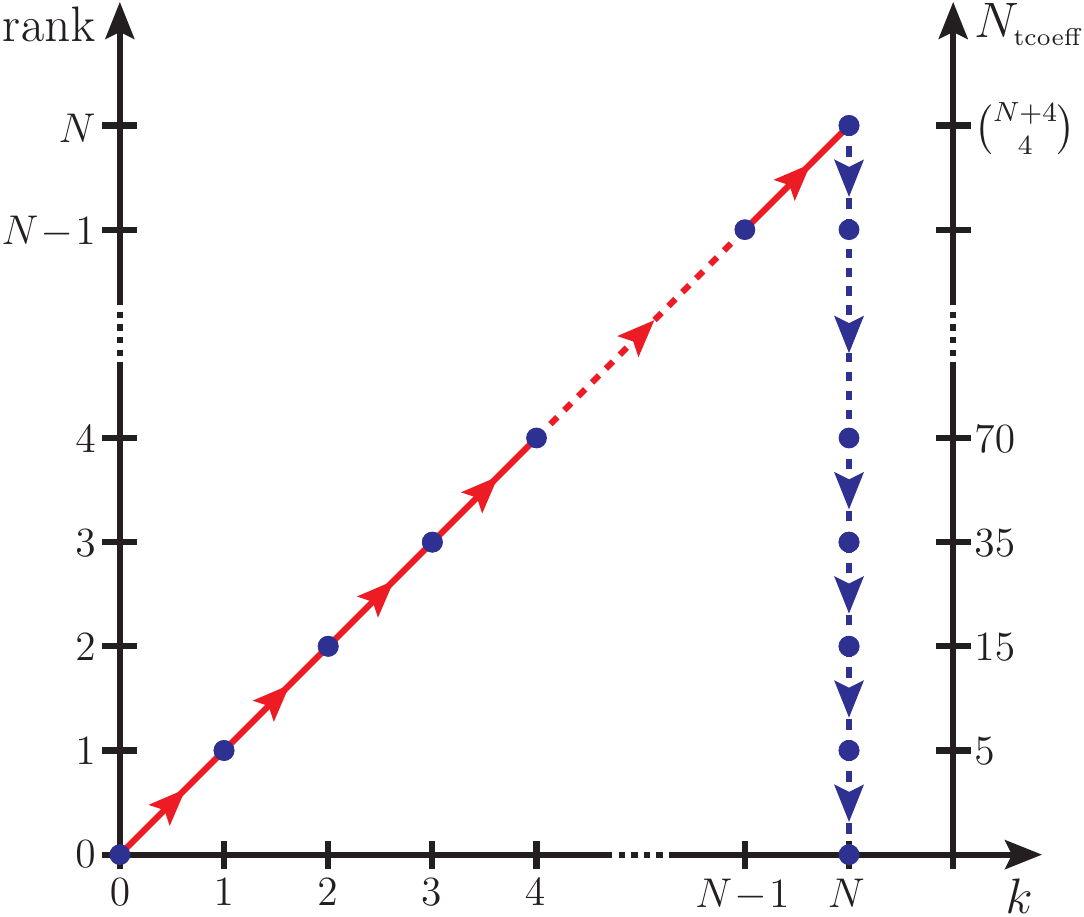} & \quad\; &
\includegraphics[height=50mm]{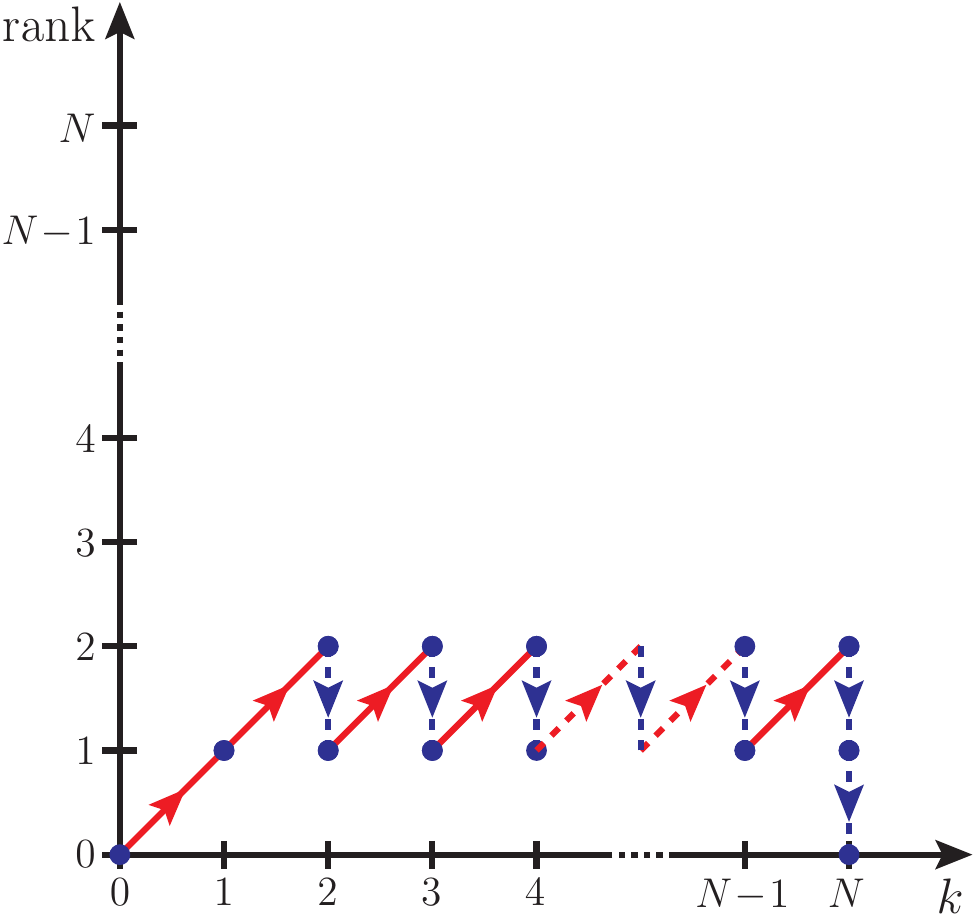}\\[3ex] (a) A posteriori reduction & & (b) On-the-fly reduction
 \end{tabular} \end{center}
\caption{Evolution of the tensor rank and the number
of open-loop tensor coefficients
(right vertical axis) as a function of the number $k$ of dressed segments
during the open-loop recursion.  
The red diagonal lines illustrate the dressing steps,
 and the blue vertical lines the reduction steps.
\label{fig:OL1OL2_r_vs_n} }
\end{figure}

In \OpenLoops{} the reduction of loop amplitudes to 
master integrals is carried out with two different methods.
Squared loop amplitudes 
and tree-loop interferences in the Higgs Effective Field Theory
(HEFT)\footnote{By HEFT we mean effective Higgs--gluon and Higgs--quark interactions in
the heavy-top limit.
}
are handled along the lines of the
original open-loop approach~\cite{Cascioli:2011va}, 
where the reduction is performed 
a posteriori of the dressing recursion.  Since every dressing step can increase
the tensor rank by one (see Fig.~\ref{fig:OL1OL2_r_vs_n}~a), 
this generates intermediate objects of high tensor rank, \ie high
complexity, with a negative impact on CPU speed.
In contrast, all other tree--loop interferences
 are computed with the on-the-fly
reduction approach%
~\cite{Buccioni:2017yxi}, 
where dressing steps are
interleaved with integrand reduction steps in such a way that the tensor rank,
and thus the complexity, remain low at all stages of the
calculation (see Fig.~\ref{fig:OL1OL2_r_vs_n}~b).

\subsubsection{A posteriori reduction}
\label{se:postred}
The a posteriori reduction to scalar integrals is done 
by means of external tools.
By default, the reduction is 
performed at the level of tensor integrals,
\bea 
T_N^{\mu_1\cdots \mu_R} &=&
\int\!\rd^D\!\momq\, 
\frac{q^{\mu_1}\cdots q^{\mu_R}}{\Dbar{0} \Dbar{1}\cdots \Dbar{N-1}}\,,
\label{eq:tensorint}
\eea
using the \Collier library~\cite{Denner:2016kdg}, which implements 
the Denner--Dittmaier reduction
techniques~\cite{Denner:2002ii,Denner:2005nn} as well as the scalar
integrals of~\cite{Denner:2010tr}.
Alternatively, the reduction can be performed at the integrand level
using  \Cuttools~\cite{Ossola:2007ax}, which implements the
OPP reduction method~\cite{Ossola:2006us}, 
in combination with the {\sc OneLOop}
library~\cite{vanHameren:2010cp} for scalar integrals.

\subsubsection{On-the-fly reduction} \label{sec:ofr}

In the on-the-fly approach, the dressing of open loops is interleaved with
reduction steps. The latter are applied in such a way that the 
tensor rank never exceeds two.

For objects with more then three loop propagators, $D_0,D_1,D_2,D_3,\dots$,
the tensor rank is reduced using an integrand-reduction identity~\cite{delAguila:2004nf}
of the form
\bea
q^\mu q^\nu  &=&   \sum\limits_{i=-1}^{3}\lb A^{\mu\nu}_{i} + B^{\mu\nu}_{i,\lambda}\,q^{\lambda} \rb D_i {}, \label{eq:qqred}
 \qquad 
\mbox{with}\qquad
D_i=\begin{cases}
             1 &\mbox{for}\; i=-1,\\
             (q + \momp{i})^2-\mass{i}^2 &\mbox{for}\;i\ge 0,
             \end{cases}
\label{eq:ofrid}
\eea
where the coefficients $A^{\mu\nu}_{i}$ and $B^{\mu\nu}_{i,\lambda}$ 
depend on the internal masses and external momenta.
The four-dimensional $D_i$ terms on the rhs of \refeq{eq:ofrid}
are cancelled against the $D$-dimensional 
loop denominators.
This gives rise to $\tilq^2$ dependent terms, $D_i/\Dbar{j}=1-\tilq^2/\Dbar{j}$, 
which are consistently taken into account and 
result into rational contributions of kind $R_1$~\cite{delAguila:2004nf,Buccioni:2017yxi}.
Note that the reduction \refeq{eq:ofrid}
and the pinching of propagators
can be carried out as soon as rank two is 
reached, irrespective of which 
loop segments are still undressed.
Every reduction step generates 
four new pinched sub-topologies, and the
proliferation of pinched objects is avoided 
by means of the merging approach described in
Section~\ref{sec:blprocesses}.

Rank-two open loops with only three loop denominators can be reduced 
on-the-fly in a similar way as open loops with more than three propagators~\cite{Buccioni:2017yxi}.
The remaining reducible integrals have the following number of propagators
$N$ and tensor rank $R$: $N\ge 5$ and $R=1,0$;
$N=4,3$ and $R=1$;
$N=2$ and $R=2,1$.
For their reduction to master integrals
we use a combination of integral reduction and OPP reduction
identities~\cite{Buccioni:2017yxi}. 
Master integrals are evaluated with \Collier \cite{Denner:2016kdg},
which is the default in double precision, 
or \OneLoop \cite{vanHameren:2010cp}, which is the default in quadruple precision.
%

\subsection{Tree--loop interference} \label{sec:blprocesses}

In the following we outline the calculation of 
tree--loop interferences~\refeq{eq:Wloop} according to the original open-loop algorithm
and with the on-the-fly approach~\cite{Buccioni:2017yxi}. The latter is used
by default in \OLtwo.
In both cases, the colour treatment
is based on the factorisation of colour structures at the level of individual loop diagrams,
$\calM_{1}(\calI,\heli)
=
\calC(\calI)\,\calA_1(\calI,\heli)$. This makes it possible
to cast the interference of loop diagrams 
with the Born amplitude into the form
\bea
\label{eq:treeloopdiagint}
2 \sum_{\col}\calM^*_0(\heli) \calM_{1}(\calI,\heli)
&=&
\calU_0(\calI,\heli)\,\calA_1(\calI,\heli)\,,
\eea
where 
$\calA_1(\calI,\heli)$ is the colour-stripped
loop amplitude, and the colour information is entirely absorbed into
 the colour-summed  interference factor
\bea
\calU_0(\calI,\heli) &=&
2\left(\sum_{\col}\calM^*_0(h)\,\calC(\calI)\right)
\,=\,
2\sum_{i,j} \left[\calA_0^{(i)}(\heli)\right]^*\,\calK_{ij}
a_j(\calI)\,,
\label{eq:initOL2}
\eea
where $a_j(\calI)$, $\calA_0^{(i)}(\heli)$, and
$\calK_{ij}$ are defined in \refeq{eq:colfactdec}--\refeq{eq:colintB}.
In this way, as detailed below, the full 
tree--loop interference can be constructed in terms of 
colour-stripped or colour-summed objects.

\begin{figure}[t]\begin{center}
$\calN_k(\mathcal{I}_{N})=
\raisebox{7mm}{\parbox{0.35\textwidth}{\includegraphics[height=15mm]{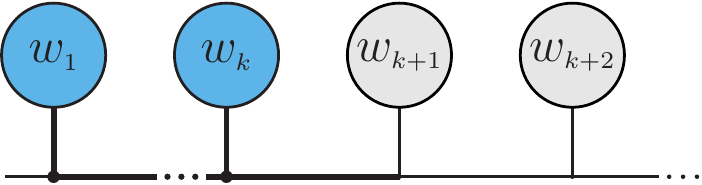}}}\quad
\calN_k(\tilde{\mathcal{I}}_{N-1})=
\raisebox{7mm}{\parbox{0.35\textwidth}{\includegraphics[height=15mm]{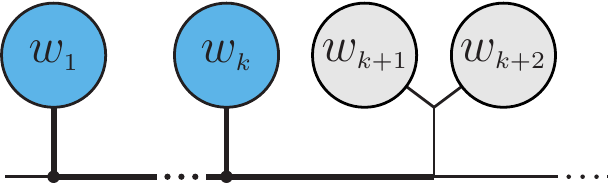}}}$
\end{center}
\caption{Example of parent-child relation between open loops.
The parent $N$-point diagram
$\calI_{N}$ and  the child $(N-1)$-point diagram $\tilde{\calI}_{N-1}$
share the first $k$ segments (blue blobs). Thus 
$\calN_k(\mathcal{I}_{N},q)$ and 
$\calN_k(\tilde{\mathcal{I}}_{N-1},q)$ are identical and 
need to be constructed only once.
}
\label{fig:parentchild} 
\end{figure}

\subsubsection{Parent-child algorithm} 
\label{se:parentchildalgo}
In the original open-loop approach, tree--loop interference
contributions of type~\refeq{eq:treeloopdiagint} are constructed as follows.

\bit
\item[(i)] The numerator of a colour-stripped  $N$-point loop diagram~\refeq{eq:OLfulldia}
is constructed as outlined in \refse{sec:loopamp}, \ie starting from
$\calN_{0}=\idop$ and applying $N$ dressing steps
of type \refeq{eq:OLrec}.

\item[(ii)] In general, open loops with higher number $N$ of loop propagators
do not need to be built from scratch, but can
be constructed starting form pre-computed 
open loops with lower $N$ exploiting  
{\it parent--child relations}~\cite{Cascioli:2011va} 
as illustrated in \reffi{fig:parentchild}.
The efficiency of the parent--child approach 
is maximised by means of {\it cutting rules}
that set the position of the cut propagator and the dressing
direction in a way that favours parent--child matching 
(for details see \cite{Cascioli:2011va,Buccioni:2017yxi}).

\item[(iii)] After the last dressing step, the loop numerator is closed by
taking the trace and, for every helicity state $\heli$,
the colour-summed Born interference~\refeq{eq:treeloopdiagint}
is built as
\bea
\calU(\calI_N,q,\heli) &=&
\calU_0(\calI_{N},\heli)
\Tr\Big[\calN(\calI_{N},q,\heli)\Big]\,.
\label{eq:colsumint}
\eea

\item[(iv)]  Helicity sums are performed,
and the set of loop diagrams with 
the same one-loop topology $t=\{D_{0},\ldots,D_{N-1}\}$, denoted $\Omega_N(t)$, 
is combined to form a 
single numerator,
\bea
\calV(t,q)&=&\sum\limits_{\heli}\sum\limits_{\calI_N\in\Omega_N(t)}
\calU(\calI_N,q,\heli)\,.
\label{eq:tensorsums} 
\eea 

\item[(v)] The corresponding loop integral,
\bea 
\calW_{01}(t) &=& \int\!\rd^D\!\momq\, \f{\calV(t,q)}{\Dbar{0} \Dbar{1}\cdots
\Dbar{N-1}}\,,
\eea
is reduced to master integrals as described in 
\refse{se:postred}, and all topologies are summed.

\eit

All operations in (i)--(v) are performed at the level of
open-loop tensor coefficients.

\subsubsection{On-the-fly algorithm} 
\label{se:ontheflyalgo}

The on-the-fly 
construction of Born-loop interferences proceeds through objects of type
\bea
\calU_k(\calI_N,q,\helicheck_k)
&=&
 \sum_{\helihat_k}\;
\calU_0(\calI_{N},\heli)\,
\calN_k(\calI_{N},q,\helihat_k)\,,
\label{eq:vOLdefwohel}
\eea
where the partially dressed open loops, $\calN_k(\calI_{N},q,\helihat_k)$, are
always interfered with the Born amplitude, summed over colours, and also over
the helicities $\helihat_k$ of all segments that are already dressed. The helicities of the remaining undressed segments
are labelled with the index $\helicheck_k$.
As outlined in the following, the algorithm interleaves dressing, merging
and reduction operations in a way that keeps the tensor rank always low and
avoids the proliferation of pinched objects 
that arise from the reduction.
For a detailed description see~\cite{Buccioni:2017yxi}.

\bit
\item[(i)] The generalised open loops~\refeq{eq:vOLdefwohel} are constructed through 
subsequent dressing steps 
\bea
\calU_k(\calI_N,q,\helicheck_k) &=&
\sum_{\heli_k}\;\calU_{k-1}(\calI_{N},q,\helicheck_{k-1})
\seg_k(q,\heli_{k})\,,
\eea
starting from $\calU_0(\calI_N,q,\helicheck_0)= \calU_0(\calI_N,\heli)$.
The summation over the helicities $\heli_k$ is performed {\it on-the-fly}
after the dressing of the related segment.
This results in a 
reduction of helicity degrees of freedom, 
and thus of the number of required operations, 
at each dressing step.

\begin{figure}[t]\begin{center}
$\calV_k(\Omega_N)=
\raisebox{7mm}{\parbox{0.35\textwidth}{\includegraphics[height=15mm]{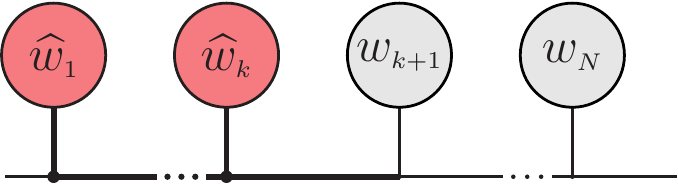}}}= \sum\limits_n\;\calU_0(\calI_N^{(n)})\;
\raisebox{7mm}{\parbox{0.35\textwidth}{\includegraphics[height=15mm]{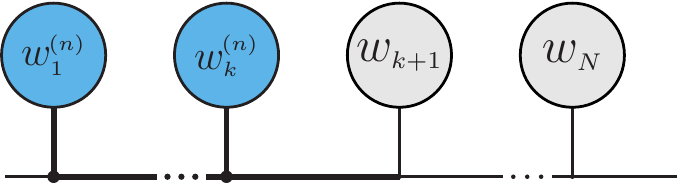}}}$
\end{center}
\caption{Schematic representation of on-the-fly merging.
Open loops with the same loop topology and the same undressed segments 
(grey blobs) are combined in a single object.}
\label{fig:merging}
\end{figure}

\item[(ii)] Before each new dressing step,
the set $\Omega_N=\{\calI_N^{(n)}\}$ of open loops 
with the same loop topology and the same
undressed segments 
is combined into a single object,
\bea 
\calV_k(\Omega_N,q,\helicheck_k)
&=&
\sum\limits_{n}\, \calU_k(\calI_N^{(n)},q,\helicheck_k)\,.
\label{eq:merging} 
\eea
In this way, the remaining dressing operations for the objects in $\Omega_N$
need to be performed only once.
This procedure, called {\it on-the-fly merging}, is illustrated 
in \reffi{fig:merging}. It 
plays an analogous role as the parent-child approach in 
\refse{se:parentchildalgo}, and its 
efficiency is maximised by means of 
{\it cutting rules} tailored to the needs of merging.

\item[(iii)] Open-loop objects of type \eqref{eq:merging} 
with  more than three loop propagators
are {\it reduced on-the-fly} using the integrand-reduction identity~\refeq{eq:ofrid}. This generates new 
open loops of the form
\bea
\frac{\calV_{k}(\Omega^k_N, \bar q)
}{\Dbar{0}\cdots \Dbar{3} \cdots \Dbar{N-1}}
&=&
\sum_{j=-1}^3
\frac{\calV_{k}(\Omega^k_N[j],\bar q)
}{\Dbar{0}\cdots
\Djslash\cdots
\Dbar{3} \cdots \Dbar{N-1}}\,,
\label{eq:OFRmasterformula}
\eea
where $\Djslash$ denotes a pinched propagator. 
This reduction is applied to rank-two objects
directly before dressing steps that would
otherwise increase the rank to three.
In order to avoid the proliferation of new objects,
pinched open loops are merged on-the-fly 
with other open loops stemming from lower-point
Feynman diagrams or from other pinched open loops~\cite{Buccioni:2017yxi}.
The numerators in~\refeq{eq:OFRmasterformula}
have the form
\bea
\calV_k(\Omega,\bar q) &=&
\sum_{s,r}
\calV^s_{k;\mu_1\dots\mu_r}(\Omega)\,
q^{\mu_1}\cdots q^{\mu_r}\,
(\tilde{q}^2)^s\,,
\label{eq:qtildepolynomials}
\eea
where $\tilq^2$ terms that arise from pinched propagators (see \refse{sec:ofr}) 
 are retained in all UV divergent
integrals and lead to $R_1$ rational terms.

\eit

Steps (i)---(iii) are iterated until the loop is entirely dressed.%
\footnote{Note that it is also possible to apply only (i)--(ii). This leads
to the same objects $\calV(t,q)$ as in \eqref{eq:tensorsums}, which can
then be reduced a posteriori.}
 
\bit

\item[(iv)] At this stage, the loops are closed by taking the trace, and the resulting loop
integrals,
\bea 
\calW_{01}(\Omega) &=& \int\!\rd^D\!\momq\,
\f{\Tr\left[\calV(\Omega, \bar q)\right]}{\Dbar{0} \Dbar{1}\cdots
\Dbar{N-1}}\,,
\eea
are reduced to master integrals upon extraction of $R_1$ terms, as described at the end of 
\refse{sec:ofr}.
Finally, all topologies are summed.

\eit

As demonstrated in \refse{se:benchmarks}, 
the on-the-fly approach yields significant efficiency
improvements wrt the original open-loop algorithm.
Moreover, based on the one-the-fly reduction
algorithm, \OLtwo has been equipped with an automated stability system that
cures Gram-determinant instabilities with unprecedented
efficiency (see \refse{sec:numstab}).

\subsection{Squared loop amplitudes} \label{sec:l2processes}

As outlined in the following, the calculation of squared loop amplitudes
\refeq{eq:Wloop2} is organised along the same lines of the parent-child
algorithm of \refse{se:parentchildalgo} but with a different colour
treatment.

\bit
\item[(i)] The numerators of  colour-stripped
loop diagrams are constructed with the dressing recursion 
\refeq{eq:OLrec} exploiting parent--child relations.

\item[(ii)] After the last dressing step, loop numerators are  closed by
taking the trace,  and
colour-stripped diagrams expressed in terms of integrals
$T_N^{\mu_1\cdots \mu_r}$~\refeq{eq:tensorint},
\bea
{\mathcal{A}}_{1}(\calI_N,\heli) \,=\,
\int\!\rd^D\!\momq\, \f{\Tr\Big[{\calN}(\mathcal{I}_{N},q,\heli)\Big]}{\Dbar{0} \Dbar{1}\cdots \Dbar{N-1}}
&=&
\sum\limits_{r}\,
\Tr\Big[\calN_{\mu_1\dots\mu_r}(\calI_N, \heli)\Big]\,
T_N^{\mu_1\cdots \mu_r}\,,
\label{eq:partnum2}
\eea
which are then computed with \Collier. While the
$\calN_{\mu_1\dots\mu_r}(\calI_N, \heli)$
coefficients need to be evaluated 
for every helicity state $h$,
the reduction is done only once -- and thus very efficiently -- at the 
level of the $h$-independent tensor integrals.

\item[(iii)]  Individual colour-stripped diagram amplitudes are combined with 
the corresponding colour structure and converted into colour vectors in the colour 
basis $\{\calC_i\}$,
\bea
\calM_1(\calI_N,\heli) &=& 
\calC(\calI_N) \calA_1(\calI_N,\heli)
\,=\,
\sum_i \calC_i\,\calA^{(i)}_1(\calI_N,\heli)\,.
\eea
Then, summing all diagrams yields the full one-loop colour vector
\bea
\calA_1^{(i)}(\heli) &=& 
\sum_{\calI} \calA^{(i)}_1(\calI,\heli)\,.
\eea

\item[(iv)] 
Finally, the helicity/colour summed squared loop amplitude is
built though the colour-interference matrix~\refeq{eq:colintB}
as
\bea
\calW_{\onelsq} &=& 
\frac{1}{\nhcs}
\sum_{\heli}\sum_{\col}
\calM_1^*(\heli)\calM_1(\heli)\,=\,
\frac{1}{\nhcs}
 \sum\limits_{\heli}\sum\limits_{i,j} K_{ij}\, \big[\calA_1^{(i)}(h)\big]^* \calA_1^{(j)}(h).
\eea

\eit

\subsection{Numerical stability} \label{sec:numstab}

The reduction of one-loop amplitudes
to scalar integrals suffers from numerical instabilities 
in exceptional phase-space regions. Such instabilities are related to small
Gram determinants of the form
\bea
\Delta_{1\dots n} &=& \Delta(p_1,\dots,p_n) \,=\, \det\big(p_i\cdot
p_j\big)_{i,j=1,\dots,n}\,,
\label{eq:Gram}
\eea
where $p_k$ are the external momenta in the
loop propagators $D_k$.
In regions where rank-two and rank-three Gram determinants become small, the
objects that result from the pinching of propagators
can be enhanced by spurious
$1/\Delta$ singularities.  At the end, when all pinched objects are
combined and the integrals evaluated, such singularities disappear.
However, this cancellation can be so severe that all significant digits are lost, 
and the amplitude output can be inflated in an uncontrolled way by orders of magnitude.
This calls for an automated system capable of detecting and curing all
relevant instabilities in a reliable way.
This is especially important for multi-particle and multi-scale NLO
calculations, and even more for NNLO applications, which 
require high numerical accuracy in regions where one external 
parton becomes unresolved, thereby inflating 
spurious poles.

In principle, numerical accuracy can be augmented 
through quadruple precision (qp) arithmetic.  But the resulting CPU
overhead, of about two orders of magnitude, is often prohibitive. 
In \OL,  numerical instabilities are thus addressed as much as possible in 
double precision (dp) using analytic methods.
In \OLone, as detailed below, numerical instabilities are
avoided by means of the~\Collier library~\cite{Denner:2016kdg} in combination with a
stability rescue system that makes use of \Cuttools~\cite{Ossola:2007ax}
in qp.
In \OLtwo, loop-induced processes are handled along the same lines,
while standard NLO calculations are carried out with the new on-the-fly reduction
algorithm,
which is equipped with its own stability system (see~\refse{se:OL2stab}).
The latter combines analytic techniques together with a new hybrid-precision
system that uses qp in a highly targeted way,
requiring only a tiny CPU overhead as compared to 
a complete qp re-evaluation.

An additional source of numerical instabilities originates from the violation
of on-shell relations or total momentum conservation of external particles,
i.e.\ due to the quality of the provided phase-space point. To this end
before amplitude evaluation on-shell conditions and momentum conservation
are checked. A warning is printed when
these conditions are violated beyond a certain relative threshold, which can
be altered via the parameter \texttt{psp\_tolerance} (default=$10^{-9}$).
Additionally, we apply a ``cleaning procedure'' which ensures kinematic constraints
of the phase-space up to double precision, rsp.\ qp where applicable.

\subsubsection{Stability rescue system}
\label{se:OL1stab}

In the original open-loop algorithm---which was used throughout in 
\OLone{} and is still used in \OLtwo{} for squared loop amplitudes and
tree--loop interferences in the HEFT---the reduction to scalar integrals is entirely based 
on external libraries, and the best option is to carry out the reduction of tensor
integrals using the~\Collier library~\cite{Denner:2016kdg}.
In the vicinity of spurious poles, \Collier cures 
numerical instabilities by means of expansions in the Gram
determinants and alternative reduction
methods~\cite{Denner:2002ii,Denner:2005nn}.
Such analytic techniques are applied in a fully automated way, and
the resulting level of numerical stability is generally very good.
Alternatively, the reduction 
can be performed at the integrand level 
using \Cuttools~\cite{Ossola:2007ax},
but this option is mainly used as rescue system in qp, since \Cuttools{}
does not dispose of any mechanism to avoid instabilities in dp.

In the calculation of tree--loop interferences,
numerical instabilities are monitored and cured by means of 
an automated rescue system based on the following strategy.

\bit

\item[(i)] The stability of tensor integrals is assessed by comparing 
the two independent \Collier implementations of the tensor reduction, 
\Coli-\Collier (default) and \DD-\Collier. This test can be applied to all phase-space points or restricted
to a certain fraction of points with the highest virtual
$K$-factor\footnote{This approach allows one 
to trigger the most extreme instabilities,
where the $K$-factor is altered by $\ord(1)$ or more.
}
Given the desired
fraction, the points to be tested are automatically selected by sampling the
distribution in the $K$-factor at runtime.

\item[(ii)] Points that are classified as unstable are re-evaluated
in qp using~{\sc CutTools} and
{\sc OneLOop}.

\item[(iii)] In \Cuttools, numerical instabilities can remain significant even in
qp.  Their magnitude is estimated through a so-called rescaling
test, where
one-loop amplitudes  are computed with rescaled masses, dimensionful couplings and momenta
and scaled back according to the mass dimensionality of the amplitude.

\eit

In this approach, the re-evaluation of the amplitude for stability tests
causes a non-negligible CPU overhead.  Moreover, additional re-evaluations
of the full amplitude in qp are very CPU intensive.  Fortunately, thanks to
the high stability of~\Collier, they are typically needed only for a tiny
fraction of phase-space points.  However, the usage of qp strongly depends
on the complexity of the process, and for challenging multi-scale
NLO calculations and NNLO applications it can become quite significant.

In the case of squared loop amplitudes, the qp rescue with \Cuttools{} is
disabled, because of the inefficiency of OPP reduction for
loop-squared amplitudes. This is due to the fact that all helicity and colour
configurations must be reduced independently.
Thus the above stability system is restricted to stage (i). Moreover, 
due to the fact that a $K$-factor is not available for loop-squared
amplitudes,
the comparison of \Coli{}-\Collier versus \DD{}-\Collier to assess numerical stability
is extended to all phase-space points.
Details on the usage of the stability rescue system can be found
in \refse{sec:stabsystem}.

\subsubsection{On-the-fly stability system\label{sec:otfss}}
\label{se:OL2stab}

The on-the-fly reduction methods~\cite{Buccioni:2017yxi}
implemented in \OLtwo
are supplemented by a new stability system, which is
based on the analysis of the analytic structure of spurious singularities
in the employed reduction identities.
In general, the reduction of loop objects with four or more propagators,
$D_0, D_1, D_2, D_3\dots$, can give rise to spurious singularities 
in the rank-three Gram determinant $\Delta_{123}$,
and in the rank-two Gram determinants  $\Delta_{12}$, $\Delta_{13}$ and $\Delta_{23}$.
In the case of the on-the-fly reduction~\eqref{eq:qqred},
the reduction coefficients associated with a $D_i$ pinch
generate spurious singularities of the form
\bea
A^{\mu\nu}_i &=& \frac{1}{\Delta_{12}} \, a^{\mu\nu}_i,\nonumber\\
 B^{\mu\nu}_{i,\lambda} &=&
\frac{1}{\Delta_{12}^2} \,{\frac{1}{\sqrt{\Delta_{123}}}}\, \Big[b_{i,\lambda}^{(1)}\Big]^{\mu\nu}
+ \frac{1}{\Delta_{12}}\,\Big[ b_{i,\lambda}^{(2)}\Big]^{\mu\nu}{}\,,
\label{eq:ofrinst}
\eea
with a clear hierarchical pattern: very strong instabilities in
$\Delta_{12}$, mild instabilities in $\Delta_{123}$, and no instability 
in $\Delta_{13}$ and $\Delta_{23}$. The on-the-fly reduction of 
objects with only three loop propagators involve only $\Delta_{12}$
and yields similar spurious singularities as in~\refeq{eq:ofrinst}, but
without the $\Delta_{123}$ term.

\paragraph{Rank-two Gram determinants} 
\label{sec:numstab_D2} Instabilities from rank-two Gram determinants are
completely avoided in \OLtwo.  In topologies with four or more propagators,
this is achieved via permutations of the loop denominators,
$(D_1,D_2,D_3)\to (D_{i_1},D_{i_2},D_{i_3})$, in the reduction identities. 
Such permutations are applied on an event-by-event basis in order to
guarantee
\bea
|\Delta_{i_1i_2}|\;=\; \max\left\{
|\Delta_{12}|,\,
|\Delta_{13}|,\,
|\Delta_{23}|\right\}\,,
\label{eq:maxgramdet}
\eea
so that the reduction is always protected from the smallest rank-two Gram determinant.

In this way, rank-two Gram instabilities are delayed to later
stages of the reduction, where three-point objects with 
a single Gram determinant $\Delta_{12}$ are encountered.
In this case, instabilities at small $\Delta_{12}$ are
cured by means of an analytic $\Delta_{12}$-expansion,
which have been introduced in~\cite{Buccioni:2017yxi} 
for the first few orders in $\Delta_{12}$ and are meanwhile 
available to any order~\cite{OL2_stability}.

Such expansions have been worked out for those topologies 
and regions that can lead to $\Delta_{12}\to 0$ in hard scattering processes.
This can happen only in $t$-channel triangle configurations, where two external
momenta $k_1,k_2$ are space-like, and $(k_1+k_2)^2=0$.
The relevant virtualities are parametrised 
as $k_1^2=-Q^2$ and $k_2^2=-(1+\delta)Q^2$, 
where $Q^2$ is a (high) energy scale,
and the Gram determinant is related to $\delta$
via $\sqrt{\Delta_{12}}= Q^2\,\delta/2$.
The corresponding three-point tensor integrals 
are expanded in $\delta$ based on
covariant decompositions of type
\bea
C^{\mu_1 \ldots\mu_r}\left(-p^2,-p^2
(1+{\delta}),0,m_0^2,m_1^2,m_2^2\right) &=&
\sum_i C_{i}(\delta)\, L^{\mu_1 \ldots \mu_r}_i\,,
\eea
where $L^{\mu_1 \ldots \mu_r}_i$ are Lorentz structures
made of metric tensors and external momenta. Their 
coefficients  $C_i(\delta)$
are reduced to scalar
tadpole, bubble and triangle integrals,
\bea
T^1_0(\delta)&=&A_0\left(m_0^2\right),\nonumber\\
T^2_0(\delta) &=& B_0\left(-p^2 (1+{\delta}),m_0^2,m_1^2\right),\nonumber\\
T^3_0(\delta) &=& C_0\left(-p^2,-p^2
(1+{\delta}),0,m_0^2,m_1^2,m_2^2\right)\,,
\eea
\ie
\bea
C_i(\delta) &=&
\sum_{N=1}^3 c^N_{i}(\delta)\, T^N_{0}(\delta),
\label{eq:tensdecomp}
\eea
where $c^{N}_{i}(\delta)$ are rational functions containing $1/\delta^K$
poles, while the $C_i(\delta)$ coefficients are regular at $\delta\to 0$.
Numerically stable $\delta$-expansions 
for $C_i(\delta)$  are obtained via
Taylor expansions of the scalar integrals. The required
coefficients,
\bea
S^N_k= \frac{1}{k!}\left(\frac{\partial}{\partial\delta} \right)^{k}
T^N_0(\delta)\bigg|_{ \delta=0},
\label{eq:deltaexpcoeff}
\eea
have been determined to any order $k$ in the form of
analytic recurrence relations~\cite{Buccioni:2017yxi}
for all  mass configurations of type
$(m_0,m_1,m_2)=(0, 0, 0), (0, m, m), (m, 0, 0), (m, M, M)$,
which cover all possible QCD amplitudes 
with massless partons and massive top and bottom quarks.
Recently, such any-order expansions have been extended to all mass
configurations that can occur at NLO EW.%
\footnote{The implementation of such NLO EW expansions is in progress and
will be completed in a future update of the code.  In the
meanwhile, Gram-determinant instabilities for which no expansion is 
implemented are cured by means of the hybrid-precision system (see below).}

To stabilise the tensor coefficients~\refeq{eq:tensdecomp}, singular terms
of the form $\delta^{-K}T^N_0(\delta)$ are separated 
via partial fractioning and replaced by
\bea
{\delta^{-K}}\,T^{N}_{0}(\delta)
&=& T^{N,K}_{0,\mathrm{sing}}(\delta)+
T^{N,K}_{0,\mathrm{fin}}(\delta),\qquad
\mbox{with}\quad
T^{N,K}_{0,\mathrm{fin}}(\delta)= 
\sum_{k=K}^{\infty} S^N_k {\delta}^{k-K}\,.
\eea
The singular parts cancel exactly when combining the
contributions from $A_0$, $B_0$ and $C_0$ functions as well as
the rational terms. Thus only the finite series
$T^{N,K}_{0,\mathrm{fin}}(\delta)$ need to be evaluated.
The fact that all tensor integrals are stabilised using only $C_0$ and $B_0$
expansions makes it possible to expand with excellent CPU efficiency up to
very high orders in $\delta$, thereby controlling a broad $\delta$-range.
In practice, the $\delta$-expansions are applied 
for $\delta<\delta_\mathrm{thr}$, with a
threshold $\delta_\mathrm{thr}$ that is large enough 
to avoid significant instabilities for $\delta>\delta_{\mathrm{thr}}$,
while below $\delta_{\mathrm{thr}}$
the expansions are carried out up to 
a relative accuracy of $10^{-16}$\,($10^{-32}$) in dp\,(qp).
By default $\delta_{\mathrm{thr}}$ is set to $10^{-2}$.

\paragraph{Rank-three Gram determinants} 

The on-the-fly reduction coefficients~\refeq{eq:ofrinst}
associated with $D_i$ pinches with
$i=1,2,3$ are proportional to $1/\sqrt{\Delta_{123}}$ and read~\cite{Buccioni:2017yxi} 
\bea
K_1&=& \frac{p_3 \cdot \left(\ell_1 - \alpha_1 \ell_2\right)}{p_3 \cdot
\ell_3}\,,\quad
K_2\,=\,\frac{p_3 \cdot \left(\ell_2 - \alpha_2 \ell_1\right)}{p_3 \cdot \ell_3}\,,\qquad
K_3 = 2\frac{\ell_1 \cdot \ell_2}{p_3 \cdot \ell_3}\,,
\eea
where $\alpha_i = p_i^2/[p_1 \cdot p_2 (1 + \sqrt{\delta})]$, and $\ell^\mu_{1,2,3}$ are
auxiliary momenta used to parametrise the loop momentum~\cite{Buccioni:2017yxi}.
In topologies with more than four propagators, $D_0,D_1, D_2, D_3,
D_4,\dots$, such rank-three Gram instabilities are avoided by 
performing the reduction in terms of four of the first five propagators, $D_{i_0} D_{i_1} D_{i_2} D_{i_3}$, which are chosen by  
maximising $|\Delta_{i_1i_2}|$, to avoid rank-two instabilities, and by 
subsequently minimising $\max\{|K_{i_1}|,|K_{i_2}|,|K_{i_3}|\}$.
In this way, small rank-three~(-two) Gram determinants can largely be avoided
until later stages of the recursion, where
box~(triangle) topologies have to be reduced.

\paragraph{OPP reduction} The OPP 
method, used for five- and higher-point objects 
of rank smaller than two, is based on the 
same auxiliary momenta $\ell_i$ mentioned above.
Related rank-two Gram instabilities are 
avoided by permuting the propagators of the resulting 
scalar boxes according to \refeq{eq:maxgramdet}.

\paragraph{IR regions}
In order to mitigate numerical instabilities in the context of NNLO
calculations, \OL implements additional improvements
targeted at phase-space regions 
where one external parton becomes soft or collinear.  Such improvements 
include:
\bit \item 
global and numerically stable implementation of all kinematic quantities,
including the basis momenta $\ell^\mu_i$ used for the
reduction, in special regions;

\item analytic expressions for renormalised self-energies
to avoid numerical cancellations between
bare self-energies and counterterms in the limit of small $p^2$.
This is relevant for self-energy
insertions into propagators that are connected to two external partons via
soft or collinear branchings.

\eit

Such dedicated treatments for unresolved regions will be documented 
in~\cite{OL2_stability} and further extended in the future.

\paragraph{Hybrid precision system} 

In order to cure residual instabilities that cannot be avoided with the
methods described above, the on-the-fly reduction is equipped with a
hybrid-precision (hp) system~\cite{OL2_stability} that monitors all potentially
unstable types of reduction identities and switches from
dp to qp dynamically when 
a numerical instability is encountered.  
This system is fully automated and acts
locally, at the level of individual operations.  This makes it possible to
restrict the usage of qp to a minimal part of the calculation, thereby
obtaining a speed-up of orders of magnitude as
compared to brute-force qp re-evaluations of the full amplitude.
Typically, the extra time spent in qp is only a modest fraction of the standard
dp evaluation time.
The main features of the hp system are as follows.
\bit

\item Quad precision is triggered and used at the level of individual
reduction steps, based on the kinematics of the actual phase-space point and
the loop topology of the individual open-loop object that is being processes
at a given stage of the recursion.

\item  Reduction steps that are identified as unstable and all consecutive
connected operations are carried out in quad precision until spurious
singularities are cancelled.  Quad precision is thus used for all subsequent
operations (dressing, merging, reduction, master integrals) that are
connected to an instability.

\item For each type of reduction step, the magnitude of potential
instabilities is estimated based on the actual kinematics and
the analytical form of the reduction identity. 
This information leads to an error estimate that is attributed to 
each processed object and is propagated and updated 
through all steps of the algorithm.

\item Quad precision is triggered when 
the cumulative error esimate for a certain object
exceeds a global accuracy threshold, which can be 
adjusted by the user (see \refse{sec:stabsystem}) depending on the required
numerical accuracy.

\eit

The hp system is based on two parallel dp/qp channels for each
generic operation (reduction, dressing, merging) and a twofold
dp/qp representation of each object that undergoes such operations.  By
default the dp channel is used, and when an instability is detected the
object at hand is moved to the qp channel, which is used for all its
subsequent manipulations.  At the end, when spurious singularities are
cancelled, qp output is converted back to dp.

The efficiency of the hp system strongly benefits from the above
mentioned analytical treatments of Gram determinants and soft regions, which avoid most of
the instabilities and delay the remaining ones to later stages of the
recursion, minimising the number of subsequent qp steps.
As a result, for one-loop calculations with hard kinematics qp is
typically needed only for a tiny fraction of the phase-space points, and for
a very small part of the calculation of an amplitude.
The usage of qp can become significantly more important in NNLO
calculations, especially when local subtraction methods are used.  In this
case, one-loop amplitudes need to be evaluated in deep IR regions, where new
types of instabilities occur for which no analytic solution is available at
the moment.  Such instabilities are automatically detected and cured by the
hp system.  This may lead, depending on the process and kinematic region,
to a significant CPU overhead. In such cases, 
the accuracy threshold parameter should be tuned
such as to achieve an optimal trade-off between performance and numerical
stability.

Technical details and usage of the on-the-fly stability system are described in 
\refse{sec:stabsystem}.

\paragraph{External libraries} Finally, \OLtwo benefits from improvements 
in \Collier~1.2.3~\cite{Denner:2016kdg}, which is used 
for dp evaluations of scalar integrals and for tensor reduction 
in loop-induced processes, as well as in
\OneLoop~\cite{vanHameren:2010cp}, which is used
to evaluate scalar integrals in qp.

\section{Automation of tree- and one-loop amplitudes in the full SM}
\label{se:automation}

\subsection{Power counting}
\label{sec:powercounting}

In the Standard Model, scattering amplitudes can be 
classified based on power counting in the strong and electroweak 
coupling constants,\footnote{For
simplicity, here we regard Yukawa and Higgs couplings as parameters of order
$e$, 
keeping in mind that a separate power counting in $\lambda_\mathrm{Y}$ and
$\lambda_{\mathrm{H}}$ is possible.
} 
$g_\rS=\sqrt{4\pi \alphaS}$ and $e=\sqrt{4\pi \alpha}$.
At LO in QCD, tree amplitudes have the simple form 
\beqar
\label{eq:QCDbornamp}
\calM_0\Big|_{\LO\; \mathrm{QCD}} &=& 
g_\rS^{n}e^{m}\calM_0^{(0)},
\eeqar
where $n$ and $m$ are, respectively, the maximally allowed power in $g_\rS$ 
and the minimally allowed power in $e$.
The total coupling power is fixed by the number of scattering particles,
$n+m=\npart-2$,
where $\npart$ is the number of scattering particles.

In the SM, the general coupling structure of scattering amplitudes 
depends on the number $\nqq$ of  
external quark--antiquark pairs.
For processes with $\nqq\le 1$, the LO QCD term~\refeq{eq:QCDbornamp} is the only tree contribution,
while processes with $\nqq\ge 2$ 
involve also sub-leading EW contributions of order
$g_\rS^p e^q$ with  $p+q=\npart-2$ and variable power $q>m$.
Such contributions reflect the freedom of
connecting quark lines either through EW or QCD interactions.
As a result, tree amplitudes consist of a tower of QCD--EW contributions,
\beqar
\label{eq:SMbornamp}
\calM_0 &=&
g_\rS^{n}e^{m}\,
\sum_{k=0}^{\ntildeqq}\,
\left(\frac{e}{g_\rS}\right)^{2k}\calM_0^{(k)},
\quad {\rm where}\quad 
\label{eq:ntildeqq}
\ntildeqq=\begin{cases}
\nqq -1 & \mbox{for} \quad \nqq \ge 1,\\
0 & \mbox{for} \quad \nqq = 0.\\
\end{cases}
\eeqar
For $\nqq\ge 2$, the Born amplitude \refeq{eq:SMbornamp}
involves $\nqq$ terms, while the squared Born amplitude
consists of a tower of $2\nqq-1$ terms,
\bea
\label{eq:SMborn2}
\calW_{\tree} = \langle \calM_0|\calM_0\rangle \,=\,
(4\pi\alphaS)^{n}
(4\pi\alpha)^{m}\,
\sum_{r=0}^{2\ntildeqq}
\left(\frac{\alpha}{\alphaS}\right)^r\,\calW^{(r)}_{\tree}\,.
\eea
Each term of fixed order in $\alphaS$ and $\alpha$
in \refeq{eq:SMborn2}
results from the interference between 
Born amplitudes of variable 
order,
\beqar
\label{eq:SMborn2coeff}
\calW^{(r)}_{\tree}
&=&
\sum_{s=\smin}^{\smax}\,
\big\langle\calM_0^{(r-s)}|\calM_0^{(s)}\big\rangle\,
\qquad\mbox{for}\quad0\le r\le 2\ntildeqq,
\eeqar
where $\smin=\max(0,r-\ntildeqq)$ and $\smax=\min(r,\ntildeqq)$.
Contributions
$\big\langle\calM_0^{(k)}|\calM_0^{(k')}\big\rangle$
with $k'\neq k$ and $k'=k$ are denoted, respectively, as
Born--Born interferences and squared Born terms.
The former are typically strongly suppressed with respect to the latter. 
This is due to the fact that physical observables are typically
dominated by contributions involving propagators that are enhanced in
certain kinematic regions.  Squared amplitudes that involve such propagators
are thus maximally enhanced.  In contrast, since the propagators of Born amplitudes
with $k'\neq k$ are typically peaked in different regions, 
Born--Born interferences tend to be much less enhanced.
In addition, the interference between diagrams with gluon and photon
propagators, which are enhanced in the same regions, turn out to be
suppressed as a result of colour interference.

\begin{figure}[t!]
\begin{center} 
\includegraphics[height=40mm]{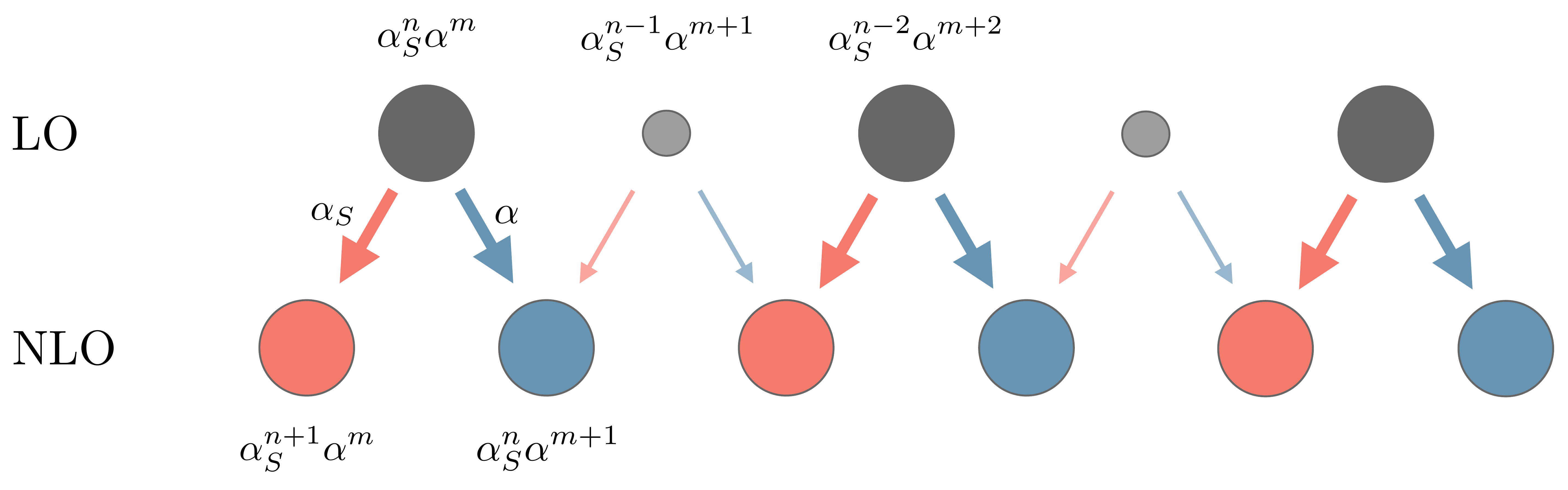} 
\end{center}
\caption{
Schematic representation of the towers of mixed QCD--EW terms 
at LO and NLO. 
The first row represents the LO tower
\refeq{eq:SMborn2}--\refeq{eq:SMborn2splitting}, which consists of an
alternating series of dominant squared Born terms (dark grey blobs) and 
sub-leading pure
interference terms (light grey blobs). 
The second row corresponds to the NLO tower 
\refeq{eq:SMloopint}--\refeq{eq:SMloopintsplittingB}.
Each LO term is connected to two NLO terms via QCD (red) 
and EW (blue) corrections, while each NLO term is 
connected to a unique squared Born term either via QCD or EW corrections.
Apart from the outer most NLO terms of pure QCD and pure EW
kind, QCD\,(EW) corrections to squared Born terms mix with 
EW\,(QCD) corrections to adjacent interference terms.
\label{Fig:SMpowercounting}
}
\end{figure}

Based on these considerations, it is interesting to note that 
each term \refeq{eq:SMborn2coeff}, with fixed order
in $\alphaS$ and $\alpha$,
contains at most one squared-Born
contribution with $r-s=s$. In fact this is possible 
only for even values of $r$.
Thus the tower \refeq{eq:SMborn2} consists of an alternating
series of $\nqq$ squared Born terms%
\footnote{In the following, for convenience,  we refer to the 
the full amplitude $\calM_0^{(2R)}$ as 
squared Born term.} with $r=2R$ and
$(\nqq-1)$ pure interference terms with $r=2R+1$,
\beqar
\label{eq:SMborn2splitting}
\calW^{(2R)}_{\tree}
&\supset&\quad
\big\langle\calM_0^{(R)}|
\calM_0^{(R)}\big\rangle\,
\;\qquad \mbox{for}\quad 0\le R \le \ntildeqq,
\nonumber\\[3mm]
\calW^{(2R+1)}_{\tree}
&\supset&\;
\mbox{interference only}
\qquad \mbox{for}\quad 0\le R \le \ntildeqq-1\,.
\eea
The tower of Born terms \eqref{eq:SMborn2} is illustrated in the upper row of \reffi{Fig:SMpowercounting}.
Squared Born terms are shown as large dark grey blobs, while interference terms are depicted as smaller 
light grey blobs.

At one loop, for processes that are not free from external QCD
partons,\footnote{In the absence of extenal quarks and
gluons, tree and one-loop amplitudes have a trivial purely EW coupling
structure,
$\calM_0=e^m\calM_0^{(0)}$ and $\calM_1=e^{m+2}\calM_1^{(1)}$.}
the leading QCD contributions have the form
\beqar
\label{eq:QCDloopamp}
\calM_1\Big|_{\NLO\; \mathrm{QCD}} &=& 
g_\rS^{n+2}e^{m}\calM_1^{(0)}\,.
\eeqar
Here NLO QCD should be understood as the
$\ord(\alphaS)$ correction wrt the LO QCD term \refeq{eq:QCDbornamp}. 
For processes with $\nqq\ge 2$,
the leading QCD terms are accompanied by a tower of 
sub-leading EW contributions, and the general form of one-loop SM amplitudes is
\beqar
\label{eq:SMloopamp}
\calM_1 &=&
g_\rS^{n+2}e^{m}\,
\sum_{k=0}^{\ntildeqq+1}\,
\left(\frac{e}{g_\rS}\right)^{2k}
\calM_1^{(k)}\,.
\eeqar
Here and in the following, the inclusion of all 
counterterm contributions of UV and $R_2$ kind
as in \refeq{eq:diaA1dbar} is implicitly understood. 
One-loop terms of fixed order in $g_\rS$ and $e$
in \refeq{eq:SMloopamp}
can be regarded either as the result of 
$\ord(g_\rS^2)$ or $\ord(e^2)$ insertions into corresponding
Born amplitudes. In this perspective, denoting 
matrix elements of fixed order as
\beqar
\calM_L^{(P,Q)}&=& 
\calM_L\Big|_{g_\rS^P e^Q}\,,
\eeqar
we can define
\bea
\label{eq:QCDEWop0}
\corrop_{\QCD}  \calM_0^{(p,q)} \equiv \calM_1^{(p+2,q)},\qquad
\corrop_{\EW}  \calM_0^{(p,q)} \equiv \calM_1^{(p,q+2)},
\eea
where $\corrop_{\QCD}$ and $\corrop_{\EW}$ should be understood as operators
that transform an $\ord(g_\rS^p e^q)$ Born matrix element 
into the {\it
complete} one-loop matrix elements of $\ord(g_\rS^{p+2} e^q)$ and $\ord(g_\rS^p
e^{q+2})$, respectively.
For processes with $\nqq\le 1$, only one Born term and two 
one-loop terms exist, and the latter can unambiguously be identified as
NLO QCD and NLO EW corrections,
\bea
\calM_1 &=&\calM_1^{(n+2,m)} + \calM_1^{(n,m+2)}
=
\corrop_{\QCD}\,\calM_0^{(n,m)}
+\corrop_{\EW}\,\calM_0^{(n,m)}
\qquad\mbox{for}\quad\nqq\le 1\,.
\eea
In contrast, processes with $\nqq\ge 2$
involve $\ntildeqq+1=\nqq$ terms of variable order
$g_\rS^P e^Q$, which can in general be regarded either as
QCD corrections to Born terms of
relative order $g_\rS^{-2}$ or EW corrections to 
Born terms of relative order $e^{-2}$, 
i.e.
\bea
\label{eq:QCDEWop1}
\calM_1^{(P,Q)} = 
\corrop_{\QCD}\,  \calM_0^{(P-2,Q)}
= 
\corrop_{\EW}\,  \calM_0^{(P,Q-2)}
\qquad\mbox{for}\quad\nqq\ge 2\,.
\eea
More precisely, one-loop terms with maximal QCD
order, $P_{\max}=n+2$, represent pure QCD corrections,
since Born terms of relative order 
$e^{-2}$ do not exist. Similarly, one-loop terms of
maximal EW order, $Q_{\max}=m+2+2\ntildeqq$,
are pure EW corrections, since Born terms 
of relative order $g_\rS^{-2}$ do not exist.
In contrast, the remaining $\nqq-2$ terms with $P<P_{\max}$
and $Q<Q_{\max}$ have a mixed QCD--EW character, in the sense that 
they involve corrections of QCD and EW type, which
coexist at the level of individual Feynman diagrams,
such as in loop diagrams where two quark lines are
connected by a virtual gluon {\it and} a virtual EW boson.
This kind of one-loop terms cannot be split into 
contributions of pure QCD or pure EW
type. Thus, in general only the full set of one-loop diagrams
containing all mixed QCD--EW  terms of order
$g_\rS^P e^Q$ represents a well defined and gauge-invariant perturbative contribution.
Keeping this in mind, as far as the terminology is concerned, it is often
convenient to refer to \refeq{eq:QCDEWop1} 
either as QCD correction wrt to 
$\ord(g_\rS^{P-2} e^{Q})$ or EW correction wrt 
$\ord(g_\rS^{P} e^{Q-2})$.

Squaring one-loop amplitudes with $\nqq\ge 2$ results in a similar tower of
$2\nqq-1$ mixed QCD--EW terms as in
\refeq{eq:SMborn2}--\refeq{eq:SMborn2coeff}.
In contrast, the interference of tree and one-loop amplitudes yields a 
tower of $2\nqq$ terms,
\bea
\label{eq:SMloopint}
\calW_{\onel} = 2\re\,\langle \calM_0|\calM_1\rangle \,=\,
(4\pi\alphaS)^{n+1}\,
(4\pi\alpha)^{m}\,
\sum_{r=0}^{2\ntildeqq+1} 
\left(\frac{\alpha}{\alphaS}\right)^r\,
\calW^{(r)}_{\onel}\,.
\eea
Each term of fixed order
in $\alphaS$ and $\alpha$
involves the interference between Born and one-loop terms of variable order,
\beqar
\label{eq:SMloopintcoeff}
\calW^{(r)}_{\onel}
&=&
2\re\,\sum_{t=\tmin}^{\tmax}\,
\big\langle\calM_0^{(r-t)}|
\calM_1^{(t)}\big\rangle\qquad\mbox{for}\quad
\qquad 0\le r\le 2\ntildeqq+1,
\eeqar
where $\tmin=\max(0,r-\ntildeqq)$ and $\tmax=\min(r,\ntildeqq+1)$.\footnote{
In~\cite{Frederix:2018nkq} the contributions $\calW_{\rm tree}^{(r)}$ and
$\calW_{\onel}^{(r)}$ are rsp.\ denoted as LO$_{r+1}$ and NLO$_{r+1}$.
}
In general, the one-loop amplitudes that enter 
\refeq{eq:SMloopintcoeff} consist of mixed QCD--EW corrections
in the sense of \refeq{eq:QCDEWop1}, \ie 
\bea
\label{eq:QCDEWop2}
\calM_1^{(k)} = 
\corrop_{\QCD}\,  \calM_0^{(k)}
=
\corrop_{\EW}\,  \calM_0^{(k-1)}
\qquad\mbox{for}\quad\nqq\ge 2\,.
\eea
In practice, as discussed above, the one-loop terms with maximal QCD or maximal EW 
order consist of pure QCD or pure EW corrections.
In \refeq{eq:SMloopint}--\refeq{eq:SMloopintcoeff} they correspond to 
$r=0$ and $r=2\ntildeqq+1$, and they read
\bea
\label{eq:pureSMloopintcoeff}
\calW^{(0)}_{\onel}
&=&
2\re\,
\big\langle\calM_0^{(0)}|
\calM_1^{(0)}\big\rangle
=
2\re\,
\big\langle\calM_0^{(0)}|
\corrop_\QCD\, \calM_0^{(0)}\big\rangle,
\nonumber\\[2mm]
\calW^{(2\ntildeqq+1)}_{\onel}
&=&
2\re\,
\big\langle\calM_0^{(\ntildeqq)}|
\calM_1^{(\ntildeqq+1)}\big\rangle
=
2\re\,
\big\langle\calM_0^{(\ntildeqq)}|
\corrop_\EW\, \calM_0^{(\ntildeqq)}\big\rangle\,.
\eea
These contributions are shown as the outer most blobs in the second row of \reffi{Fig:SMpowercounting}.
They emerge as pure $\ord(\alphaS)$ and pure $\ord(\alpha)$  corrections as indicated by the red and blue arrows respectively.
The remaining $(2\nqq-2)$ terms cannot be regarded as pure QCD or pure EW
corrections.  Nevertheless, due to the fact that the squared Born tower is
an alternating series consisting of $\nqq$ squared Born terms and $(\nqq-1)$ pure
interference terms, see \refeq{eq:SMborn2}--\refeq{eq:SMborn2splitting}, the
tree--loop interference \refeq{eq:SMloopint}
corresponds to an alternating series of
$\nqq+\nqq$ terms that can be interpreted, respectively,
as QCD and EW corrections with respect to squared
Born terms.
Specifically, the terms \refeq{eq:SMloopintcoeff} with
even indices, $r=2R$ with $0\le R\le \nqq-1$, can be written in the form
\beqar
\calW^{(2R)}_{\onel}
&=&
2\re\,
\sum_{t=\tmin}^{\tmax}
\big\langle\calM_0^{(2R-t)}|
\corrop_{\QCD}\,
\calM_0^{(t)}\big\rangle\,,
\eea
where the terms with $t=R$,
\bea
\big\langle\calM_0^{(R)}|
\corrop_{\QCD}\,
\calM_0^{(R)}\big\rangle
\;\subset\;
\calW^{(2R)}_{\onel}\,,
\eea
represent QCD corrections to squared Born amplitudes.
In contrast, the alternative representation
\beqar
\calW^{(2R)}_{\onel}
&=&
2\re\,
\sum_{t=\tmin}^{\tmax}\,
\big\langle\calM_0^{(2R-t)}|
\corrop_{\EW}\,
\calM_0^{(t-1)}\big\rangle\,,
\eea
where $2R-t \neq t-1$ for all $t$,
shows that EW corrections arise only in connection with 
interference Born terms, which are 
typically strongly sub-leading.
Vice versa, for terms with
odd indices, $r=2R+1$ with $0\le R\le \ntildeqq$, the representation
\bea
\calW^{(2R+1)}_{\onel}
&=&
%
2\re\,
\sum_{t=\tmin}^{\tmax}\,
\big\langle\calM_0^{(2R+1-t))}|
\corrop_{\EW}\,
\calM_0^{(t-1)}\big\rangle\,\nonumber\\
%
\eea
involves terms with $t=R+1$, 
\bea
\big\langle\calM_0^{(R)}|
\corrop_{\EW}\,
\calM_0^{(R)}\big\rangle
\;\subset\;
\calW^{(2R+1)}_{\onel}\,,
\eeqar
which represent EW corrections to squared 
Born amplitudes, while writing
\bea
\label{eq:SMloopintsplittingB}
\calW^{(2R+1)}_{\onel}
&=&
2\re\,
\sum_{t=\tmin}^{\tmax}\,
\big\langle\calM_0^{(2R+1-t))}|
\corrop_{\QCD}\,
\calM_0^{(t)}\big\rangle\,,
\eea
where $2R+1-t\neq t$ for all $t$,
shows that QCD correction effects enter only through pure interference Born terms
and are typically suppressed.

In summary, apart from the leading QCD and EW terms,
NLO SM contributions at a given order
$\alphaS^{n+1-r}\alpha^{m+r}$ cannot be regarded as pure QCD or pure EW
corrections.
Nevertheless, 
the orders $r=2R$ and $2R+1$ are typically dominated,
respectively, by QCD and EW corrections to the squared Born amplitude
$\calW_{\tree}^{2R}\sim \big\langle\calM_0^R|\calM_0^R\big\rangle$. 
Thus, keeping in mind that all relevant EW--QCD mixing and interference
effects must always be included, each NLO order can be labelled in a natural and unambiguous 
way either as QCD or EW correction as illustrated in
\reffi{Fig:SMpowercounting}.

As detailed in Section~\ref{se:processselection},
\OpenLoops supports the calculation of tree and one-loop contributions 
of any desired order in $\alphaS$ and $\alpha$.
In practice, scattering probability densities at different orders in $\alphaS$ and $\alpha$,
\beqar
\calW_{LL'}^{(P,Q)}&=& 
\calW_{LL'}\Big|_{\alphaS^P \alpha^Q}\,,
\eeqar
are treated as 
separate subprocesses.
Squared Born terms $\calW_{\tree}^{(p,q)}$ and squared one-loop
terms $\calW^{(p,q)}_{\onelsq}$ are
selected by specifying the QCD order $p$ or the EW order $q$.
Fixing $q$ selects also the related NLO QCD tree--loop interferences,
$\calW_{\onel}^{(p+1,q)}$, while fixing $p$ yields their
NLO EW counterpart, $\calW_{\onel}^{(p,q+1)}$.
Alternatively, tree-loop interferences of order $\alphaS^P\alpha^Q$
can be selected directly through the corresponding one-loop powers $P$ or $Q$.
%

\subsection{Input schemes and parameters}
\label{sec:ewschemes}

\def\reg{\mathrm{reg}}
\def\thetaw{\theta_{\mathrm w}}
\def\ngamma{N_\gamma}
\def\ngammastar{N_{\gamma^*}}
\def\alphaoff{\alpha_{\mathrm{off}}}
\def\alphamz{\alpha(M_Z^2)}
\def\alphamu{\alpha\vert_{\GF}}

\def\zgamma{Z_\gamma}
\def\zgammaon{Z^{(\mathrm{on})}_\gamma}
\def\zgammaoff{Z^{(\mathrm{off})}_\gamma}
\def\zgammaonoff{Z^{(\mathrm{on/off})}_\gamma}

\def\kgamma{K_\gamma}
\def\kgammaonoff{K^{(\mathrm{on/off})}_\gamma}
\def\kgammaon{K^{(\mathrm{on})}_\gamma}
\def\kgammaoff{K^{(\mathrm{off})}_\gamma}

\def\rgamma{R_\gamma}
\def\rgammaon{R^{(\mathrm{on})}_\gamma}
\def\rgammaoff{R^{(\mathrm{off})}_\gamma}
\def\rgammaonoff{R^{(\mathrm{on/off})}_\gamma}
\def\rgammaonbare{R^{(\mathrm{on})}_{0,\gamma}}
\def\rgammaoffbare{R^{(\mathrm{off})}_{0,\gamma}}
\def\rgammaonoffbare{R^{(\mathrm{on/off})}_{0,\gamma}}

\def\Qactive{\mathcal{Q}_{\mathrm{active}}}
\def\Qdecoupled{\mathcal{Q}_{\mathrm{decoupled}}}
\def\nqtot{N_{q,\mathrm{loop}}}
\def\nqactive{N_{q,\mathrm{active}}}
\def\nqmassless{N_{q,m=0}}
\def\NF{N_{\mathrm{F}}}
\def\LRS{\mathrm{LRS}}

In this section we discuss the different input schemes and the 
SM input parameters that are used for the calculation of scattering
amplitudes in \OpenLoops.
All parameters are initialised with physical default 
values, and can be adapted by the user 
by calling the \Fortran routine \texttt{set\_parameter} 
or the related C/\cpp functions 
as detailed in 
\refapp{app:native:parameters}.
\refta{app:inputparameters} in \refapp{app:input_parameters} 
summarises input parameters and switchers that can be controlled through
\texttt{set\_parameter}.
Parameters with mass dimension should be entered in 
GeV units. The values of specific parameters in \OL can be obtained by calling the routine 
\texttt{get\_parameter}, and
the full list of parameter values can be printed to a file 
by calling the function \texttt{printparameter} (see \refapp{app:native:parameters}).

\paragraph{Masses and widths} 
The \OL parameters \texttt{mass(PID)} and
\texttt{width(PID)} correspond, respectively, to the on-shell mass $M_i$
and the width $\Gamma_i$ of the particle with PDG particle number
PID (see \refta{particleid}). Masses and widths are treated as independent
inputs.
For unstable particles, when $\Gamma_i>0$,
the complex-mass scheme~\cite{Denner:2005fg} is used.
In this approach,  
particle masses are replaced throughout by the
complex-valued parameters
\beq
\begin{split}
\label{eq:complexmasses}
\mu^2_i\,=&\;\;M_i^2-\ri\Gamma_iM_i\,.
\end{split}
\eeq
This guarantees a gauge-invariant description of 
resonances and related off-shell 
effects.
By default, $\Gamma_i=0$ and $\mu_i = M_i \in \mathbb{R}$ 
for all SM particles, \ie unstable particles are treated as on-shell states,
while setting $\Gamma_i> 0$ for one or more unstable particles 
automatically activates the complex-mass scheme for the
particles at hand.
By default, $M_i>0$ only for $i=W,Z,H,t$.

For performance reasons,
the public \OpenLoops libraries are
typically generated with
$m_e=m_\mu=m_\tau=0$ and 
$m_u=m_d=m_s=m_c=0$, while generic mass parameters
$m_q$ are used for the heavy quarks $q=b,t$.
By default, heavy-quark masses are set to $m_b=0$ and $m_t=172$\,GeV,
but their values can be changed by the user as desired.
Dedicated process libraries with additional 
fermion-mass effects (any masses at NLO QCD and finite $m_\tau$ at NLO EW) can be easily generated upon request.
For efficiency reasons, when $m_Q$ is set to zero for a certain heavy quark,
whenever possible amplitudes that involve $Q$ as external particle are 
internally mapped to corresponding (faster) massless amplitudes.
To this end the desired fermion masses have to be specified before any
process is registered, see Section~\ref{se:processselection}.

\paragraph{Strong coupling} The values of the strong coupling
$\alphaS(\mur^2)$ and the related renormalisation scale $\mur$ can be
controlled through the parameters \texttt{alphas} and \texttt{muren},
respectively.  These parameters can be set dynamically on an event-by-event
basis,\footnote{For historical reasons their default values are
$\mur=100$\,GeV and $\alphaS=0.1258086856923967$.} 
and \OLtwo implements an automated scale-variation system that makes it
possible to re-evaluate the same scattering amplitude at multiple values of
$\mur$ and/or $\alphaS(\mur^2)$ with high efficiency (see
\refse{sec:amplitudecalls}).

\paragraph{Number of colours}

By default, in \OL colour effects and related interferences are included
throughout, \ie
scattering amplitudes are evaluated by retaining the exact dependence on the
number of colours $N_c$.
In addition, dedicated process libraries with large-$N_c$ expansions can be
generated by the authors upon request.
When available, leading-colour amplitudes
can be selected at the level of process registration (see \refse{se:processselection})
via the parameter $\texttt{leading\_colour}=1$ (default=0).

\paragraph{EW gauge couplings}
The U(1) and SU(2) gauge couplings $g_1, g_2$ are derived from 
\be
\label{eq:gaugecouplings}
g_1=\frac{e}{\cos\thetaw},\qquad
 g_2=\frac{e}{\sin\thetaw}, 
\ee
where $e=\sqrt{4\pi\alpha}$ and
$\thetaw$ denotes the weak mixing angle. The latter is always defined
through the ratio of the weak-boson masses~\cite{Sirlin:1980nh},
\be
\label{eq:weakmixing}
\cos^2\thetaw=\frac{\mu^2_W}{\mu^2_Z}.
\ee
If $\Gamma_{W}=\Gamma_Z=0$, then $\cos\thetaw =M_{W}/M_Z$ is 
real valued. But in general the mixing angle is complex valued.
For the electromagnetic coupling three different definitions are supported:
\begin{itemize}
\item[(i)] {\bf $\alpha(0)$-scheme:} as input for $\alpha$ the parameter
\texttt{alpha\_qed\_0} is used, which 
corresponds to the QED coupling in the $Q^2\to 0$ limit.
This scheme is appropriate for pure QED interactions
at scales $Q^2\ll M_{W}^2$, and for the production of on-shell
photons (see below).

\item[(ii)] {\bf $\GF$-scheme}: 
the input value of $\alpha$ 
is derived from the matching condition
\be 
\label{eq:GFmatching}
\Big\vert
\frac{8}{\sqrt{2}} \GF
\Big\vert^2
=
\Big\vert
\frac{g_2^2}{\mu_W^2}
\Big\vert^2\,,
\ee
which relates squared matrix elements for the muon decay 
in the Fermi theory to 
corresponding $W$-exchange matrix elements in the low-energy limit.
This results into\footnote{In the literature, 
$\alphamu$ 
is often defined as $\sqrt{2}/\pi\, \GF 
\re\left(\mu_W^2 \sin\thetaw^2\right)$, where 
the truncation of the imaginary part is 
an ad-hoc prescription aimed at keeping 
$\alpha\in \mathbb{R}$ in the complex-mass scheme.
However, 
from the matching condition \refeq{eq:GFmatching} 
it should be clear that \refeq{eq:alphaGmu} is 
the natural way of defining $\alphamu$
as real-valued parameter.
}
\be 
\label{eq:alphaGmu}
\alphamu = \frac{\sqrt{2}}{\pi} \, \GF \Big\vert
\mu_W^2 
\sin^2\thetaw
\Big\vert \,.
\ee
As input for $\alphamu$ the parameter \texttt{Gmu} is used, which
corresponds to the Fermi constant $\GF$. 
The \mbox{$\GF$-scheme} resums large logarithms associated with 
$\alphamz$ as well as universal $M_t^2/M_W^2$ enhanced
corrections associated with the $\rho$ parameter.
This guarantees an optimal description of 
the strength of the SU(2) coupling, \ie $W$-interactions,
at the EW scale.

\item[(iii)] {\bf $\alphamz$-scheme}:  as input for $\alpha$ the parameter
\texttt{alpha\_qed\_mz} is used, which 
corresponds to the QED coupling at $Q^2=M_Z^2$.
This scheme is appropriate for hard EW interactions around the EW scale,
where it guarantees an optimal description of the strength of QED interactions 
and a decent description of the strength of 
weak interactions.

\end{itemize}

The choice of $\alpha$-input scheme is controlled by 
the \OL parameter \texttt{ew\_scheme} as detailed 
in 
\refta{tab:ewschemes}, where also the default input values are specified. 
Note that $\alpha(0)$ and $\alphamz$ are described by means of two
distinct parameters in \OpenLoops. Depending on the selected scheme, the
appropriate parameter should be set.

\begin{table}
\centering
\begin{tabular}{|l|l|l|l|l|ll|} \hline
\texttt{ew\_scheme}  & scheme & input parameter  & default input  &
value of $\alpha$  \\  \hline
0                & $\alpha(0)$   & \texttt{alpha\_qed\_0} & $1/137.035999074$
& idem  \\
1   (default)& $\GF$     & \texttt{Gmu} & $1.16637\cdot 10^{-5}$\,GeV$^{-2}$ & 1/132.34890452162441  \\
2     & $\alphamz$     & \texttt{alpha\_qed\_mz} & $1/128$ & idem  \\ \hline
\end{tabular}
\caption{Available EW input schemes 
and corresponding values of the $\texttt{ew\_scheme}$ selector.
For each scheme the default values of the corresponding input parameter is
indicated. Note that instead of $\alphamz=1/127.94$~\cite{Tanabashi:2018oca}
we use $1/128$. 
Assuming the default weak-boson mass values $M_W=80.399$\,GeV,
$M_Z=91.1876$\,GeV and
$\Gamma_W=\Gamma_Z=0$. For the weak mixing angle,
$\sin^2\thetaw=0.22262651564387248$ in all three schemes, while the
derived value of $\alphamu$ is reported in the table.
}
\label{tab:ewschemes}
\end{table}

\paragraph{External photons}

The high-energy couplings $\alphamu$ and $\alphamz$ are appropriate for
the interactions of EW gauge bosons with virtualities of the order of the EW
scale.  In contrast, the appropriate coupling for external high-energy
photons is $\alpha(0)$~\cite{Andersen:2014efa}.
More precisely, for photons of virtuality $Q_\gamma^2$ the coupling
$\alpha(Q_\gamma^2)$ should be used.  For initial- or final-state 
{\it on-shell}
photons this corresponds to $\alpha(0)$.  However, in photon-induced
hadronic collisions, initial-state photons inside the hadrons effectively 
couple as {\it off-shell} partons with virtuality $Q^2_\gamma = \mu_F^2$,
 where $\muf$ is the
factorisation scale of the parton distribution functions
(see Appendix A.3~of\cite{Kallweit:2017khh}),
Thus, at high $\mu_F^2$ the
high-energy couplings $\alphamu$ or $\alphamz$ should be used.

Based on these considerations, 
for processes 
with $n$ on-shell and 
$n_*$ ``off-shell'' hard external photons plus a possible unresolved photon,
\be
A\to B+ n \gamma + n_* \gamma^*\; (+\gamma)\,,
\ee
the scattering probability densities $\calW=\calW_{00},\calW_{01},\calW_{11}$
are automatically rescaled as\footnote{In the case of NLO EW contributions
$\calW_{01}$, the rescaling factors are renormalised according to 
\refeq{eq:gammarescren}.}
\be
\label{eq:LSZgamma}
\calW \;\rightarrow\;
\Big[\rgammaon\Big]^n\,
\Big[\rgammaoff\Big]^{n_*} 
\,\calW\,,
\ee
with LSZ-like coupling correction factors
\be
\label{eq:lszfactors}
\rgammaon = \frac{\alpha(0)}{\alpha}
\qquad\mbox{and}\quad
\rgammaoff = \frac{\alphaoff}{\alpha}\,. 
\ee
Here $\alpha$ should be understood as the QED coupling in the
input scheme selected by the user, while
the value of $\alpha(0)$ correspond to the parameter
$\texttt{alpha\_qed\_0}$ and is independent of the scheme
choice.
The coupling of off-shell external photons
and the resulting $\rgammaoff$ factor
are set internally as
\be
\label{eq:alphaoff}
\alphaoff=
\begin{cases}
\alphamu  & \mbox{if}\; 
\alpha = \alpha(0),\\
\alpha & \mbox{if}\; 
\alpha = \alphamu\;\mbox{or}\;\alpha=\alphamz\\
\end{cases}
\qquad
\Rightarrow
\qquad
\rgammaoff =
\begin{cases}
\frac{\alphamu}{\alpha(0)}  & \mbox{if}\; 
\alpha = \alpha(0),\\
1 & \mbox{otherwise}.
\end{cases}
\ee
In this way $\alphaoff$ is guaranteed to be 
a high-energy coupling.
Note that unresolved photons, \ie additional photons 
emitted at NLO EW, need to be treated in a different way.
In this case, in order to guarantee the correct cancellation of IR
singularities, real and EW corrections should be 
computed with the same QED coupling.
This implies that the coupling $\alpha$ of unresolved photons
should not receive any $\rgamma$ rescaling.

The relevant information to determine the 
number of on-shell and off-shell external photons
in \refeq{eq:LSZgamma}
should be provided by the user on a process-by-process
basis.  To this end, when registering a process  with external photons
(see
\refse{se:processselection}), 
unresolved photons should be labelled
with the standard PDG identifier \texttt{PID} = 22, while for on-shell and
off-shell hard photons, respectively, \texttt{PID} = $2002$ and 
\texttt{PID} = $-2002$ should be used.
In order to guarantee an optimal choice of $\alpha$, external photons
should be handled according to the following classification. 
\bit
\item Unresolved photons (\texttt{iPDG} = 22):
extra photons (absent at LO) in 
NLO EW bremsstrahlung.

\item Hard photons of on-shell type (\texttt{iPDG} = 2002):
standard hard final-state photons that do not undergo $\gamma\to f\bar f$
splittings at NLO EW, 
or initial-state photons at photon colliders;

\item Hard photons of off-shell type (\texttt{iPDG}$= -2002$):
hard final-state photons 
that undergo $\gamma\to f\bar f$
splittings at NLO EW,
or
initial-state photons from QED PDFs in high-energy
hadronic collisions.

\eit
Here ``hard'' should be understood as the 
opposite of ``unresolved'', \ie it 
refers to all photons that are present 
as external particles starting from LO.

By default, the $\rgammaon$ and $\rgammaoff$ rescaling factors in 
\refeq{eq:LSZgamma}--\refeq{eq:lszfactors}
are applied to all on-shell and off-shell photons. They can be deactivated independently of
each other by setting, respectively, 
\texttt{onshell\_photons\_lsz=0} (default=1) and/or  
\texttt{offshell\_photons\_lsz=0} (default=1).

\paragraph{Yukawa and Higgs couplings}
 The interactions of 
Higgs bosons with massive fermions is described by the 
Yukawa couplings
\be
\label{eq:SMyuk}
\lambda_f = \frac{\sqrt{2}\,\muyuk{f}}{v}\qquad \mbox{with}\quad
v=\frac{2\mu_W \sin\thetaw }{e}\,.
\ee
Here $v$ corresponds to the vacuum expectation value, while
$\muyuk{f}$ is a Yukawa mass parameter. 
At LO and NLO QCD, the complex-valued Yukawa masses can be freely adapted 
through the parameters
\texttt{yuk(PID)} and \texttt{yukw(PID)}, 
which play the role of real Yukawa masses $\massyuk{i}$ 
and widths  $\widthyuk{i}$. More explicitly,  in analogy with
\refeq{eq:complexmasses}, 
\beq\begin{split}
\label{eq:yukcomplexmasses}
\muyuk{f}^2\,=&\;\;\massyuk{f}^2-\ri\widthyuk{f}\, \massyuk{f}\,.
\end{split}
\eeq
At NLO QCD, as discussed in \refse{sec:qcdrenorm}, 
Yukawa couplings can be renormalised in the 
$\msbar$ scheme or, alternatively, as 
on-shell fermion masses.

By default, according to the SM relation between Yukawa couplings and
masses, the Yukawa masses $\muyuk{f}$ are set equal to the complex masses
$\mu_f$ in \refeq{eq:complexmasses}.
More precisely, each time that \texttt{mass(PID)} and \texttt{width(PID)}
are updated, the corresponding Yukawa mass parameters \texttt{yuk(PID)} and
\texttt{yukw(PID)} are set to the same values.  Thus, modified Yukawa masses should
always be set after physical masses.
This interplay, 
can be deactivated by setting \texttt{freeyuk\_on}=1 (default=0).
In this case, \texttt{yuk(PID)} and
\texttt{yukw(PID)} are still initialised with the same default values as mass
parameters, but are otherwise independent.
This switcher acts in a similar way on the Yukawa renormalisation scale
$\mu_{f,\rY}$ in~\refeq{eq:qcdyukmassct}. 
At NLO EW, modified Yukawa masses are not allowed.\footnote{More precisely,
Yukawa masses are always renormalised like
physical masses at $\ord(\alpha)$. Moreover, when 
$\muyuk{f}\neq \mu_f$  for any particle 
during process registration
NLO EW process libraries cannot be loaded
and if $\muyuk{f}\neq \mu_f$ is set at a later stage a warning is printed.}

The triple and quartic Higgs self-couplings are implemented as
\be
\label{eq:SMlambda}
\lambda_H^{(3)} = \kappa_H^{(3)}\,\frac{3 \mu_H^2}{v}\,,\qquad 
\lambda_H^{(4)} = \kappa_H^{(4)}\,\frac{3 \mu_H^2}{v^2}\,,
\ee
where $\mu_H$ denotes the Higgs mass.
By default  $\kappa^{(3)}_{H}$=$\kappa^{(4)}_{H}=1$, 
consistently with the SM. 
At NLO QCD, and also at NLO EW for processes that are independent
of $\lambda_H^{(3,4)}$ at tree level, 
the Higgs self-couplings can be modified through the 
naive real-valued rescaling parameters
\texttt{lambda\_hhh}$\equiv \kappa^{(3)}_{H}$
and \texttt{lambda\_hhhh}$\equiv \kappa^{(4)}_{H}$.

Wherever present, the
imaginary parts of $\mu_f$, $\mu_H$, $\mu_W$ and $\sin\thetaw$ are consistently 
included throughout in \refeq{eq:SMyuk}--\refeq{eq:SMlambda}.

\paragraph{Higgs effective couplings}
Effective Higgs interactions in the $M_t\to \infty$ limit 
are parametrised in such a way that the
Feynman rule for the vertices with two gluons and $n$ Higgs bosons read
\begin{align}
V^{\mu\nu}_{ggH^n} \,=\,
\lambda_{ggH^n}\,
\left(g^{\mu\nu} p_1\cdot p_2 - p_1^\nu p_2^\mu\right)\,,
\qquad {\rm with} \quad  
\lambda_{ggH^n}\,=\,\frac{1}{n}\frac{g_\rS^2}{4\pi^2} \left(\frac{-\ri e}{6 \mu_W
\sin\thetaw}\right)^n\,, 
\label{eq:HEFTcoupling}
\end{align}
where $p_1^\mu$ and $p_2^\nu$ are the incoming momenta of the gluons.
The power counting in the coupling constants is done in $e$ and $g_\rS$
as in the SM.
In the Higgs Effective Field Theory, only QCD corrections are currently
available.

\paragraph{CKM matrix}
The  \OpenLoops program 
can generate scattering amplitudes with a generic CKM
matrix $V_{ij}$.  However, for efficiency reasons, most process libraries
are generated with a trivial CKM matrix, $V_{ij}=\delta_{ij}$.  Process
libraries with a generic CKM matrix are publicly available for selected
processes, such as charged-current Drell-Yan production in association with
jets, and further libraries of this kind can be generated upon request.
When available, such libraries can be
used by setting $\texttt{ckmorder=1}$
before the registration of the process at hand
(see \refse{se:processselection}).
In this case the default values of $V_{ij}$ remain equal to 
$\delta_{ij}$, but the real and imaginary parts of 
the CKM matrix can be set to any desired value 
by means of the input parameters \texttt{VCKMdu},
\texttt{VCKMsu}, \texttt{VCKMbu}, \texttt{VCKMdc}, \texttt{VCKMsc},
\texttt{VCKMbc}, \texttt{VCKMdt}, \texttt{VCKMst}, \texttt{VCKMbt} for 
$\re(V_{ij})$ and
\texttt{VCKMIdu}, \texttt{VCKMIsu}, etc.~for $\mathrm{Im}(V_{ij})$.
%

\subsection{Renormalisation}
\label{se:renormalisation}

Divergences of UV and IR
type are regularised in $D=4-2\eps$ dimensions and are expressed as 
poles of the form 
$\ceps\, \mudim^{2\eps}/\eps^{n}$, 
where
$\mudim$ is the scale of dimensional regularisation, and
\bea
\label{eq:cepsilon}
\ceps &=& \frac{(4\pi)^\epsilon}{\Gamma(1-\epsilon)}
= 1+\eps\left[\ln\left(4\pi\right)-\gamma_\rE\right]+\ord(\eps^2)
\eea
is the conventional $\msbar$ normalisation factor.
For a systematic bookkeeping of the different kinds of divergences,
UV and IR poles are parametrised in terms of 
independent dimensional factors ($\epsuv,\epsir$) and scales
($\mudimuv,\mudimir$).
Thus, one-loop amplitudes involve three types of poles,
\bea
\label{eq:dimregpoles}
\frac{\ceps \left(\mudimuv^2\right)^{\epsuv}}{\epsuv}&=& 
\ceps\left[\frac{1}{\epsuv}+\ln(\mudimuv^2)\right]+\ord(\epsuv)\,,\nonumber\\
\frac{\ceps \left(\mudimir^2\right)^{\epsir}}{\epsir}&=& 
\ceps\left[\frac{1}{\epsir}+\ln(\mudimir^2)\right]+\ord(\epsir)\,,\nonumber\\
\frac{\ceps \left(\mudimir^2\right)^{\epsir}}{\epsir^2}&=& 
\ceps\left[\frac{1}{\epsir^2}+\frac{1}{\epsir}\ln(\mudimir^2)+\frac{1}{2}\ln^2(\mudimir^2)\right]+\ord(\epsir)\,.
\eea
Renormalised one-loop amplitudes computed by \OpenLoops are free of UV divergences. 
Yet, bare amplitudes with explicit
UV poles
can also be obtained (see Section~\ref{sec:polesandscales}).
The remaining IR divergences 
are universal and can be cancelled through appropriate 
subtraction terms (see Section~\ref{sec:irsubtration}). 

For the renormalisation of UV divergences
we apply the following generic transformations of 
masses, fields and coupling parameters, 
\beqar
\mu^2_{i,0} &=& \mu_i^2+ \delta \mu_i^2,\\
\varphi_{i,0} &=& \left(1+\frac{1}{2}\delta Z_{\varphi_i
\varphi_j}\right)\varphi_j,\\
g_{i,0}   &=& g_i + \delta g_i = \left(1+\delta Z_{g_i} \right) g_i,
\eeqar
where $\mu^2_{i,0}$, $\varphi_{i,0}$, $g_{i,0}$ denote 
bare quantities, and $\delta \mu_i^2$, $\delta Z_{\varphi_i
\varphi_j}$, $\delta Z_{g_i}$ the respective counterterms.

For unstable particles, as discussed in \refse{sec:ewrenorm}, \OpenLoops
implements a flexible combination of the on-shell
scheme~\cite{Denner:1991kt} and the complex-mass
scheme~\cite{Denner:2005fg}.  In this approach, the width parameters
$\Gamma_i$ of the various unstable particles can be set to non-zero or zero
values independently of each other. Depending on this choice, the
corresponding particles are consistently renormalised as resonances with
complex masses or as on-shell external states with real masses.

In the following, we discuss the various 
counterterms needed at NLO QCD and NLO EW.
In general, as discussed in \refse{sec:powercounting},
one-loop contributions of 
$\ord(\alphaS^P\alpha^Q)$
can require $\ord(\alphaS)$ counterterm insertions in 
Born terms of $\ord(\alphaS^{P-1}\alpha^Q)$
as well as $\ord(\alpha)$ counterterm insertions in 
Born terms of $\ord(\alphaS^{P}\alpha^{Q-1})$.

\subsubsection{QCD renormalisation}
\label{sec:qcdrenorm}
The SM parameters that involve one-loop counterterms of $\ord(\alphaS)$ 
are the strong coupling, the quark masses, and the related Yukawa couplings.

 \paragraph{Strong coupling} The renormalisation of the strong coupling
constant is carried out in the $\msbar$ scheme,
and can be matched in a flexible way to the different flavour-number schemes that are commonly
used in NLO QCD calculations.
To this end, the full set of light and heavy quarks  that 
contribute to one-loop amplitudes and counterterms is split into a subset of
active quarks  ($q\in \Qactive$) and a remaining subset of decoupled
quarks ($q\notin \Qactive$).
Active quarks with mass $m_q\ge 0$ are assumed to contribute to the
evolution of $\alphaS(\mur^2)$ above threshold. Thus they 
are renormalised via $\msbar$ subtraction at the scale $\mu=\max(\mur,m_q)$.
The remaining heavy quarks ($q\notin\Qactive$)
are assumed to contribute only to loop amplitudes and counterterms, but not
to the running of $\alphaS(\mur^2)$. Thus, they are renormalised in the so-called 
decoupling scheme, which corresponds to a subtraction at 
zero momentum transfer.

\def\deltaonebar{\bar \Delta^1_{\eps_i}}
\def\deltaonebaruv{\bar \Delta^1_{\epsuv}}
\def\deltaonebarir{\bar \Delta^1_{\epsir}}

The explicit form of the $g_\rS$ counterterm reads
\bea \label{eq:gsCT}
\frac{\delta{g_\rS}}{g_\rS}&=& \frac{\alphaS}{4\pi}\left\{ -\frac{11}{6} C_A
\left[\frac{\ceps}{\epsuv}+\ln\left(\frac{\mudimuv^2}{\mur^2}\right)
\right]
+\frac{2}{3} T_F 
\sum_{q}
\left[
\frac{\ceps}{\epsuv} 
+L_q(\mudim,\mur,\mu_q)
\right]
\right\}\,,
\eea
where $C_A=3$ and $T_F=1/2$, while $\mur$ and $\mudimuv$ are the
renormalisation and dimensional regularisation scales for UV divergences, respectively. 
The logarithmic terms associated with quark loops read
\bea
\label{eq:quarkflavourlogs}
L_q(\mudim,\mur,\mu_q) &=&
\begin{cases}
\ln\left(\frac{\mudim^2}{\mur^2}\right) &  \quad 
\mbox{if $q\in \Qactive$ and $\mur>m_q$\,,}
\\[3.5mm]
&  \quad \mbox{if $q\in \Qactive$  
and $\mur<m_q$ or}\\[-3.5mm]
\re
\ln\left(\frac{\mudim^2}{\mu_q^2}\right)
&
\\[-3.5mm]
&
\quad \mbox{if $q\notin \Qactive$\,.} \\
\end{cases}
\eea

The number of active and decoupled quarks included in \refeq{eq:gsCT}
is determined as explained in the following.

\paragraph{Choice of flavour-number scheme} In NLO QCD calculations,
the logarithms of $\mur$ in the counterterm
\refeq{eq:gsCT}--\refeq{eq:quarkflavourlogs}
should cancel the leading-order 
$\mur$ dependence associated with $\alphaS(\mur^2)$.
To this end, the number $\nqactive$ of active quark flavours in 
\refeq{eq:gsCT} should be set equal to the number $\NF$ 
corresponding to the flavour-number scheme of the calculation at hand.
More precisely, 
when using a running $\alphaS(\mur^2)$ 
with $\NF$ quark flavours, the user%
\footnote{Note that 
$\alphaS(\mur^2)$ and $\mur$ are separate 
input parameters controlled by the user, 
\ie \OpenLoops does not control the evolution of $\alphaS(\mur^2)$
but only the related counterterm.
Thus it is the role of the user to set
$\nqactive$ to the correct value $\NF$.
} 
should set
$\nqactive=\NF$.
In variable-flavour number schemes,
$\NF$ corresponds to the maximum number of 
quark flavours in the evolution, and typically
$\NF=4,5$ or 6.~\footnote{In case the running of $\alphaS$ is obtained from \textsc{LHAPDF}
the information about the number of quark flavours contributing to  the evolution of $\alphaS$ is 
 available in the PDF info file as the tag $\texttt{NumFlavors}$ for \textsc{LHAPDF} versions~$\ge 6.0$.}

In practice, the number of active quarks in \OpenLoops is determined as
\be
\nqactive=\max(\NF,\nqmassless)\,,
\ee
where $\NF$ corresponds to the desired flavour-number scheme
and can be specified by the user through the parameter \texttt{nf$\_$alphasrun}, while
$\nqmassless$ is determined from the number of quarks with $m_q=0$ at runtime.
By default \texttt{nf$\_$alphasrun=0}, and all massless quarks are treated as
active, while massive quarks are decoupled.  In contrast,
if \texttt{nf$\_$alphasrun} is set to a
value $\NF>\nqmassless$, the first $\NF$ massless or massive quarks
are treated as active above threshold, and only the
remaining heavy quarks are decoupled.
For example, when $m_b=0$ the default value
of $\nqactive$ is 5, and \texttt{nf$\_$alphasrun} should be set to 6
in case a 6-flavour $\alphaS$ is used.
In contrast, for $m_b\neq 0$ the default value of $\nqactive$ is 4, and
\texttt{nf$\_$alphasrun} should be set to $\NF$ in case a $\NF$-flavour
$\alphaS$ with $\NF>4$ is used.

\paragraph{Total number of quark flavours}
By default, most public \OpenLoops libraries involve quark-loop 
contributions with $\nqtot=6$ quark flavours.
Such libraries can be used for NLO calculations in any flavour-number scheme
with $\NF=\nqtot$ or $\NF<\nqtot$.  In the latter case, heavy-quark loop
contributions that do not contribute to the evolution of $\alphaS(\mur^2)$ are
consistently accounted for by the $\nqtot-\NF$ decoupled quarks in the
one-loop matrix elements.

Extra libraries without top-quark loops ($\nqtot=5$) can be easily generated
upon request and are publicly available for selected processes.
When available, libraries with $\nqtot<6$ can be used by setting the
parameter $\texttt{nf}$ (default=6) to the desired value of $\nqtot$ at the moment of the process
registration.%

\paragraph{Quark masses} At NLO QCD, quark masses can be renormalised in the on-shell
scheme (default) or in the $\msbar$ scheme. The general form of mass
counterterms is
\bea
\label{eq:qcdmassct}
\frac{\delta \mu_q}{\mu_q} &=&
-\frac{3\,\alphaS}{4\pi}\,C_{F}
\left[
\frac{\ceps}{\epsuv}+\ln\left(\frac{\mudimuv^2}{\mu_q^2}\right)
+X(\mu_q,\Lambda_q)
\right]\,,
\eea
where $C_F=3/4$, and logarithms of the
complex mass $\mu_q$ are complex valued when $\Gamma_q>0$.
The scheme-dependent finite part reads
\bea
\label{eq:massCTX}
X(\mu_q,\Lambda_q)=\begin{cases}
\frac{4}{3} &\;\mbox{in the on-shell scheme ($\Lambda_q=0$)}, \\[2mm]
\ln\left(\frac{\mu_q^2}{\Lambda_q^2}\right)
&\;\mbox{in the $\msbar$ scheme ($\Lambda_q>0$). }
\end{cases}
\eea
Here $\Lambda_{q}$ denotes the $\msbar$ renormalisation scale for the mass
of the quark $q$.  This scale is controlled by the (real-valued) parameter
\texttt{LambdaM(PID)}, which plays also the role of scheme setter for the
mass counterterm of the quark at hand.  For $\Lambda_{q}=0$ (default) the
on-shell scheme is used, while setting $\Lambda_{q}>0$ activates
the $\msbar$ scheme.

\paragraph{Yukawa couplings} 
According to \refeq{eq:SMyuk},
Yukawa couplings are defined in terms of related Yukawa masses.
Their ratio $\muyuk{q}/\lambda_q=v/\sqrt{2}$
depends only on the vacuum
expectation value, 
which does not receive $\ord({\alphaS})$ corrections.
This implies the trivial counterterm relation
\bea
\frac{\delta\lambda_q}{\lambda_q}&=&
\frac{\delta \muyuk{q}}{\muyuk{q}}\,.
\eea
Similarly as for the quark masses $\mu_q$, also Yukawa masses 
can be renormalised on-shell (default) or via $\msbar$
subtraction. The counterterms read
\bea
\label{eq:qcdyukmassct}
\frac{\delta \mu_{q,\rY}}{\mu_{q,\rY}} &=&
-\frac{3\,\alphaS}{4\pi}\,C_{F}
\left[
\frac{\ceps}{\epsuv}+\ln\left(\frac{\mudimuv^2}{\mu_{q,\rY}^2}\right)
+X(\mu_{q,\rY},\Lambda_{q,\rY})
\right]\,,
\eea
with $X(\mu_{q,\rY},\Lambda_{q,\rY})$ as defined in \eqref{eq:massCTX}. 
The $\msbar$ renormalisation scale $\Lambda_{q,\rY}$ for the Yukawa mass
of the quark $q$ is controlled by the independent 
parameter \texttt{LambdaY(PID)}. 
By default $\Lambda_{q,\rY}=0$, and the on-shell counterterm is used,
while setting $\Lambda_{q,\rY}>0$ activates the $\msbar$ renormalisation.
Similarly as for Yukawa masses~\refeq{eq:yukcomplexmasses}, the values of
$\Lambda_{q,\rY}$ are automatically synchronised with $\Lambda_{q}$ when the latter is
changed, but not vice versa.  Thus the order in which
\texttt{LambdaM(PID)} and \texttt{LambdaY(PID)} are set matters.
As for Yukawa masses, this interplay can be deactivated by setting
\texttt{freeyuk\_on}=1 (default=0).

\paragraph{Wave functions} The QCD counterterms for gluon and quark wave
functions read
\bea
\delta Z_{g} &=&  \frac{\alphaS}{4\pi}
\bigg[
\frac{5}{3}C_A\,\Delta(0)
-\frac{4}{3} T_F \sum_q \,\Delta(\mu_q)
\bigg]\,,\\
\delta Z_{q} &=& - \frac{\alphaS}{4\pi}  C_F  \bigg\{
\Delta(\mu_q)+\left[2\left( \frac{\ceps}{\epsir} + \re\ln\left(\frac{\mudimir^2}{\mu_q^2}\right)\right)  +4\right] \,\Theta(M_q)
\bigg\}\,,
\eea
where $\mudimir$ is the dimensional regularisation scale for IR divergences, $\Theta(M)$  is the step function with $\Theta(0)=0$ and
\be
\label{eq:deltamuq}
\Delta(\mu_q)=
\begin{cases}
 \frac{\ceps}{\epsuv}- \frac{\ceps}{\epsir} + \ln\left(\frac{\mudimuv^2}{\mudimir^2}\right) & \mbox{for}\;
\mu_q=0\,,\\
 \frac{\ceps}{\epsuv}+\re\ln\left(\frac{\mudimuv^2}{\mu_q^2}\right)
& \mbox{otherwise}\,.\\
\end{cases}
\ee

\paragraph{Higgs effective couplings}
The QCD counterterm associated with the Higgs effective vertex~\refeq{eq:HEFTcoupling}
reads
\begin{align}
\frac{\delta g_{ggH^n}}{g_{ggH^n}} = 2 \frac{\delta g_\rS}{g_\rS} + \delta Z_g+\frac{11}{4 \pi}\alpha_\rS\,,
\end{align}
where the last term originates from the two-loop matching of the Higgs effective coupling~\cite{Spira:1995rr,Dawson:1998py}.
For double- (and multi-) Higgs production at the same order as the NLO QCD corrections also
double-operator insertions with the same total number of Higgs bosons contribute. In \OL these contributions are automatically included
 as pseudo-counterterms together with the virtual amplitudes.

\paragraph{Renormalisation and regularisation scales}
At the level of the user interface, 
the UV and IR regularisation
scales are treated as a common scale
$\mudim=\mudimuv=\mudimir$, and the
logarithms of $\mudimuv^2/\mudimir^2$ in  \refeq{eq:deltamuq} 
are set to zero.
In the literature, also the logarithms of $\mudimuv^2/\mur^2$ in \refeq{eq:gsCT}--\refeq{eq:quarkflavourlogs}
are often omitted by assuming $\mudimuv=\mur$ in the $\msbar$ scheme.
On the contrary, in \OpenLoops the values of $\mudim$ and $\mur$ 
are controlled by two independent 
parameters, \texttt{mureg} and \texttt{muren}, respectively.
Their default values are $\mudim=\mur=100$\,GeV. For convenience it 
is  possible to
simultaneously set $\mur=\mudim=\mu$ by means of the auxiliary \OpenLoops
parameter \texttt{mu}. 
As described in \refse{sec:amplitudecalls}, 
variations of $\mur$ and $\alphaS(\mur^2)$ can be carried out in a 
very efficient way in \OLtwo.

\subsubsection{EW renormalisation}
\label{sec:ewrenorm}

The renormalisation of UV divergences in the EW sector is based
on the scheme of~\cite{Denner:1991kt} for on-shell particles, and 
on the complex-mass scheme~\cite{Denner:2005fg} for the treatment of 
off-shell unstable particles.
In \OLtwo these two schemes are combined into a flexible renormalisation
scheme that makes it possible to deal with processes such as $pp\to t \bar t
\ell^+ \ell^-$, where some unstable particles ($t,\bar t$) are treated as
on-shell external states, while other ones ($Z$) play the role of intermediate
resonances.
This is achieved through a refined definition of field-renormalisation
constants, and by adapting the mass-renormalisation prescriptions
for unstable particles on a particle-by-particle basis,
depending on whether the individual width parameters
$\Gamma_i$ are set to non-zero values or not by
the user.
As explained in the following, the $\ord(\alpha)$ renormalisation in \OpenLoops 
involves also a non-standard treatment of $\Delta\alphamz$
and special features related to external photons.

\paragraph{Counterterms for complex masses}
The propagators of unstable particles with $\Gamma_i\neq0$ 
are renormalised in the complex-mass
scheme~\cite{Denner:2005fg}, where the renormalised self-energy is defined
as 
\beqar
\label{eq:genericcmsmassctA}
\hat \Sigma^{i}(p^2) &=& \Sigma^{i}(p^2)-\delta\mu_i^2\,
\qquad\mbox{with}\qquad
\delta \mu_i^2 =
\Sigma^{i}\left(p^2\right)\Big|_{p^2=\mu_i^2}\,.
\eeqar
The counterterm $\delta \mu_i^2$ associated with the complex
mass \refeq{eq:complexmasses} corresponds to a subtraction of the full
complex-valued self-energy at $p^2=\mu_i^2$.
In particular, the counterterm $\delta \mu^2_i$ includes also the imaginary part
of the self-energy, which is related to the width through
\be
\label{eq:ImSigma} 
\mathrm{Im}\, \Sigma^{i}(M^2_i) = \Gamma_i M_i\,, 
\ee
and is already included in the imaginary part of $\mu^2_i$.  Thus 
the subtraction of $\mathrm{Im}\,\Sigma$
in the complex-mass scheme is mandatory in order to
avoid double counting.
Since the renormalised self-energy \refeq{eq:genericcmsmassctA} vanishes at
$p^2\to \mu_i^2$, the tree-level and one-loop propagators have the same
resonant form $1/(p^2-M_i^2+\ri\Gamma_i M_i)$, where width effects are
controlled by the user-defined width parameter $\Gamma_i$.

For convenience, the relevant 2-point integrals with complex-valued momenta
$p^2=\mu^2_i=M^2-\ri\Gamma_i M_i$ can be obtained through a first-order
expansion in $\Gamma_i/M_i$ around $p^2=M_i^2$~\cite{Denner:2005fg}.
In this context, self-energy graphs involving massless photons require a
special treatment due to the presence of a threshold at $p^2=\mu^2$.  In
this case, the correct expansion of the scalar two-point function reads
\beqar
\label{eq:B0complexexp}
B_0(p^2,\mu^2,0) \Bigg|_{p^2= M^2-\ri\Gamma M}&=&  
B_0(M^2,M^2,0)-\ri \Gamma M  B_0'(M^2,\mu^2,0)
-\frac{\ri \Gamma}{M}+ \ord\left(\frac{\Gamma^2}{M^2}\right),
\eeqar
where the additional $-\ri\Gamma/M$ term accounts for the non-analytic
behaviour at $p^2=\mu^2$.
The related expansion formula for generic self-energies reads
\beqar
\label{eq:genericcmsexpansion}
\Sigma^{i}\left(\mu_i^2\right)\Bigg|_{p^2= M^2-\ri\Gamma M}&=&   
\Sigma^{i}\left(M^2_i\right) 
-\ri\Gamma_i M_i\,
\Sigma'^{i}\left(M^2_i\right)
+\ri c_i M_i^2
+ \ord\left(\Gamma^2\right),
\eeqar
where the non-analytic expansion coefficient is given by
\beqar
\label{eq:nonanalyticcorr}
c_i &=& \frac{\alpha}{\pi}Q_i^2\frac{\Gamma_i}{M_i}\,,
\eeqar
and depends only on
the electromagnetic charge $Q_i$ of the particle at hand.
This is due to the fact that~\refeq{eq:nonanalyticcorr} originates only from
photon-exchange diagrams and is related to the presence of an infrared
singularity in $B_0'$ at $p^2\to \mu_i^2$.

The expanded mass counterterms for Higgs ($i=H$) and vector bosons
($i=V=W,Z$) read
\beqar
 \label{eq:cmsmassctS}
 \delta \mu_{H}^2 &=&  \Sigma^{H}\left(\mu_H^2\right) 
= \Sigma^{H}\left(M^2_H\right) 
-\ri\Gamma_H M_H\,
\Sigma'^{H}\left(M^2_H\right)
 \,,\\[1mm]
\delta \mu_V^2 &=& \Sigma^{V}_{\rT}\left(\mu_V^2\right) 
=  \Sigma^{V}_{\rT}\left(M_V^2\right) 
-\ri\Gamma_V M_V\,
\Sigma'^{V}\left(M^2_V\right) 
+\ri c_V M_V^2\,,
\label{eq:cmsmassctV}
\eeqar
where $\Sigma_{\rT}$ denotes the transverse part of the 
gauge-boson propagator. The renormalisation of fermion 
masses depends on the following combination 
of left-handed (L), right-handed (R) and scalar
(S) self-energy contributions,
\beqar
 \Sigma^{f}_{\LRS}\left(p^2\right) =  
\Sigma^{f}_\rL \left(p^2\right) 
+ \Sigma^{f}_\rR  \left(p^2\right) 
+ 2\, \Sigma^{f}_{\mathrm{S}} \left(p^2\right)\,,
\eeqar
and the expanded counterterm reads
\beqar
\delta \mu_f &=& \frac{\mu_f}{2} \, \Sigma^{f}_{LRS} \left(\mu_f^2\right)
=  \frac{\mu_f}{2} \, \left[ 
\Sigma^{f}_{\LRS} \left(M_f^2\right)
-\ri M_f\Gamma_f\,
\Sigma'^{f}_{\LRS} \left(M_f^2\right)  +\ri c_f \right] \,.
\label{eq:cmsmassctF}
\eeqar

\paragraph{Counterterms for real masses}
When $\Gamma_i$ is set to zero, 
unstable and stable particles are described as on-shell 
states with a real-valued mass parameter, $\mu_i=M_i$.
In this case a conventional on-shell renormalisation is applied,
\beqar
\label{eq:genericcmsmassctB}
\hat \Sigma^{i}(p^2) &=& \Sigma^{i}(p^2)-\delta M_i^2\,
\qquad\mbox{with}\qquad
\delta \mu_i^2 =\delta M_i^2 =
\widetilde\re\,\Sigma^{i}\left(p^2\right)\Big|_{p^2=M_i^2}\,.
\eeqar
Here the subtraction is restricted to the real part of the self-energy,
while the $\mathrm{Im}\,\Sigma$ contribution must be retained, since it is not
included in the renormalised parameter $M_i^2$.
More precisely,  the $\widetilde\re$ operator in \refeq{eq:genericcmsmassctB}
truncates only the imaginary parts associated with the
UV-finite absorptive parts of two-point integrals,%
\footnote{\label{truncation} In practice, the truncation of absorptive
contributions is implemented at the level of the scalar two-point
integrals through 
\beqar
\widetilde\re\, B_0(p^2,m_1,m_2) =     
\begin{cases}
      \re\, B_0(p^2,m_1,m_2), & \text{if}\ p^2 > m_1^2+m_2^2, \\
      B_0(p^2,m_1,m_2), & \text{otherwise}\,,
    \end{cases} 
\eeqar
and in the same way for $B'_0$.
For the derivative of self-energies also the
following formulas for $B_1$ and $B'_1$ functions are used
 \begin{align}
 \widetilde\re\, B_1(p^2,m_1,m_2) &=
\frac{m_2^2-m_1^2}{2p^2}\left[\widetilde\re\,
B_0(p^2,m_1,m_2)-B_0(0,m_1,m_2) \right] 
-\frac{1}{2}\widetilde\re\, B_0(p^2,m_1,m_2)\,,\\
  \widetilde\re\, B'_1(p^2,m_1,m_2) &=
-\frac{m_2^2-m_1^2}{2p^4}\left[\widetilde\re\,
B_0(p^2,m_1,m_2)-B_0(0,m_1,m_2) \right]
+\frac{m_2^2-m_1^2-p^2}{2p^2}\widetilde\re\, B'_0(p^2,m_1,m_2)\,.
 \end{align}
}
while, in order to ensure a consistent cancellation of UV divergences,
all other imaginary parts associated with 
complex-valued couplings or complex masses inside the loop
are kept throughout. 
The explicit on-shell mass counterterms for Higgs bosons, 
vector bosons and fermions read
\beqar
\label{eq:oshmassct}
 \delta \mu_H^2 &=&  \delta M_H^2 \,=\, \widetilde\re\, \Sigma^{H} \left(M_H^2\right)\,,\\
\label{eq:osvmassct}
\delta \mu_V^2 &=& \delta M_V^2 \,=\, \widetilde \re\, \Sigma^{V}_\rT\left(M_V^2\right)\,,\\
\delta \mu_f &=&  \delta M_f \,=\,  \frac{M_f}{2} \widetilde\re\, \Sigma^{f}_{\LRS}\left(M_f^2\right)
\,.
\label{eq:osfmassct}
\eeqar

\paragraph{Yukawa couplings} 
At NLO EW, Yukawa couplings \refeq{eq:SMyuk} are always related to fermion
masses as predicted by the SM.  Thus Yukawa masses and physical fermion
masses, as well as the respective counterterms, are equal to each other.
This implies
\bea
\frac{\delta \lambda_f}{\lambda_f} &=&  
\frac{\delta \muyuk{f}}{\muyuk{f}}\,=\,
\frac{\delta \mu_f}{\mu_f}\,.
\eea
For the renormalisation of the fermion masses $\mu_f$ only 
the on-shell scheme, or its complex-mass scheme variant, are supported.

\paragraph{Wave functions} The wave-function renormalisation constants
(WFRCs) $\delta Z_{ij}$ are defined in a way that one-loop propagators do
not mix, and their residues are normalised to one.  Thus renormalised
amplitudes correspond directly to $S$-matrix elements and do not require
additional LSZ factors.  On the one hand, due to the presence of absorptive
contributions and complex parameters, in the complex-mass scheme the $\delta Z_{ij}$ constants can
acquire complex values.  
On the other hand, the WFRCS for on-shell particles
are usually defined as real parameters~\cite{Denner:1991kt}.
As explained in detail below, 
in \OpenLoops these two approaches are reconciled
by implementing WFRCs in a way that is consistent
with~\cite{Denner:1991kt} when the width parameters $\Gamma_i$ are set to
zero for all particles, while imaginary $\delta Z_{ij}$ contributions are
taken into account wherever they are strictly needed for the consistency of the complex-mass
scheme at $\ord(\alpha)$.


At NLO, the renormalisation of the field $\varphi_i$ associated with a
certain external leg yields 
\beq
\bigg|\Big(\delta_{ij}+\frac{1}{2}\sum_j\delta Z_{ij}
\Big)\calM^{(j)}_0\bigg|^2
=
\Big(1+\re \left(\delta Z_{ii}\right)\Big)\Big|\calM^{(i)}_0\Big|^2
+
\sum_j\re\Big[\Big(\calM_0^{(i)}\Big)^*\delta Z_{ij}\calM^{(j)}_0\Big]
+\ord(\alpha^2).
\eeq
Since the imaginary parts of the diagonal WFRCs $\delta Z_{ii}$ 
contribute only at $\ord(\alpha^2)$, in \OpenLoops we omit them 
by defining
\begin{align}
\label{eq:diagwfrcs}
\delta Z_{AA} &= -\re\, \Sigma'^{A}_{T}\left(0\right)\,,     
&\delta Z_{ZZ} =& -\re\,\Sigma'^{ZZ}_{T}\left(M_Z^2\right)\,,
\nonumber\\
\delta Z_{WW} &=-\re\, \Sigma'^{W}_{T}\left(M_W^2\right)\,,   
&\delta Z_{H} =& -\re\, \Sigma'^{H}\left(M_H^2\right)\,.  
\end{align}
In contrast, the non-diagonal WFRCs associated with $\gamma$--$Z$ mixing are defined as
\beqar
\label{eq:nondiagwfrcs}
\delta Z_{ZA} &=& 2\, \frac{\widetilde\re\,\Sigma_\rT^{AZ}(0)}{\mu_Z^2}\,,
\\ 
\delta Z_{AZ} &=& -2\, \widetilde\re\,\frac{\Sigma_\rT^{AZ}\left(\mu_Z^2\right)}{\mu_Z^2} 
= -2\, \frac{\widetilde\re\Sigma_\rT^{AZ}\left(M_Z^2\right)}{\mu_Z^2} 
+2\ri \frac{\Gamma_Z}{M_Z} \Sigma'^{AZ}_{T}\left(M_Z^2\right)\,,
\eeqar
where $\Sigma_\rT^{AZ}(Q^2)$ denotes the transverse part of the $\gamma$--$Z$
mixing energy.
Here the imaginary part of $\mu_Z$ in the denominator is retained in order to
ensure UV cancellations in the complex-mass scheme, while absorptive parts
are truncated\footnote{Note that $\widetilde \re\, \Sigma_\rT^{AZ}(0)=
\Sigma_\rT^{AZ}(0)$ since $\Sigma_\rT^{AZ}(0)$ is free from absorptive parts.} in
order to match the conventional on-shell scheme when all $\Gamma_i$ are set
to zero. For $\delta Z_{AZ}$ the mixing energy at $p^2=\mu_Z^2$ 
is expressed through an expansion around $p^2=M_Z^2$ neglecting terms of 
$\ord(\Gamma^2/M^2)$. However, in practice this expansion is
irrelevant, since $\delta Z_{AZ}$ only contributes for processes with external $Z$-bosons, where $\Gamma_Z=0$
is required.

At NLO EW, the independent renormalisation 
of left- and right-chiral fields corresponds 
to a diagonal renormalisation matrix in chiral space,
\bea
\label{eq:fermionwfrcs}
\delta Z_f &=& \delta Z_{f_\rR}\omega_\rR+\delta Z_{f_\rL}\omega_\rL
\qquad
\mbox{with}
\qquad  
\omega_{\rR,\rL}=\frac{1}{2}(1\pm \gamma_5)\,.
\eea
For massless fermions, the matrix
\refeq{eq:fermionwfrcs} is diagonal also in helicity space, and
imaginary parts can be amputated 
similarly as for the diagonal WFRCs \refeq{eq:diagwfrcs}.
In contrast, for massive fermions the matrix
\refeq{eq:fermionwfrcs} mixes left- and right-handed 
helicity states. Thus, in this case imaginary parts are
treated in a similar way as for the non-diagonal WFRCs~\refeq{eq:nondiagwfrcs}.
Thus the explicit form of the fermionic WFRCs $\delta Z_{f_{\rR,\rL}}$
reads
\beqar
\delta Z_{f_{\lambda}} &=& 
\begin{cases}
-\re\,\Sigma'^{f}_{\lambda}\left(0\right)\,,
& \quad \mbox{for}\quad M_f = 0\,, \\[3mm]
-\widetilde\re\,\Sigma^f_{\lambda}(M_f^2)
-M_f^2 \,\widetilde\re\,\Sigma'^{f}_{\LRS}(M_f^2)
& \quad\mbox{for}\quad M_f>0\,. 
\end{cases}
\label{eq:EWfermionWFRC}
\eeqar

\paragraph{Variations of the complex-mass scheme} Certain aspects of the
complex-mass scheme at $\ord(\alpha)$ can be changed using the 
parameter \texttt{complex\_mass\_scheme} as detailed in the
following.

\bit

\item[(i)] \texttt{complex\_mass\_scheme=1} (default) corresponds to the implementation
described above: the 
complex-mass counterterms 
\refeq{eq:cmsmassctS}--\refeq{eq:cmsmassctF} are used when $\Gamma_i>0$, 
and the on-shell mass counterterms \refeq{eq:oshmassct}--\refeq{eq:osfmassct} are used when
$\Gamma_i=0$, while for WFRCs the generic formulas
\refeq{eq:diagwfrcs}--\refeq{eq:EWfermionWFRC} are applied.
As discussed above,
this flexible approach guarantees a 
consistent one-loop description of processes like
$pp\to t \bar t \ell^+ \ell^-$,
where unstable particles 
occur both as internal resonances
and as on-shell external states.

\item [(ii)] \texttt{complex\_mass\_scheme=0} keeps 
the complex masses \refeq{eq:complexmasses} unchanged
but deactivates the complex-mass scheme at the level of all
$\ord(\alpha)$ counterterms: for mass counterterms
the on-shell formulas
\refeq{eq:oshmassct}--\refeq{eq:osfmassct} are used throughout;
moreover, the $\widetilde \re$ operations in \refeq{eq:oshmassct}--\refeq{eq:osfmassct}
and \refeq{eq:diagwfrcs}--\refeq{eq:EWfermionWFRC} are
replaced by a complete truncation of the imaginary parts at the level of the
full counterterms.
This option is implemented for validation purposes.
Depending on the process,
it can result in incomplete pole cancellations or other inconsistencies,
in particular when internal or external particles with $\Gamma_i>0$ are
present.

\item [(iii)] \texttt{complex\_mass\_scheme=2} corresponds to the implementation of the
complex-mass scheme in \Recola~\cite{Actis:2016mpe}. In this case
all mass counterterms are evaluated with the complex-mass scheme
formulas \refeq{eq:cmsmassctS}--\refeq{eq:cmsmassctF},
while all $\re$ and $\widetilde \re$ operators are removed from 
\refeq{eq:diagwfrcs}--\refeq{eq:EWfermionWFRC}, \ie 
all imaginary parts of WFRCs are kept exact. 
\eit

\paragraph{Light-fermion contributions to $\Delta\alphamz$}

The $\ord(\alpha)$ corrections to processes with on-shell external photons
involve the renormalisation constant $\delta Z_{AA}$ defined in \refeq{eq:diagwfrcs}, which is related to
the photon vacuum polarisation $\Pi^{\gamma\gamma}(Q^2)$ at $Q^2\to 0$ via
\beq
\label{eq:dZAA}
\delta Z_{AA}  =  -\re\, \Sigma'^{AA}_{T}\left(0\right) =
-\Pi^{\gamma\gamma}(0)\,.
\eeq
Terms involving $\Pi^{\gamma\gamma}(0)$ occur 
also in the $\alpha(0)$ counterterm \refeq{eq:alphazeroewct},
which contributes to any process that is parametrised in terms of
$\alpha(0)$ at tree level.
In the presence of $\Pi^{\gamma\gamma}(0)$ terms, high-energy cross sections
become sensitive to large logarithms of the
light-fermion masses, $m_f=\{m_e$, $ m_\mu$, $ m_\tau$, $
m_u$, $ m_d$, $ m_s$, $ m_c$, $ m_b\}$.  In  \OpenLoops such
a dependence is systematically avoided by 
replacing $\Pi^{\gamma\gamma}(0)$ through $\Delta\alphamz$ 
via
\bea
\label{eq:PiAAzero}
\Pi^{\gamma\gamma}(0)
&=& \Pi^{\gamma\gamma}_{\rm heavy}(0)
+ \Pi^{\gamma\gamma}_{\rm light}\left(M_Z^2\right) + 
\left[\Pi^{\gamma\gamma}_{\rm light}(0)
-\Pi^{\gamma\gamma}_{\rm light}\left(M_Z^2\right)\right]\nonumber\\
&=& \Pi^{\gamma\gamma}_{\rm heavy}(0)
 + \Pi^{\gamma\gamma}_{\rm light}\left(M_Z^2\right) + \Delta \alphamz\,.
\eea
Here $\Pi^{\gamma\gamma}(Q^2)$ is split into a ``heavy'' contribution
due to $W$-boson and top-quark loops, plus a remnant ``light'' contribution.
The latter is subtracted at $Q^2=M_Z^2$. In this way the 
sensitivity to light-fermion masses 
is isolated in $\Delta \alphamz$,
which describes the running of $\alpha$ from $Q^2=0$ to $M_Z^2$.
The explicit light-fermion mass dependence is avoided by 
expressing $\Delta \alphamz$
as
\beq
\label{eq:DalphaMZ}
\Delta \alphamz = 
1-\frac{\alpha(0)}{\alphamz}\,,
\eeq
where $\alpha(0)$ and $\alphamz$ are evaluated using the numerical 
values of the parameters $\texttt{alpha\_qed\_0}$ and
\texttt{alpha\_qed\_mz}  introduced in \refse{sec:ewschemes}.
By default, \refeq{eq:DalphaMZ} is used throughout apart 
for the $\Delta\alphamz$ terms associated with external off-shell photons.
In that case, as discussed in the context of eq.~\refeq{eq:dzgammaoff},
the following explicit expression with dimensionally
regularised mass singularities is used,
\beq
\label{eq:Delta_alphadimreg}
\begin{split}
\Delta \alpha^{(\reg)} (\MZ^2)
\,=&\;
\Pi_{\rm light}^{\gamma\gamma}(0)-\Pi_{\rm light}^{\gamma\gamma}(\MZ^2)
\\
\,=&\;
\frac{\alpha}{2\pi}\;\gamma_\gamma 
\left[\frac{\ceps}{\epsir}+\ln\left(\frac{\mudimir^2}{\MZ^2}\right)
+\frac{5}{3}\right]
-\frac{\alpha}{3\pi}
\sum_{f\in F_{\mathrm{m}}} 
N_{\mathrm{C},f}\, Q^2_f 
\left[\ln\left(\frac{m_f^2}{\MZ^2}\right)+\frac{5}{3}\right]\,.
\end{split}
\eeq
Here $\gamma_\gamma=\gamma^{\QED}_\gamma$ is the anomalous dimension 
defined in \refta{tab:anomalousdim},
and $F_{\mathrm{m}}$ is the set of light
fermions with $0<m_f<\MZ$.
For later convenience, 
we also define the $\Delta \alpha$ conversion term
\beqar
\label{eq:Dalpharegshift}
\mathcal{D}\alpha^{(\reg)}(M_Z^2)&=& 
\Delta\alpha^{(\reg)}(M_Z^2)-\Delta\alpha(M_Z^2)\,.
\eeqar

Concerning $\Delta\alphamz$ contributions to 
processes that do not involve external off-shell photons, 
if the $\alpha$-input scheme is chosen as recommended in \refse{sec:ewschemes},
all $\Delta\alphamz$ terms drop out in renormalised 
 matrix elements, and the 
treatment of $\Delta\alphamz$ is irrelevant.
Instead, for alternative choices of the $\alpha$-input scheme 
that yield $\Delta \alphamz$ corrections,
the prescription \refeq{eq:DalphaMZ} becomes relevant and 
guarantees sound physical results irrespectively of the $m_f$ input values, 
\ie also in the case  of  vanishing light-fermion masses, where 
$\Pi^{\gamma\gamma}(0)$ is formally divergent.

\paragraph{EW coupling counterterms}
The renormalisation of the EW gauge couplings \refeq{eq:gaugecouplings} is
implemented through counterterms for the photon coupling $e$ and
the weak mixing angle $\thetaw$.
The latter is defined in terms of the weak-boson masses by imposing the 
relation \refeq{eq:weakmixing} to all orders. The resulting counterterm
reads
\beqar
\frac{\delta\cos^2 \thetaw}{\cos^2 \thetaw}
&=& 
{}-\frac{\delta\sin^2 \thetaw}{\cos^2 \thetaw}
\,=\,
\frac{\delta \mu^2_W}{\mu_W^2}-\frac{\delta \mu_Z^2}{\mu_Z^2}\,.
\eeqar
Here, for $\Gamma_{W,Z}>0$ and $\Gamma_{W,Z}=0$, 
the mass counterterms $\delta \mu^2_{W,Z}$ are computed according to 
\refeq{eq:cmsmassctV} and \refeq{eq:osvmassct}, respectively.
As discussed in \refse{sec:ewschemes}, in \OpenLoops the 
photon coupling $e$ can be defined according to three different schemes,
which correspond to different renormalisation conditions.
The form of the related counterterm $\delta Z_e$ in the various 
schemes is as follows.

\begin{itemize}
\item[(i)] {\bf $\alpha(0)$-scheme:}
the parameter $\alpha$ is identified with the strength of the photon 
coupling at $Q^2\to 0$. The resulting counterterm reads
\beqar
\label{eq:alphazeroewct}
\delta Z_e \vert_{\alpha(0)} &=& 
-\frac{1}{2}\,\mathrm{Re}\,\left(\delta Z_{AA} + \frac{s_W}{c_W} \delta Z_{ZA}\right)
\nonumber\\
&=&
\frac{1}{2}\,\mathrm{Re}\,\left[\Pi^{\gamma\gamma}_{\rm heavy}(0)
+ \Pi^{\gamma\gamma}_{\rm light}\left(M_Z^2\right) + \Delta \alphamz
- \frac{2 s_W}{c_W} \frac{\Sigma_\rT^{AZ}(0)}{\mu_Z^2}\right]\,.
\eeqar

\item[(ii)] {\bf $\GF$-scheme:}
%
the QED coupling is related to the Fermi constant 
through \refeq{eq:alphaGmu}. This relation can be connected 
to the $\alpha(0)$-scheme via
\beq
\frac{\alphamu}{\big\vert s_W^2 \mu_W^2\big\vert }
=
\frac{\sqrt{2}G_\mu}{\pi} = 
\alpha(0)\left\vert \frac{1+\Delta r}{s_W^2 \mu_W^2}\right\vert\,,
\eeq
where $\Delta r$ represents the radiative corrections to the muon decay, 
\ie to the Fermi constant, in the $\alpha(0)$-scheme~\cite{Denner:1991kt}.
This leads to the $G_\mu$-scheme counterterm
\beqar
\label{eq:alphamuct}
\delta Z_e \vert_{\gmu} &=& 
\delta Z_e \vert_{\alpha(0)} -  \frac{1}{2}\mathrm{Re}\left(\Delta r\right)
=\frac{1}{2}\,\mathrm{Re}\, \Bigg\{
\frac{\delta s_W^2}{s_W^2}
+\frac{\delta \mu_W^2-\Sigma^{W}_\rT(0)}{\mu_W^2}
\nonumber\\
&&{}-\frac{\alpha}{\pi s_W^2}\left[
\frac{\ceps}{\epsuv}+
\ln\left(\frac{\mu^2_{\mathrm{UV}}}{\mu_Z^2}\right)
+
\frac{3}{2}+
\frac{7-12 s_W^2}{8 s_W^2}
\ln\left(\frac{\mu_W^2}{\mu_Z^2}\right)
\right]
\Bigg\}\,.
\eeqar
Note that, since $\alphamu$ is effectively defined at the EW scale,
its counterterm \refeq{eq:alphamuct}
does not depend on $\Pi^{\gamma\gamma}(0)$.


\item[(iii)]{\bf $\alphamz$-scheme:} the photon coupling is defined as the
strength of the pure QED interaction at $Q^2=M_Z^2$. This corresponds to
the counterterm
\beq
\delta Z_e \vert_{\alphamz} = 
\delta Z_e \vert_{\alpha(0)} -\frac{\Delta\alphamz}{2}
=
\frac{1}{2}\,\mathrm{Re}\,
\left[\Pi^{\gamma\gamma}_{\rm heavy}(0)
         +\Pi^{\gamma\gamma}_{\rm light}(M_Z^2)
         - \frac{2 s_W}{c_W} \frac{\Sigma_\rT^{AZ}(0)}{\mu_Z^2}
\right]\,.
\eeq
Also in this case $\Pi^{\gamma\gamma}(0)$ drops out.

\end{itemize}

In \OpenLoops the appropriate counterterm $\delta Z_e$ is selected
automatically based on the choice of the $\alpha$-input scheme.  The latter
is controlled by the parameter \texttt{ew\_scheme} as detailed in
\refta{tab:ewschemes}.

\paragraph{External photons}
In processes with external photons, the renormalisation of $e$ is
automatically adapted to the coupling rescaling factors
\refeq{eq:LSZgamma}--\refeq{eq:lszfactors} for on-shell and off-shell
external photons.
To this end, the coupling $e$ is renormalised in two steps.  
First, each factor $e$ that is present at tree level is renormalised with a
standard $\delta Z_e$ counterterm corresponding to the $\alpha$-scheme selected
by the user. Then, a finite renormalisation 
of the rescaling factors 
\refeq{eq:LSZgamma}--\refeq{eq:lszfactors} is applied,
\beqar
\rgammaonoffbare
&=&
\rgammaonoff\left(1+\delta\zgammaonoff\right)\,,
\label{eq:gammarescren}
\eeqar
which yields an extra counterterm 
$\delta\zgammaonoff$ for each coupling $\alpha$ associated with external 
photons.
Combined with the standard photon-coupling and wave-function counterterms $2 \delta Z_e+\delta
Z_{AA}$,
this results in a renormalisation factor
\beqar
\label{eq:kgammaonoff}
\delta\kgammaonoff
&=&
2\,\delta Z_e+\delta\zgammaonoff+\delta Z_{AA}\,,
\eeqar
for each external photon.

\bit
\item[(i)] For {\bf on-shell photons} the coupling $\alpha(0)$ is used. Thus,
\beqar
\delta\zgammaon
&=&
2\left[\delta Z_e \vert_{\alpha(0)} - \delta Z_e\right],
\eeqar
and  $\delta\kgammaon=2\,\delta Z_e\vert_{\alpha(0)}+\delta Z_{AA}$ yields the correct 
coupling counterterm $\delta Z_e\vert_{\alpha(0)}$.
Note that, as a result of the choice of a low-energy coupling,
the $\Delta\alphamz$ contributions to $\delta Z_{AA}$ and
$\delta Z_e\vert_{\alpha(0)}$ cancel out in
$\delta\kgammaon$.

\item[(ii)] For {\bf off-shell photons} the high-energy coupling $\alphaoff$
defined in \refeq{eq:alphaoff} is used.
As a result, the $\Delta\alphamz$ contribution to $\delta Z_{AA}$
remains uncancelled, and the renormalised scattering amplitude
depends on large logarithms of the light-fermion masses.
In photon-induced hadronic collisions,
such logarithmic mass singularities are 
cancelled by collinear singularities associated with 
virtual \mbox{$\gamma \to f \bar f$} splitting contributions 
to the photon-PDF counterterm~\cite{Kallweit:2017khh} 
(see Section~\ref{sec:irsubtration}).  
The latter are typically handled in dimensional regularisation with massless
light fermions, which results in collinear singularities of the form $1/\epsir$.
For consistency, the same
regularisation must be used also for the related light-fermion contributions
from $\Delta\alphamz$.
To this end, the finite renormalisation factor for
off-shell photons is defined as
\beqar
\label{eq:dzgammaoff}
\delta\zgammaoff
&=&
2\,\left[\delta Z_e \vert_{\alphaoff} - \delta Z_e\right]
-\mathcal{D}\alpha^{(\reg)}(M_Z^2),
\eeqar
where the counterterm $\delta Z_e \vert_{\alphaoff}$ corresponds to the
renormalisation scheme associated with $\alphaoff$ according to
\refeq{eq:alphaoff}, while $\mathcal{D}\alpha^{(\reg)}(M_Z^2)$, defined in
\refeq{eq:Dalpharegshift}, converts $\Delta \alpha(M_Z^2)$ into its
dimensionally regularised variant \refeq{eq:Delta_alphadimreg}.
The resulting overall renormalisation
factor for off-shell photons reads
\beqar
\label{eq:fullphotonfactor}
\delta\kgammaoff
&=& 2\delta Z_e \vert_{\alphaoff} +\delta Z^{(\reg)}_{AA}\,,
\eeqar
with
\beqar
\delta Z^{(\reg)}_{AA} &=&
\delta Z_{AA}- \mathcal{D}\alpha^{(\reg)}(M_Z^2)
=
-\left[\Pi^{\gamma\gamma}_{\rm heavy}(0)
+ \Pi^{\gamma\gamma}_{\rm light}\left(M_Z^2\right) + \Delta^{(\reg)} \alphamz\right]\,.
\eeqar

\eit

In \OpenLoops, the counterterms $\delta \zgammaonoff$ 
are automatically adapted to the settings that control the
type of external photons and their tree-level couplings as summarised in
\refta{tab:photontreatment}.

For the various $\Delta \alpha(M_Z^2)$ terms that enter the factors $\delta
Z_e$, $\delta Z_{AA}$ and $\delta \zgammaonoff$ associated with
external photons, depending on the type of photon, either the numerical
expression \refeq{eq:DalphaMZ} or the dimensionally regularised form
\refeq{eq:Delta_alphadimreg} are used as explained above.
Alternatively, it is possible to  enforce
the usage of $\alpha^{(\reg)}(M_Z^2)$
in all terms associated with external
photons by setting \texttt{all\_photons\_dimreg=1} (default=0).

\begin{table}
\renewcommand{\arraystretch}{1.3}
\centering
\begin{tabular}{|c|c|c|c|c|c|} \hline
photon type & \texttt{iPDG}  & switcher (1=on, 0=off) &coupling &  $\Delta \alpha$ &  $\gamma \to f \bar f$\\ \hline
unresolved& 22 &&  $\alpha$ & $\Delta\alphamz$ & off\\
on-shell& 2002 & \footnotesize{\texttt{onshell\_photons\_lsz}} &  $\alpha(0)$ 
& $\Delta\alphamz$ & off\\
off-shell& -2002 & \footnotesize{\texttt{offshell\_photons\_lsz}}    & $\alphaoff$ 
& $\Delta\alpha^{\reg}(M_Z^2)$  & on\\ \hline
\end{tabular}
\caption{PDG identifiers for photons and switchers that control the
coupling factors and renormalisation constants 
for the different types of external photons introduced in 
\refse{sec:ewschemes}.
The high-energy coupling $\alphaoff$ is defined in \refeq{eq:alphaoff}.  
If the switchers are set to zero (default=1)
the standard user-defined coupling $\alpha$ is used, and
the related $\delta Z^{(\mathrm{on/off})}$ factors are deactivated.
As indicated in the last column, 
contributions from collinear $\gamma \to f \bar f$ splittings
are included in Catani--Seymour's $\mathbf{I}$-operator (see
\refse{sec:irsubtration})
only for off-shell photons.}
\label{tab:photontreatment}
\end{table}

\subsection{Infrared subtraction}
\label{sec:irsubtration}
\def\calB{\mathcal{B}}
\def\calV{\mathcal{V}}
\def\calQ{\mathcal{Q}}
\def\calT{\mathcal{T}}
\def\calTSC{\mathcal{T}^{\mathrm{SC}}}
\def\pout{p_{\mathrm{out}}}
\def\qout{q_{\mathrm{out}}}
\def\Pout{P_{\mathrm{out}}}
\def\Qout{Q_{\mathrm{out}}}
\def\pin{p_{\mathrm{in}}}
\def\qin{q_{\mathrm{in}}}
\def\Pin{P_{\mathrm{in}}}
\def\Qin{Q_{\mathrm{in}}}
\def\pout{p}
\def\qout{q}
\def\Pout{P}
\def\Qout{Q}
\def\pin{p}
\def\qin{q}
\def\Pin{P}
\def\Qin{Q}
\def\pborn{p_{\mbox{\tiny Born}}}
\def\qborn{q_{\mbox{\tiny Born}}}

One-loop matrix elements with on-shell external legs involve divergences of
IR (soft and collinear) origin, which take the form
of double and single $1/\epsir$ poles in \mbox{$D=4-2\epsir$}
dimensions.
In \OpenLoops such divergences 
can be subtracted through an automated implementation of
Catani--Seymour's $\mathbf{I}$-operator that accounts for QCD
singularities~\cite{Catani:1996vz,Catani:2002hc} as well as for
singularities of QED
origin~\cite{Dittmaier:1999mb,Dittmaier:2008md,Gehrmann:2010ry,Kallweit:2017khh,Schonherr:2017qcj}.
The singular part of the
$\mathbf{I}$--operator is universal and can be used 
to check the cancellation of IR poles in any one-loop calculation.
Moreover, the full $\mathbf{I}$--operator 
provides a useful
building block for NLO calculations based on Catani--Seymour's dipole subtraction.

In addition to the $\mathbf{I}$-operator, as documented in
\refse{sec:amplitudecalls} and \refapp{app:native:iop}, \OpenLoops
provides also routines for more general building blocks of IR divergences,
namely colour- and gluon-helicity correlated Born matrix elements for QCD
singularities, and corresponding charge- and photon-helicity correlations
for QED singularities.

In \OL  it is possible to calculate the $\mathbf{I}$-operator
contributions that are required for the NLO corrections to 
conventional processes with $\calM_0\neq 0$
and for loop-induced processes.
The relevant \OpenLoops functions are
\texttt{evaluate\_iop} 
and \texttt{evaluate\_iop2} 
(see \refapp{app:native:iop}).
At a certain order $\alphaS^{P}\alpha^{Q}$,
their output corresponds to
\bea
\calW^{(P,Q)}_{\treeiop} &=&
\langle
M_0| \mathbf{I}(\{p\};\epsir) | M_0\big\rangle 
\bigg|_{\alphaS^{P}\alpha^Q}\,,\qquad
\calW^{(P,Q)}_{\onelsqiop}=
\langle M_1| \mathbf{I}(\{p\};\epsir) | M_1\big\rangle 
\bigg|_{\alphaS^{P}\alpha^Q}\,,
\label{eq:ioperator}
\eea 
where the $\mathbf{I}$-operator 
consists of the following IR insertions of order $\alphaS$ and
$\alpha$ into LO contributions of order $\alphaS^{P-1}\alpha^{Q}$
and $\alphaS^{P}\alpha^{Q-1}$, 
\bea
\langle
M_i| \mathbf{I}(\{p\};\epsir) | M_i\big\rangle 
\bigg|_{\alphaS^{P}\alpha^Q}
&=&  
-\frac{\alphaS}{2\pi} 
\ceps 
 \sum_{j\in\mathcal{S}
}  
\sum_{\substack{k\in\mathcal{S}
\\ k\neq j}
}  
\calV^{\rm QCD}_{jk}(\epsir)\, 
\langle M_i|\,
\calT^{\QCD}_{jk}
\,|M_i\big\rangle
\bigg|_{\alphaS^{P-1}\alpha^Q} 
\nonumber \\
&&{}- \frac{\alpha}{2\pi}\ceps  
\sum_{j\in\mathcal{S}
}  
\sum_{\substack{k\in\mathcal{S}
\\ k\neq j}
}  
\calV^{\rm QED}_{jk}(\epsir)\, 
\langle M_i|\, \calT^{\QED}_{jk}\,
|M_i\big\rangle\bigg|_{\alphaS^{P}\alpha^{Q-1}} \,.
\label{eq:ioperatorB}
\eea 
Here, helicity/colour sums and symmetry factors are as in~\refeq{eq:Wtree}--\refeq{eq:Wloop2}.
The indices $j$ and $k$ 
represent so-called emitter and spectator partons,
respectively.  They are summed over the full set
$\mathcal{S}=\mathcal{S}_{\rm in}\cup \mathcal{S}_{\rm out}$ of initial
($\mathcal{S}_{\rm in}$) and final-state ($\mathcal{S}_{\rm out}$) partons.
By default both  $\alphaS$ and $\alpha$ insertions are activated, but
for processes with less than two external $q\bar q$ pairs 
only one of them contributes.
Via the switch 
\texttt{ioperator\_mode} (default=0) either only $\alphaS$ (\texttt{ioperator\_mode}=1) or 
only $\alpha$ insertions (\texttt{ioperator\_mode}=2) can be selected.
The $\ord(\alphaS)$ contribution involves the colour correlator
\bea
\calT^{\QCD}_{jk}=\begin{cases}
\frac{T^a_j\,T^a_k}{T_j^2} & \mbox{if $j$ and $k$ are gluons or
(anti-)quarks,}
\\[2mm]
0 & \mbox{otherwise,}\\
\end{cases}
\label{eq:colcorrel}
\eea
where $T^a_i$ denotes the SU(3) generator\footnote{Here all SU(3) generators 
as well as electromagnetic charges should be understood in terms of 
incoming charge flow.}
acting on the external leg $i$,
and $T_j^2 = T^a_j T^a_j$. The corresponding 
charge correlator at $\ord(\alpha)$ is defined as
\bea
\calT^{\QED}_{jk}=\begin{cases}
\frac{Q_j\,Q_k}{Q_j^2} & \mbox{if $j$ and $k$ are charged (anti-)fermions or $W^\pm$
bosons,} 
\\
- \frac{1}{n_{{\rm I},j}}  
& \mbox{if $j$ is an off-shell photon and}\; 
k\in \mathcal{S}_{\rm in}\backslash\{j\},\\
0  & \mbox{if $j$ is an on-shell photon or any other neutral parton.}
\end{cases}
\label{eq:chargecorrel}
\eea
Here $Q_i$ denotes the electromagnetic charge of parton $i$, while
$n_{{\rm I},j}$ is the number of initial-state partons in 
$\mathcal{S}_{\rm in}\backslash\{j\}$.
By definition, on-shell photons do not undergo collinear splittings at NLO.
Thus, $\calT^{\QED}_{jk}$ vanishes when the emitter 
$j$ is an on-shell photon.
Vice versa, off-shell photons are subject to final-state $\gamma\to f\bar{f}$
and initial-state $f\to f\gamma$ splittings at NLO.
The related $-1/n_{{\rm I},j}$ term in \refeq{eq:chargecorrel}
is such that the recoil of the collinear radiation
is shared by all initial-state partons that belong to 
$\mathcal{S}_{\rm in}\backslash\{j\}$~\cite{Kallweit:2017khh}.

The functions $\calV_{jk}(\epsir)$ in \refeq{eq:ioperatorB}
contain single and double poles in $\epsir$. They 
depend on the kinematic quantities $s_{jk}=\vert 2p_jp_k \vert$ and
\bea
v_{jk}=\sqrt{1-4\frac{M_j^2 M_k^2}{s_{jk}^2}}\,,\qquad
q_{jk}^2=s_{jk}+M_j^2+M_k^2\,,\qquad
\Omega_{jk}^{(i)}=\frac{(1-v_{jk})s_{jk}+2M^2_{i}}{(1+v_{jk})s_{jk}+2M^2_{i}}\,.
\eea
Using the  constants defined in \refta{tab:anomalousdim},
they can be written as~\cite{Catani:2002hc}
\bea
\calV_{ij}^{\QCD/\QED}(\epsir) &=&
\calQ_j^{2,\QCD/\QED}\bigg\{
\frac{1}{2 v_{jk}}\bigg[\sum_{i=j,k}
  V^{(i)}_{{\rm S},jk}(\epsir, M_i)\bigg]
+ V_{{\rm NS},jk}^{\QCD/\QED}
-\frac{\pi^2}{3}\bigg\}
\nonumber\\
&&{}
+\gamma^{\QCD/\QED}_j\,\bigg[U_{j}(\epsir,M_j)
+\ln\left(\frac{\mudimir^2}{s_{jk}}\right)\bigg]
+K^{\QCD/\QED}_j\,.
\eea
%
%
\begin{table}
\renewcommand{\arraystretch}{1.3}
\centering
\begin{tabular}{c|c|c|c|c} 
interaction & $j$ & $\calQ_j^{2,\QCD/\QED}$ & $\gamma_j^{\QCD/\QED}$ & $K_j^{\QCD/\QED}$ \\\hline
QCD & quark & $C_F$ & $\frac{3}{2} C_F$ &
$(\frac{7}{2}-\frac{\pi^2}{6})C_F$\\
QCD & gluon& $C_A$ & $\frac{11}{6}C_A - \frac{2}{3}T_R N_f$	 &     $(\frac{67}{18} - \frac{\pi^2}{6})C_A -
\frac{10}{9}T_RN_f$\\
QED & fermion or $W^\pm$ & $Q_j^2$ & $\frac{3}{2}Q_j^2$ & $(\frac{7}{2}-\frac{\pi^2}{6})
Q_j^2$\\
QED & $\gamma$ & $0$  &$-  \frac{2}{3}\Big[ N_C \big(N_{f,u}\, Q_u^2$ & $\frac{5}{3}\gamma_{\gamma}^{\rm QED}$ \\
&&& $+N_{f,d}\,
Q_d^2\big)+  N_{f,l}\,Q_l^2\Big]$&
\end{tabular}
\caption{Here $N_{f,u},N_{f,d},N_{f,l}$ are the numbers of massless 
up-type quarks,  down-type quarks and leptons, respectively, while
$N_f=N_{f,u}+N_{f,d}$.
Since massive external legs induce only soft singularities, 
external  $W^\pm$-bosons are treated in the same way as 
massive fermions with mass $M_W$ and charge $\pm 1$.
}
\label{tab:anomalousdim}
\end{table}
The singularities are contained in the functions
\bea
U_{j}(\epsir,M_j)=\begin{cases}
\frac{1}{\epsir}+1 & \mbox{if $M_j=0$}\,, \\[2mm]
\frac{2}{3}\frac{1}{\epsir}-\frac{1}{3}\ln\left(\frac{\mudimir^2}{M_j^2}\right)
-\frac{1}{3} & \mbox{if $M_j>0$}\,,\\
\end{cases}
\eea
and
\bea
V^{(i)}_{{\rm S},jk}(\epsir, M_i)&=&
\begin{cases} 
\frac{1}{\epsir^2}+ \frac{1}{\epsir}\ln\left(\frac{\mudimir^2 q_{jk}^2}{s^2_{jk}}\right)
+ \frac{1}{2}\ln^2\left(\frac{\mudimir^2 q_{jk}^2}{s^2_{jk}}\right) & \mbox{for
$M_i=0$}\,, \\
\ln\left(\Omega^{(i)}_{jk}\right)\left[
\frac{1}{\epsir}
+\ln\left(\frac{\mudimir^2 q_{jk}^2}{s^2_{jk}}\right)
-\frac{1}{2}\ln\left(\Omega^{(i)}_{jk}\right)
\right]-\frac{\pi^2}{6}
& \mbox{for $M_i>0$}\,. \\
\end{cases}
\eea
%
The functions $V_{{\rm{NS}},jk}$ are free from poles and vanish 
for $M_j=M_k=0$. For gluon and photon emitters
%
%
\begin{align}
  \label{eq:Vnsg}
  V^{\QCD/\QED}_{{\rm NS}, jk}\Bigg|_{j=g,\gamma}
=
\hat\gamma^{\QCD/\QED}_{j} 
\left[\ln\left(\frac{s_{jk}}{q_{jk}^2}\right) -
2\ln\left(\frac{q_{jk}-M_k}{q_{jk}}\right)- \frac{2M_k}{q_{jk}+M_k} \right] +
\frac{\pi^2}{6}-\Li_2\left(\frac{s_{jk}}{q^2_{jk}}\right)\,,
\end{align}
with $\hat\gamma^{\QCD}_{g} = \frac{\gamma^\QCD_g}{C_A}$
and\footnote{
Due to our recoil conventions 
for (off-shell) photon emitters in
\refeq{eq:chargecorrel},
$\hat\gamma^{\QED}_{\gamma}$
contributions are only relevant for massive initial-state spectators.}
$\hat\gamma^{\QED}_{\gamma} = \gamma^\QED_\gamma$.
For quarks, charged leptons and $W^{\pm}$ emitters we have
\bea
V^{\rm QCD/QED}_{{\rm NS}, jk}\Bigg|_{j=q,\ell,W}
\hspace{-22mm}&&\hspace{18mm}
\,=\,
\frac{\gamma_j^{\rm
\QCD/\QED}}{\mathcal{Q}_j^{2,\QCD/\QED}}\ln\left(\frac{s_{jk}}{q_{jk}^2}\right)
+\frac{1}{v_{jk}}\bigg[\ln(\Omega_{jk})\ln(1+\Omega_{jk})+2\Li_2(\Omega_{jk})
-\frac{\pi^2}{6}
\nonumber\\
&&{}
-\Li_2(1-\Omega^{(j)}_{jk})
-\Li_2(1-\Omega^{(k)}_{jk})
\bigg] 
 +
\ln\left(\frac{q_{jk}-M_k}{q_{jk}}\right)-2\ln\left(\frac{\left(q_{jk}-M_k\right)^2-M_j^2}{q^2_{jk}}\right)
\nonumber\\
&&{}-\frac{2M_j^2}{s_{jk}}\ln\left(\frac{M_j}{q_{jk}-M_k}\right)-\frac{M_k}{q_{jk}-M_k}+\frac{2M_k\left(2M_k-q_{jk}\right)}{s_{jk}}+\frac{\pi^2}{2}\,,
\eea
where $\Omega_{jk}=\frac{(1-v_{jk})}{(1+v_{jk})}$.
%

\section{Overview of the program} \label{sec:program}

This section describes various aspects that are relevant for the usage of
\OpenLoops in the context of external programs.  Once installed and linked to
an external program, \OpenLoops can
be controlled through its native interfaces for \Fortran and C/\cpp codes, or
using the standard BLHA interface~\cite{Binoth:2010xt,Alioli:2013nda}.
In the following, we introduce various functionalities of the \OpenLoops
interfaces, such as the registration of processes, the setting of
parameters, and the evaluation of different types of matrix elements.
In doing so we will always refer to the names of the relevant \Fortran
interface functions.  The corresponding C functions are named in the same
way with an extra \texttt{ol\_} prefix.

Further technical aspects, such as the signatures of the interfaces,
can be found in \refapp{app:native} and \refapp{app:otherinterfaces}.  
As discussed there, the multi-purpose 
Monte Carlo programs 
\Munich/\Matrix{}~\cite{Grazzini:2017mhc}, 
\Sherpa{}~\cite{Gleisberg:2008ta,Bothmann:2019yzt},
\Herwig{}~\cite{Bellm:2015jjp},
\Powheg{}~\cite{Alioli:2010xd}, \Whizard{}~\cite{Kilian:2007gr} and
\Geneva~\cite{Alioli:2012fc} dispose of built-in 
interfaces that control all relevant \OpenLoops 
functionalities in a largely automated way requiring only 
little user intervention.
Besides the \Fortran and C/\cpp interfaces the \OL package also contains a {\sc Python}
wrapper and a command line tool. Further details and examples 
of the {\sc Python} interface
are given in Appendix~\ref{app:python}.

The \OpenLoops{} program itself is written in \Fortran and consists of process-independent
 main code and process-dependent code provided in the form of process libraries, which can
be downloaded and automatically installed within the \OpenLoops{}
program for a wide range of processes in the Standard Model (SM) and Higgs
effective theory (HEFT), as detailed in the following. The process libraries
are automatically generated based on a (private)  process
generator implemented in \Mathematica.

\subsection{Download and Installation}
\label{se:installation}
\subsubsection{Installation of the main program}
\label{se:maininstallation}

This section describes the installation of the process-independent part of
the \OpenLoops program, which is denoted as base code.  
The calculation of specific scattering amplitudes 
requires additional process-specific libraries, denoted as process code.
Their installation is discussed in
\refse{se:proclibinstallation}.

\paragraph{Prerequisites}%
To install \OL a \Fortran compiler (\texttt{gfortran} 4.6 or later,
or \texttt{ifort}) and \texttt{Python}\,2.7 or 3.5 or later are needed.

\paragraph{Download}%
The process-independent part of the \OL program is available on the \texttt{Git} repository
\texttt{gitlab.com/openloops/OpenLoops}.
The latest release version can be found in the  master branch and downloaded via
\begin{verbatim}
git clone https://gitlab.com/openloops/OpenLoops.git
\end{verbatim}
Older and newer versions are available as git tags.
The latest beta version available in the branch ``public\_beta''
that can be downloaded via
\begin{verbatim}
git clone -b public_beta https://gitlab.com/openloops/OpenLoops.git
\end{verbatim}
Current and older \OL versions can be also be downloaded from
the {\sc hepforge} webpage
\\[4mm]
\url{http://openloops.hepforge.org}
\\[4mm]
where the user can also find a detailed list of the available process libraries
and extra documentation, as well as an up-to-date version of this paper.

\paragraph{Installation} The compilation of the process-independent \OpenLoops library 
is managed by the \texttt{SCons} build system\footnote{A version of \texttt{SCons} (``\texttt{scons-local}'') 
is shipped with \OL, but a system-wide installation may be used as well.}
and is easily carried out by running
\begin{verbatim}
./scons
\end{verbatim}
in the \OL directory.  By default, \texttt{Scons} utilises all available CPU
cores, while running \texttt{./scons\;-j<n>} restricts the number of
employed cores to \texttt{<n>}.
The compiled library is placed in the  \texttt{lib} subdirectory.\footnote{An
installation routine to move the library to a different location is
currently not available.}

The default compiler is \texttt{gfortran}, alternatively \texttt{ifort} can be used.
To change the compiler and set various other options, rename the sample
configuration file \texttt{openloops.cfg.tmpl} in the \OL directory to
\texttt{openloops.cfg} and set the options in there.
The sample configuration file lists various available options and describes
their usage.

\subsubsection{Installation of process libraries}
\label{se:proclibinstallation}

The calculation of scattering amplitudes for specific processes requires the
installation of corresponding process libraries.  The available collection
of \OpenLoops process libraries supports the calculation of QCD and EW
corrections for a few hundred different partonic
reactions, which cover essentially all interesting processes at the
LHC, as well as several lepton-collider processes.
This includes $pp\to$\,jets, $t\bar t$+jets, 
$V$+jets, $VV$+jets, $HV$+jets, $H$+jets and various other classes of
processes with a variable number of extra jets.
Process libraries for a large variety of
loop-induced processes such as $gg\to \ell\ell\ell\ell$+jets, $gg\to
HV$+jets, $gg\to HH(H)$+jets, etc.~are also available.

New processes libraries, especially with EW corrections, are continuously
added to the collection by the authors.  Moreover, extra processes libraries
can be easily made available upon request, either through an online form at 
\url{https://openloops.hepforge.org/process_library.php}
or by contacting the authors.
In particular this allows for the generation of
dedicated process libraries tailored to specific user requirements.
For example, it is possible to generate
dedicated process libraries with special filters for the selection of
certain classes of diagrams/topologies or various approximations related to the
treatment of heavy-quark flavours, the expansion in the number of colours,
the selection of resonances, non-diagonal CKM matrix elements, and so on.

\paragraph{Download and Installation}

The web page\\[4mm] 
\url{https://openloops.hepforge.org/process_library.php}\\[4mm]
%
provides a complete list of process libraries available in the public
process repository, with a description
of their content and the relevant process-library names to be used for download. 
The needed process libraries can be downloaded and compiled via
\begin{verbatim}
./openloops libinstall <processes> <options>
\end{verbatim}
where \verb|<processes>| is either a predefined process collection (see below)
or a list of white-space or comma separated names of
process libraries.
A single process library typically contains the full set of 
parton-level scattering amplitudes that is needed for 
the calculation of a certain family of hadron-collider processes,
either at NLO QCD or including EW corrections.
For instance, the libraries named 
\texttt{ppllll} and
\texttt{ppllll\_ew} include, respectively, 
the NLO QCD and NLO EW matrix elements for the 
production of four leptons, 
\ie the processes $pp\to \ell_i^+\ell_i^-\ell_k^+\ell_k^-$,
$\ell_i^+\ell_i^-\ell_k^+\nu_k$,
$\ell_i^+\ell_i^-\bar\nu_k\ell^-_k$,
$\ell_i^+\ell_i^-\bar\nu_k\nu_k$,
$\ell_i^+\nu_i\ell^+_k\nu_k$,
$\ell_i^+\nu_i\bar\nu_k\ell^-_k$,
$\bar\nu_i\ell_i^-\bar\nu_k\ell^-_k$,
$\ell_i^+\nu_i\bar\nu_k\nu_k$,
and $\bar\nu_i\ell_i^-\bar\nu_k\nu_k$,
with lepton flavours $i\neq k$ or $i=k$.

Each process library includes all relevant LO and NLO 
ingredients for the  partonic processes at hand, \ie all Born, one-loop and
real-emission amplitudes at the specified order.
More precisely, NLO~QCD libraries contain LO contributions of 
a given order $\alphaS^p\alpha^q$ and corrections of
order $\alphaS^{p+1}\alpha^q$, while
NLO~EW libraries contain the full tower of LO and NLO 
contributions apart from the NLO terms with the highest possible 
order in $\alphaS$.
Real-emission matrix elements are available throughout, but are 
not installed by default. This can be changed by using the option \texttt{compile\_extra=1} 
(default=0)
when installing the process.
This option can also be set in the \texttt{openloops.cfg} file in order to enable
real corrections for every process installation.

With the \texttt{libinstall} command
it is also possible to install pre-defined or user-defined process collections.
The pre-defined collection \texttt{lhc.coll} covers the most relevant LHC processes.%
\footnote{The collection \texttt{all.coll} makes it possible to download the
full set of available processes libraries at once.  However, due to the
large overall number of processes and the presence of several complex
processes, this is requires a very large amount of disk space and very long
CPU time for compilation.  Thus \texttt{all.coll} should not be used for
standard applications.}
In particular, it includes matrix elements for $V+$jets, $VV+$jets, $t\bar
t+$jets, $HV+$jets and $H+$jets (for finite and infinite $m_t$), where $V$
stands for photons as well as for the various leptonic decay products of
off-shell $Z$ and $W^\pm$ bosons.
Additional user-defined collections can be created as plain text files with
the file extension \texttt{.coll}, listing the desired process-library names,
one per line.

\paragraph{Updates}
When a new version of \OpenLoops is available, it is recommended to
update both the base code and the process code.\footnote{In general, 
base code and process code can be combined in a rather flexible way,
but care must be taken that they remain mutually consistent. The API
compatibility between base code and process code is typically guaranteed
across many sub-versions, both in the forward and backward directions.
To this end, all mutually consistent versions are labelled with the same
(internal) API version number, and \OpenLoops accepts to use only combinations of
process code and base code that belong to the same API version.
}
If \OL was installed from \texttt{Git}, this is easily achieved by running
\begin{verbatim}
./openloops update
\end{verbatim}
while \texttt{git pull \&\& ./scons} would update only the base code.
Instead, if \OL was not installed from \texttt{Git}, the installed processes can be updated by running
\begin{verbatim}
./openloops update --processes
\end{verbatim}
while the base code should be updated manually.

\subsection{Selection of processes and perturbative orders}
\label{se:processselection}

The \OpenLoops program supports the calculation of scattering probability densities
for a variety of processes at different orders in 
$\alphaS$ and $\alpha$.
Before starting the calculations, the user should register all needed
scattering amplitudes, which are automatically labelled with integer
identifiers for the bookkeeping of the various partonic channels and
perturbative orders.
As described in detail below, each desired matrix element should be registered
in two steps. First, the 
user should select the desired order in the QCD and EW couplings, model parameters and
specify possible approximations. In the second step, called process
registration, the user should  specify the list of 
external scattering particles, and select one of the available 
types of perturbative contributions.
The three possible types, denoted in the following as
amplitude types (\texttt{amptype}), are specified in~\refta{amptypes}
together with the list of corresponding objects of LO and NLO kind 
that can be
evaluated in \OpenLoops.
As explained in the following, the 
classification into LO and NLO kinds is relevant for 
the selection of the desired order in $\alphaS$ and $\alpha$.
Note that squared-loop objects are classified as LO quantities, since 
they are assumed to describe loop-induced processes.

\begin{table}[t]
\renewcommand{\arraystretch}{1.4}
  \begin{center}
    \begin{tabular}[t]{|c|c|c|c|}\hline
      \texttt{amptype} & amplitude type & LO output & NLO output \\ \hline
1 & tree--tree 
& $\calW_{\tree}^{(p,q)}$, $\calC^{(p,q)}_{00,\ssLO}$, $\calB^{(p,q)}_{00,\ssLO}$,
& 
\\[2mm]
11 & tree--loop  
& $\calW_{\tree}^{(p,q)}$
%
%
& $\calW_{\onel}^{(P,Q)}$,
$\calW_{\treeiop}^{(P,Q)}$,
$\calC^{(P,Q)}_{01,\ssNLO}$, $\calB^{(P,Q)}_{01,\ssNLO}$
\\[2mm]
12 & loop--loop  
& $\calW_{\onelsq}^{(p,q)}$, $\calC^{(p,q)}_{11,\ssLO}$, $\calB^{(p,q)}_{11,\ssLO}$, 
& $\calW_{\onelsqiop}^{(P,Q)}$
\\ \hline
    \end{tabular}
  \end{center}
  \caption{Values of \texttt{amptype} to register different types of
  perturbative contributions and corresponding probability
  densities that can be computed by \OL. Objects of LO and 
  NLO kind are evaluated at order 
  $\alphaS^p\alpha^q$  and   $\alphaS^P\alpha^Q$, respectively, 
  according to the values $p,q,P,Q$ of the LO and NLO power selectors
in \refta{tab:powerselectors}.
The symbols $\calB$ and 
$\calC$ stand for the various spin and colour/charge correlators defined 
in \refse{sec:correlators}. 
}
\label{amptypes}
\end{table}


\paragraph{Selection of QCD and EW power}
As discussed in \refse{sec:powercounting}, 
the general form of scattering probability densities in the SM
is a tower of terms of order $\alphaS^p\alpha^q$ 
with fixed perturbative order $p+q$ but variable powers $p,q$ in
the QCD and EW couplings.
In \OpenLoops, contributions with different orders in $\alphaS$ and $\alpha$
should be registered as separate (sub)processes.
Under each \texttt{amptype}, the 
various objects that can be calculated 
are classified into output of LO and NLO kind as 
indicated in~\refta{amptypes}. All
objects of LO type are evaluated at 
a certain power 
$\alphaS^p\alpha^q$, while all NLO objects 
are evaluated at a related power
$\alphaS^P\alpha^Q$. 
The desired powers $p,q,P,Q$, and the relation between 
($p,q$) and $(P,Q)$, can be controlled 
in four alternative ways 
by setting one of the power selectors listed in \refta{tab:powerselectors}.
\bit
\item[(a)] Setting $\texttt{order\_ew}=q$ selects contributions of fixed EW
order, \ie 
LO terms of $\ord(\alphaS^p\alpha^q)$
and NLO QCD corrections of $\ord(\alphaS^{p+1}\alpha^q)$.
In this case, the QCD order $p$ is automatically fixed according to $p+q=\npart-2$. 

\item[(b)] Similarly, $\texttt{order\_qcd}=p$ selects a fixed QCD order,
\ie LO terms of $\ord(\alphaS^p\alpha^q)$ 
and NLO EW corrections of $\ord(\alphaS^{p}\alpha^{q+1})$.
In this case, $q$ is automatically derived from $p+q=\npart-2$. 

\item[(c)] Alternatively, NLO terms of $\ord(\alphaS^{P}\alpha^Q)$
can be selected by setting 
$\texttt{loop\_order\_qcd}=P$ {\it or} $\texttt{loop\_order\_ew}=Q$.
This option is supported only for the evaluation of tree-loop interferences
(\texttt{amptype}=11). In that case, the output includes also 
the dominant underlying Born contribution of $\ord(\alphaS^{p}\alpha^q)$,
which is chosen between $\ord(\alphaS^{P}\alpha^{Q-1})$ 
and $\ord(\alphaS^{P}\alpha^{Q-1})$ as indicated in \reffi{Fig:SMpowercounting}.
When the loop order $P$ or $Q$ is specified, the complementary order
$Q$ or $P$ is fixed internally according to $P+Q=\npart-1$.
\eit
The desired order parameter should be set through the 
\texttt{set\_parameter}
%
routine before the registration of the process at hand.
As explained above, it is sufficient to specify the QCD or the EW order, and
{\it only one} of the order selectors in \refta{tab:powerselectors} should
be used.  If more than one order parameter is set by the user only the last
setting before registration is considered.

Before registering a process, also 
various approximations can be specified 
by setting  \OpenLoops parameters such as
\texttt{nf}, to control the number of active quarks,
\texttt{ckmorder}, to activate non-diagonal CKM matrix elements, etc.
A list of such parameters can be found in Tab.~\ref{app:tab1}
(see Appendix~\ref{app:input_parameters}).

\newcommand{\FOOTnotesize}[1]{#1}

\begin{table}
\renewcommand{\arraystretch}{1.3}
\centering
\begin{tabular}{|c||c|c||c|c|} \hline
power selection $\backslash$ derived powers
& \multicolumn{2}{c||}{LO power $\alphaS^p\alpha^q$}
& \multicolumn{2}{c|}{NLO power $\alphaS^P\alpha^Q$}\\\hline
$\texttt{order\_qcd}=\pin$ 
& \FOOTnotesize{$\pin$} 
& \FOOTnotesize{$\npart-\pin-2$}
& \FOOTnotesize{$\pin$}  
& \FOOTnotesize{$\qout+1$}  
\\
$\texttt{order\_ew}=\qin$  
& \FOOTnotesize{$\npart-\qin-2$} 
& \FOOTnotesize{$\qin$} 
& \FOOTnotesize{$\pout+1$}  
& \FOOTnotesize{$\qin$}  
\\
$\texttt{loop\_order\_qcd}=\Pin$ 
& \FOOTnotesize{$\pborn$} 
& \FOOTnotesize{$\qborn$} 
& \FOOTnotesize{$\Pin$}
& \FOOTnotesize{$\npart-\Pin-1$}
\\
$\texttt{loop\_order\_ew}=\Qin$  
& \FOOTnotesize{$\pborn$} 
& \FOOTnotesize{$\qborn$} 
& \FOOTnotesize{$\npart-\Pin-1$}
& \FOOTnotesize{$\Qin$}
\\ 
\hline
\end{tabular}
\caption{Selection of the orders $\alphaS^p\alpha^q$ and $\alphaS^P\alpha^Q$
for the LO and NLO objects defined in \refta{amptypes}.
Each selector takes one of 
the powers $p,q,P,Q$ as input and derives all other
powers as indicated in columns 2--5.
The QCD and EW coupling powers at LO and NLO are related through $p+q=\npart-2$ and
$P+Q=\npart-1$, where $\npart$ is the number of external particles. 
The $\texttt{loop\_order}$ selectors are supported only for 
\texttt{amptype=11}. They return the desired loop--tree
interference of $\ord(\alphaS^P\alpha^Q)$
together with the dominant underlying squared Born term of
$\ord(\alphaS^{p}\alpha^q)$
whose powers,
$(p,q)$ = $(\pborn,\qborn)$ = $(P-1,Q)$ or $(P,Q-1)$,
are selected in a unique way 
as indicated in \reffi{Fig:SMpowercounting}.
}
\label{tab:powerselectors}
\end{table}

\paragraph{Process registration}
Each (sub)process should be registered 
by means of the native interface function\footnote{The
registration procedure through the BLHA is explained 
in~\refapp{app:blha}.}
\texttt{register\_process},
which automatically assigns a unique process identifier,
as detailed in \refapp{app:native:registration}.
The syntax to specify the external particles of a generic $n\to m$ 
scattering process with $n\ge 1$ is
\begin{align}
{\rm PID}_{i,1} \dots {\rm PID}_{i,n} \,\texttt{->}\,  {\rm PID}_{f,1} \dots {\rm PID}_{f,m}
\label{eq:processdef}
\end{align}
The particle identifier (PID) can be specified either using the PDG
numbering scheme~\cite{Tanabashi:2018oca} or the string identifiers 
listed in Tab.~\ref{particleid}.

\begin{table}[t]
\begin{center}
    \begin{tabular}[t]{c|cccccccccccccc} \hline
      particle & $q_d/\tilde q_d$ & $q_u/\tilde q_u$ & $q_s/\tilde q_s$ & $q_c/\tilde q_c$ & $q_b/\tilde q_b$ & $q_t/\tilde q_t$ \\
      PID & \texttt{1/-1} & \texttt{2/-2} & \texttt{3/-3}  & \texttt{4/-4}  & \texttt{5/-5} &   \texttt{6/-6} \\
      string-PID & \texttt{d/d$\sim$} & \texttt{u/u$\sim$} & \texttt{s/s$\sim$} & \texttt{c/c$\sim$} & \texttt{b/b$\sim$}& \texttt{t/t$\sim$}\\ \hline
      
      particle & $l_e/\tilde l_e$ & $\nu_{e}/\tilde \nu_{e}$ & $l_{\mu}/\tilde l_{\mu}$ & $\nu_{\mu}/\tilde \nu_{\mu}$ & $l_{\tau}/\tilde l_{\tau}$ & $\nu_{\tau}/\tilde \nu_{\tau}$   \\ 
        PID & \texttt{11/-11} & \texttt{12/-12} & \texttt{13/-13} & \texttt{14/-14} & \texttt{15/-15} & \texttt{16/-16}\\
        string-PID &  \texttt{e-/e+} & \texttt{ve/ve$\sim$} & \texttt{mu-/mu+} & \texttt{vm/vm$\sim$} & \texttt{ta-/ta+}& \texttt{vt/vt$\sim$}\\ \hline
      
            particle & $g$ &  $\gamma$ & on-/off-$\gamma$& $Z$ & $W^{\pm}$ & Higgs\\
            PID & \texttt{21} & \texttt{22} & \texttt{2002/-2002} & \texttt{23} & \texttt{24/-24} & \texttt{25}\\
        string-PID & \texttt{g} & \texttt{a} & \texttt{aon/aoff} & $\texttt{z}$ & $\texttt{w+/w-}$ & \texttt{h}\\
      
      \hline
    \end{tabular}
  \end{center}
  \caption{Particle identifiers (PID) for process specification in \OL.  The numerical
  and string PID representations can be mixed. As explained in
  \refse{sec:ewschemes}, for an optimal treatment of 
the coupling of on-shell and off-shell hard external photons
the special PIDs $\pm$2002 
should be used.}
\label{particleid}
\end{table}

Together with the external particles, also a specific type of perturbative output
(\texttt{amptype}) should be selected. As summarised in Tab.~\ref{amptypes}, the
available options correspond to the various scattering probability densities  
defined in \refeq{eq:Wtree}--\refeq{eq:Wloop2}, \ie squared tree amplitude ($\calW_{\tree}$)
tree--loop interference ($\calW_{\onel}$), and  squared one-loop amplitude
($\calW_{\onelsq}$), but each \texttt{amptype} supports also the calculation of
various related objects.

\subsection{Evaluation of scattering amplitudes}
\label{sec:normalisation}
\label{sec:amplitudecalls}
\label{sec:polesandscales}

In this section we introduce various \OpenLoops interface functions for the
evaluation of 
the scattering probability densities~\refeq{eq:Wtree}--\refeq{eq:Wloop2},
the $\mathbf{I}$-operators~\refeq{eq:ioperator},
and some of their building blocks.
The input required by the various interface functions consists of 
a phase-space point together with the integer
identifier for the desired (sub)process. The output
is always returned according to the normalisation conventions of eqs.~\refeq{eq:Wtree}-\refeq{eq:Wloop2},
\ie symmetry factors, external colour and helicity sums, and average factors
are included throughout.
This holds also for the interface functions discussed in 
\refses{sec:correlators}{sec:colourbasis}.
The syntax of the various interfaces is detailed in \refapp{app:native}.


In general, the output depends on the values of all relevant physical and technical
input parameters (see \refses{sec:ewschemes}{se:renormalisation}) at
the moment of calling the actual \OpenLoops interface routine.  All parameters and 
settings are
initialised with physically meaningful default values, which can be updated at 
any moment by means of \texttt{set\_parameter}.
In principle, all parameters can be changed before any new amplitude
evaluation.
As explained below,
thanks to a new automated scale-variation system,
scattering amplitudes can be 
re-evaluated multiple times with different values
of $\mur$ and $\alphaS$ in a very efficient way.

The calculation of the probability densities~\refeq{eq:Wtree}--\refeq{eq:Wloop2}
is supported by the following interfaces.

\paragraph{Squared Born amplitudes}\hspace{-3.5mm}  
$\calW_{\tree}=\langle \calM_0|\calM_0\big\rangle$
are evaluated by the function \texttt{evaluate\_tree}.

\paragraph{Tree--loop interferences}\hspace{-3.5mm} $\calW_{\onel}=2\,\re\,  \langle
\calM_0|\calM_1\big\rangle$
are evaluated by \texttt{evaluate\_loop}, which 
yields a UV renormalised result. The output 
is returned in the form of an array 
$\{\mathcal{W}^{(0)}_{\onel},\mathcal{W}^{(1)}_{\onel},\mathcal{W}^{(2)}_{\onel}\}$
consisting of the coefficients of the Laurent expansion,
\begin{align}
\label{eq:laurentseries}
 \mathcal{W}_{\onel} = \ceps 
\left(\frac{\mathcal{W}^{(2)}_{\onel}}{\epsilon^2}
  + \frac{\mathcal{W}^{(1)}_{\onel}}{\epsilon} + \mathcal{W}^{(0)}_{\onel}\right)
  + \mathcal{O}(\epsilon)\,,
\end{align}
where $\eps=\epsuv=\epsir$. In general, the $\mathcal{W}^{(1)}$ 
residues receive contributions from IR and UV divergences, but 
UV-renormalised results contain only IR poles.
By default, the  normalisation factor $\ceps$ is
defined as in \refeq{eq:cepsilon}, which 
corresponds to the BLHA convention~\cite{Binoth:2010xt}. 
Alternatively, by setting \texttt{polenorm=1} (default=0)
it can be changed into\footnote{This corresponds to 
the normalisation convention used by the
\Collier \cite{Denner:2016kdg} library.}
\bea
\label{eq:cepsiloncoli}
\tilde\ceps = (4\pi)^\epsilon\Gamma(1+\epsilon)
= \ceps +\frac{\pi^2}{6}\eps^2 + \ord(\eps^3)\,,
\eea
which results in a modified Laurent series,
$\widetilde\calW_{\onel} =\calW_{\onel}-
\calW^{(2)}_{\onel}\,\frac{\pi^2}{6}$.
The output of \texttt{evaluate\_loop} consists of the sum of a bare
contribution with four-dimensional loop numerator, a standard UV
counterterm, a counterterm of type $R_2$ and, optionally, also the contribution of
the related $\mathbf{I}$-operator~\refeq{eq:ioperator},
\bea
\label{eq:loopingredients} 
\calW_{\onel} &=& 
\calW_{\onel,\fourdim}+
\calW_{\onel,\ct}+
\calW_{\onel,R_2}\;
\left(+
\calW_{\treeiop}\right)\,.
\eea
The $\mathbf{I}$-operator
can be activated by setting \texttt{iop\_on=1} (default=0).
The counterterm and the $R_2$ contributions can be 
deactivated by setting, respectively,  
\texttt{ct\_on=0} (default=1) and 
\texttt{r2\_on=0} (default=1).
The various divergent building blocks of \refeq{eq:loopingredients} 
are Laurent series of the form \refeq{eq:laurentseries}. 
For efficiency reasons,
in \OL they are constructed 
as single-valued objects 
\begin{align}
\label{eq:laurentseriesB} 
\mathcal{W}_{\onel,k}(\Delta_2,\Delta_1) = 
\mathcal{W}^{(2)}_{\onel,k}\,\Delta_2
+\mathcal{W}^{(1,\ir)}_{\onel,k}\,\Delta_{1,\ir}
+\mathcal{W}^{(1,\uv)}_{\onel,k}\,\Delta_{1,\uv}
+\mathcal{W}^{(0)}_{\onel,k}\,,
\end{align}
where the IR and UV poles are replaced by numerical
constants%
\footnote{The values of $\Delta_2$, $\Delta_{\ir,1}$ and $\Delta_{\uv,1}$
are controlled internally by \OpenLoops. For validation purposes
they can be changed using 
the parameters \texttt{pole\_IR2},
\texttt{pole\_IR1} and \texttt{pole\_UV1}, respectively. 
However such modifications may jeopardise the calculation of
UV and IR divergent quantities.} 
($\ceps/\epsir^2\to \Delta_2$,
$\ceps/\epsir\to \Delta_{\ir,1}$,
$\ceps/\epsuv\to \Delta_{\uv,1}$)
and 
$\mathcal{W}^{(1,\ir)}_{\onel,k}+\mathcal{W}^{(1,\uv)}_{\onel,k}
=\mathcal{W}^{(1)}_{\onel,k}$.
A posteriori, the three coefficients
$\mathcal{W}^{(i)}_{\onel}$ can be reconstructed through three 
evaluations of \refeq{eq:laurentseriesB} with different $\Delta_{i}$ values. 
However, the most efficient approach it to restrict the calculation of 
the most CPU expensive objects to their finite parts by setting all
$\Delta_{i}=0$ (default),
and to reconstruct the
poles by exploiting the fact that UV and IR subtracted results are finite.
In practice, when the $\mathbf{I}$-operator is active, all poles are simply set to
zero in \refeq{eq:loopingredients}, and only finite parts are computed.  Also
when the $\mathbf{I}$-operator is switched off in \refeq{eq:loopingredients}, 
only the finite part of the right-hand-side of \refeq{eq:loopingredients} is explicitly 
computed, while IR poles are reconstructed from the $\mathbf{I}$-operator, \ie 
\bea
\calW^{(i)}_{\onel}\Big|_{i=1,2} &=&
\begin{cases}
-\calW^{(i)}_{\treeiop} & \mbox{for \texttt{iop\_on=0} (default)}\,,\\
0 & \mbox{for \texttt{iop\_on=1}}\,.
\end{cases}
\eea
The explicit calculation of all poles in $\calW_{\onel}$ 
through multiple evaluations of \refeq{eq:laurentseriesB} can be enforced 
by setting \texttt{truepoles\_on=1} (default=0).
Thus, the correct cancellation of UV and IR poles can be explicitly checked
by calling \texttt{evaluate\_loop} with
\texttt{truepoles\_on=1} and \texttt{iop\_on=1}.

The individual building blocks of $\calW_{\onel}$
can be evaluated by various dedicated interfaces:
\bit

\item[(i)] 
The {\bf bare loop amplitudes}
$\calW_{\onel,\fourdim}$, 
with four-dimensional numerator, are evaluated by
\texttt{evaluate\_loopbare}, which returns a Laurent series similar to
\refeq{eq:laurentseries}.   
As for \texttt{evaluate\_loop},
pole residues are derived from the related UV and IR counterterms (default)
or explicitly reconstructed, depending on the value of \texttt{truepoles\_on}.

\item[(ii)] The {\bf UV counterterms}
$\calW_{\onel,\ct}$ 
are evaluated by \texttt{evaluate\_loopct},
which returns a Laurent series similar to \refeq{eq:laurentseries}.  
In this case, UV pole coefficients are always 
obtained via  two-fold evaluation.
The more efficient function \texttt{evaluate\_ct} restricts the
calculation of the counterterm to its finite part
$\calW^{(0)}_{\onel,\ct}$\,.

\item[(iii)] The {\bf $R_2$ rational part}
$\calW_{\onel,R_2}$ is free from UV and IR divergences. It is
evaluated by \texttt{evaluate\_r2}, which returns a 
single-valued output.

\item[(iv)] {\bf Tree--tree $\mathbf{I}$-operator insertions},
$\calW_{\treeiop}=\langle M_0| \mathbf{I}(\{p\};\epsir) | M_0\big\rangle
$, are evaluated by the function \texttt{evaluate\_iop}.
The output is a Laurent series similar to \refeq{eq:laurentseries}.

\item[(v)] The poles of all divergent building blocks
of \refeq{eq:loopingredients} can be accessed with a single call of
\texttt{evaluate\_poles}, which returns the residues of the
$1/\epsuv$, $1/\epsir$ and $1/\epsir^2$ poles for each 
building block.
In this case, irrespectively of the value of \texttt{truepoles\_on}, all
residues are always computed explicitly.

\eit

Note that, for efficiency reasons, the
combination \refeq{eq:loopingredients} should always be computed 
via a call of \texttt{evaluate\_loop} rather than separate calls for 
its building blocks.

\paragraph{Squared loop amplitudes}\hspace{-3.5mm} $\calW_{\onelsq}=\langle \calM_1|\calM_1\big\rangle$
are evaluated by the function \texttt{evaluate\_loop2}.
Since we assume that it is used for loop-squared processes,
which are free from UV and IR divergences at LO,
\texttt{evaluate\_loop2} returns a single-valued finite output.
The calculation of {$\mathbf{I}$-operator insertions in loop-squared
amplitudes},
$\calW_{\onelsqiop}=\langle M_1| \mathbf{I}(\{p\};\epsir) | M_1\big\rangle
$, is supported by \texttt{evaluate\_loop2iop}.
Since we assume loop-induced processes, the output is
a Laurent series of type \refeq{eq:laurentseries}
with poles up to order $1/\eps^2$. 
In general, $\calW_{\onelsq}$ and $\calW_{\onelsqiop}$ are
evaluated using only the finite part of $\calM_1$,
and possible UV and IR poles are simply amputated at the
level of $\calM_1$.

\paragraph{Efficient QCD scale variations}
\OLtwo implements a new automated system for the efficient assessment of QCD
scale uncertainties.  This system is designed for the case where scattering
amplitudes are re-evaluated multiple times with different values of $\mur$
and $\alphaS$, while all other input and kinematic parameters are kept
fixed.
This type of variations are automatically detected by keeping track, on a
process-by process basis, of the pre-evaluated phase-space points, and
possible variations of parameters.
For each new phase-space point, matrix elements are
computed from scratch and stored in a cache, which is used for
$(\mur,\alphaS)$ variations.  In that case, the previously computed bare
amplitude is reused upon appropriate rescaling of $\alphaS$, and only the
$\mur$-dependent QCD counterterms are explicitly recomputed.
This mechanism is implemented for both types of loop contributions
\refeq{eq:Wloop}--\refeq{eq:Wloop2}.

\subsection{Colour- and spin-correlators}
\label{sec:correlators}

This section presents interface functions for the evaluation of
colour- and helicity-correlated quantities that are needed in the context of
NLO and NNLO subtraction methods, both for tree- and loop-induced processes. 
For efficiency reasons, colour/spin correlations are always computed
in combination with the related squared tree or loop matrix elements, in such
a way that the former are obtained with a minimal CPU overhead.

\paragraph{Colour and charge correlators} 
The exchange of soft gluons/photons between two external legs, $j$ and $k$,
gives rise to colour/charge correlations of the form
\bea
\colcorr{p,q}{jk}{LL}{\ssLO\, \ssQCD} &\,=\, &
\langle \calM_L|T^a_j T^a_k | \calM_L \big\rangle 
\bigg|_{\alphaS^{p}\alpha^q},\qquad
\label{eq:treecolcordef}
\\
\colcorr{p,q}{jk}{LL}{\ssLO\,\ssQED} &\,=\,&
\langle \calM_L|Q_j Q_k | \calM_L \big\rangle 
\bigg|_{\alphaS^{p}\alpha^{q}}\,,
\label{eq:treechargedef}
\eea
where $T^a_i$ and $Q_i$ denote SU(3) and charge operators 
acting on the $i$-th external particle.%
\footnote{As usual, the corresponding SU(3)$\times$U(1) quantum numbers 
should be understood in terms of incoming charge flow, in such a way that 
$\sum_k T^a_k | \calM \big\rangle=\sum_k Q_k | \calM \big\rangle=0$.}
\mbox{Tree--tree} correlators correspond to $LL=00$ in
\refeq{eq:treecolcordef}--\refeq{eq:treechargedef}
and can be evaluated by the interface functions \texttt{evaluate\_ccmatrix} and \texttt{evaluate\_ccewmatrix}, which
return the full matrices $C_{00}^{(p,q|jk)}$  as two-dimensional arrays.
Alternatively, the $N(N-1)/2$ independent colour correlators in \refeq{eq:treecolcordef} can be obtained
 in the form of one-dimensional arrays using \texttt{evaluate\_cc}.
Loop--loop correlators
($LL=11$) can be evaluated in a similar way using  the functions 
\texttt{evaluate\_ccmatrix2}, \texttt{evaluate\_ccewmatrix2} and \texttt{evaluate\_cc2}.

In $\texttt{amptype}=11$ mode, 
also the  tree--loop colour correlators
\bea
\colcorr{P, Q}{jk}{01}{\ssNLO\, \ssQCD} &=& 
2\re\,\langle \calM_0|T^a_j T^a_k | \calM_1 \big\rangle 
\bigg|_{\alphaS^{P}\alpha^Q, \rm{finite}}
\label{eq:treeloopcolcordef}
\eea
are available. They are evaluated by the functions
 \texttt{evaluate\_loopccmatrix} and \texttt{evaluate\_loopcc}, 
 which return only the finite part, \ie a term corresponding to $\calW_{01}^{(0)}$ in the Laurent series 
\refeq{eq:laurentseries}.

\paragraph{Spin-colour correlators} 
The emission of soft-collinear radiation off external gluons/photons
generates also spin-correlation effects.
For their description we use the notation
\bea
\label{eq:spinconotation}
\langle \lambda,j | \calM\rangle
&=&
%
%
\eps^{\mu}_\lambda (p_j)\,
\langle \mu, j | \calM \rangle\,,
\eea
where $\cal M$ is a generic helicity amplitude, and 
$j$ is a gluon or photon emitter with helicity $\lambda$.
The helicity states of all other external particles 
are kept implicit.
With this notation, unpolarised squared matrix elements
can be expressed as
\bea
\label{eq:squaredhelamp}
\langle \calM | \calM\rangle &=&
\sum_{\lambda}\,
\langle \calM | \lambda, j \rangle\,
\langle \lambda, j    | \calM\rangle
\,=\,
-
\langle \calM | \mu, j \rangle\,
\langle \mu, j | \calM\rangle\,,
\eea
where the normalisation conventions
of Eqs.~\refeq{eq:Wtree}-\refeq{eq:Wloop2}
are implicitly understood.
Spin-correlation effects arise as terms of type
$\langle \calM | P_j | \calM\rangle$ with spin correlators
of the form
\bea
P_j \,=\, P_j^{\mu\nu}\, |\mu,j \rangle \langle \nu,j|\,.
\eea
They can be evaluated in a convenient way 
in terms of the spin-correlation tensor
\bea
\calB_j^{\mu\nu}&=&
\langle \calM | \mu,j\rangle \,
\langle \nu,j | \calM\rangle
\,=\,
\sum_{\lambda,\lambda'}\,
\langle \calM | \lambda,j \rangle\,
\eps^{\mu}_\lambda (p_j)\,{\eps^{*\,\nu}_{\lambda'}} (p_j)\,
\langle \lambda',j    | \calM\rangle\,,
\label{eq:tensorspincorr}
\eea
which allows one to write
\bea
\langle \calM | P_j | \calM\rangle &=&
\langle \calM | \mu, j \rangle\,
P_j^{\mu\nu}
\langle \nu, j   | \calM\rangle\,=\,
P_j^{\mu\nu}\,\calB_{j,\mu\nu}\,.
\eea

Alternatively, spin correlations can be implemented in a 
more efficient way by exploiting the fact that,
in NLO calculations, they arise only
through operators of the form
\bea
G_j=  g^{\mu\nu} |\mu,j \rangle \langle \nu,j|
\,\quad\mbox{and}\quad
P_j(\kperp)= 
-\left(\frac{\kperp^\mu\kperp^\nu}{\kperp^2}
\right)\,|\mu,j \rangle \langle \nu,j|
\,=\,
- \frac{1}{\kperp^2}|\kperp,j \rangle \langle \kperp,j|,
\eea
where $\kperp^\mu$ is a certain vector\footnote{Explicit expression for
$\kperp^{\mu}$ in the dipole subtraction formalism are for example listed in
Tab.~1 of~\cite{Gleisberg:2007md} for all relevant splittings.}
%
with $\kperp\cdot p_j=0$.
%
%
Since $\langle \calM | G_j | \calM\rangle = 
-\langle \calM | \calM\rangle$, all 
non-trivial spin-correlation effects can be
encoded into the scalar quantity
\bea
\calB_{j}(\kperp)
&=& 
\langle \calM | P_j(\kperp) | \calM\rangle
\,=\,
-
\frac{\kperp^\mu\kperp^\nu}{\kperp^2}\,
\calB_{j,\mu\nu}
\,=\,
- \frac{1}{\kperp^2}
\langle \calM |\kperp , j \rangle\,
\langle \kperp,j | \calM\rangle\,,
\label{eq:scalarspincorr}
\eea
where $\langle \kperp,j | \calM\rangle$
corresponds to the helicity amplitude \refeq{eq:spinconotation}
with $\eps^\mu_\lambda(p_j)$ replaced by $\kperp^\mu$.

In NLO calculations, spin correlations arise in combination 
with colour correlations through  
operators of the type $T^a_jT^a_k
|\kperp,j\big\rangle \big\langle \kperp,j |$,
where $j$ and $k$ are called emitter and 
spectator.
In \OL, such spin-colour correlators are implemented in the form
\be
\spincorrB{p,q}{jk}{LL,\ssLO}(\kperp)= 
-\frac{1}{\kperp^2}
\langle \calM_L
| \calTSC_{jk} 
|\kperp,j\big\rangle \big\langle \kperp,j 
| \calM_L \big\rangle 
\bigg|_{\alphaS^{p}\alpha^q}\,
\mbox{with}\quad
\calTSC_{jk}=\begin{cases}
T^a_j T^a_k & \mbox{if $j$ is a gluon}\,,\\
1 & \mbox{if $j$ is a photon}\,,\\
0 & \mbox{otherwise}\,,\\
\end{cases}
\label{eq:treecolspincorscalar}
\ee
which corresponds to the scalar representation~\refeq{eq:scalarspincorr}.
Tree--tree ($LL=00$) and loop--loop ($LL=11$) correlators of this kind 
are evaluated by the functions 
\texttt{evaluate\_sc} and \texttt{evaluate\_sc2}, respectively.
An alternative implementation
with the form of the spin-colour-correlation tensor \refeq{eq:tensorspincorr},
\bea
\spincorr{p,q}{jk}{\mu\nu}{LL,\ssLO} &=& 
\langle \calM_L
| \calTSC_{jk} 
|\mu,j\big\rangle \big\langle \nu,j 
| \calM_L \big\rangle 
\bigg|_{\alphaS^{p}\alpha^q}\,,
%
%
\label{eq:treecolspincortensor}
\eea
is available through the functions \texttt{evaluate\_sctensor} (for $LL=00$) and
\texttt{evaluate\_sctensor2} (for $LL=11$). 
Furthermore the spin-correlation tensor according to the \Powheg~\cite{Alioli:2010xd} convention, \ie without
colour insertions
\bea
\spincorr{p,q}{j}{\mu\nu}{LL,\ssLO} &=& 
\langle \calM_L
|\mu,j\big\rangle \big\langle \nu,j 
| \calM_L \big\rangle 
\bigg|_{\alphaS^{p}\alpha^q}\,,
%
%
\label{eq:treespincortensor}
\eea
is available via the functions \texttt{evaluate\_stensor} (for $LL=00$) and
\texttt{evaluate\_stensor2} (for $LL=11$). 
All implementations~\refeq{eq:treecolspincorscalar}-\refeq{eq:treespincortensor}
 are well suited for the subtraction of IR singularities
with the Catani--Seymour~\cite{Catani:1996vz,Catani:2002hc} and
FKS~\cite{Frixione:1995ms} methods.
The tensor representations \refeq{eq:treecolspincortensor}-\refeq{eq:treespincortensor} are more general, while 
the scalar form~\refeq{eq:treecolspincorscalar} is more efficient, but should be used only
if $\kperp\cdot p_j=0$ is fulfilled.\footnote{\OpenLoops automatically
amputates possible non-orthogonal parts of $\kperp$ by projecting 
$\kperp^\mu$ onto 
$\eps^\mu_{\pm}(p_j)$.}

In $\texttt{amptype}=11$ mode,
also the tree--loop spin correlators
\bea
\spincorrB{P,Q}{jk}{01,\ssNLO}(\kperp)= 
-\frac{2}{\kperp^2}\,\re\,
\langle \calM_0
| \calTSC_{jk} 
|\kperp,j\big\rangle \big\langle \kperp,j 
| \calM_1 \big\rangle 
\bigg|_{\alphaS^{P}\alpha^Q,\mathrm{finite}}\,,
\label{eq:treeloopcolspincortensor}
\eea
\bea
\spincorr{P,Q}{jk}{\mu\nu}{01,\ssNLO} &=& 
2\,\re\,\langle \calM_0
| \calTSC_{jk} 
|\mu,j\big\rangle \big\langle \nu,j 
| \calM_1 \big\rangle 
\bigg|_{\alphaS^{P}\alpha^Q,\mathrm{finite}}\,
\label{eq:treeloopspincortensorB}
\eea
and
\bea
\spincorr{P,Q}{j}{\mu\nu}{01,\ssNLO} &=& 
2\,\re\,\langle \calM_0
|\mu,j\big\rangle \big\langle \nu,j 
| \calM_1 \big\rangle 
\bigg|_{\alphaS^{P}\alpha^Q,\mathrm{finite}}\,
\label{eq:treeloopcolspincortensorB}
\eea
are available. They are evaluated by the functions
\texttt{evaluate\_loopsc}, \texttt{evaluate\_loopsctensor} and \texttt{evaluate\_loopsctensor} respectively, which return only the finite part, similarly as for~\refeq{eq:treeloopcolcordef}.

\subsection{Tree-level amplitudes in colour space}
\label{sec:colourbasis}
\newcommand{\barj}[1]{\bar{j}_{#1}}
\def\procarrow{\hspace{-1mm}$\scriptstyle\to$\hspace{-1mm}}

Besides calculating squared matrix elements, \OpenLoops also provides full
tree-level colour information at the amplitude level.  
Such information is relevant in the context 
of parton-shower matching in order
to determine the probabilities with which a parton shower should start from a specific colour
configuration. Moreover it can be used to determine colour correlations with more than two colour insertions, 
as needed within NNLO subtraction schemes.

\newcommand{\qflow}[1]{{\scriptstyle (\alpha_{#1},0)}}
\newcommand{\gflow}[2]{{\scriptstyle (\alpha_{#1},\tilde\alpha_{#2})}}
\newcommand{\aflow}[1]{{\scriptstyle (0,\tilde\alpha_{#1})}}
\newcommand{\noflow}{{\scriptstyle (0,0)}}
\begin{table}
\renewcommand{\arraystretch}{1.4}
\centering
\begin{tabular}{r|ccccccccccc}\hline
{external particles}
& $q$    & $\bar q$ &  \procarrow  & $\gamma$ & ${q}$  & ${\bar q}$  & ${Z}$  &
${g}$ & $g$ & $g$ &
\\[0mm]\hline
{integer labels} 
& 1    &  2 &
& 3  & 4  & 5 & 6 & 7 & 8  & 9 & \\[-2mm]
&  $\alpha_1$  & $\beta_1$ &   & & $\beta_2$ & $\alpha_2$ &  & $\sigma_{1}$ &
$\sigma_2$ & $\sigma_3$ & \\[-2mm]
&   & &  & & & &  & $\alpha_{3}$ &
$\alpha_4$ & $\alpha_5$ & \\[0mm]\hline
{SU(3) indices}
   &  $i_1$  & $\barj{2}$ & & & $\barj{4}$ &
$i_5$ &  &
$a_{7}$ & $a_8$ &
$a_9$ & \\[0mm]\hline
{colour flow} & $\qflow{1}$ & $\aflow{1}$ & & $\noflow$  & $\aflow{2}$ & $\qflow{2}$ & $\noflow$
& $\gflow{3}{3}$ & $\gflow{4}{4}$ & $\gflow{5}{5}$\\
\end{tabular}
\caption{Particle and colour numbering scheme.  The external particles are
labelled through consecutive integers $1,2,\dots,\npart$ according to the
ordering~\refeq{eq:processdef} specified through the process registration.
The symbols $\sigma_k$ are used in~\refeq{eq:colbasis}--\refeq{eq:colbasisC}
to represent the integer labels of external gluons, while $a_{\sigma_k}$ 
are the corresponding colour indices.
Similarly, \mbox{$\alpha_k\,(\beta_l)$} represent the integer labels of 
incoming quarks (antiquarks) or outgoing antiquarks (quarks), and their
colour indices are \mbox{$i_{\alpha_k}\,(\barj{\beta_l})$}.
%
For the process considered in the table, $q\bar q \to \gamma q  \bar q {Z} g  g g$,
we have 
$(\alpha_1,\alpha_2)$\,=\,$(1,5)$,
$(\beta_1,\beta_2)$\,=\,$(2,4)$,
$(\sigma_1,\sigma_2,\sigma_3)$\,=\,$(\alpha_3,\alpha_4,\alpha_5)$\,=\,$(7,8,9)$.
The last row illustrates the notation of the colour-flow basis. In this case, as
explained in the text, the antiquark indices $\beta_k$ are replaced by 
a permutation $\tilde\alpha_k=\pi(\alpha_k)$ of the quark indices
according to the actual colour flow. Moreover,
gluons are represented by a pair of indices ($\alpha_k,\tilde\alpha_k$)
corresponding to a quark--antiquark pair.
}
\label{tab:colornumbering}
\end{table}

\paragraph{Colour vector}
As indicated in \eqref{eq:colourvector}, any tree-level amplitude is
represented as a vector $\{\calA_0^{(i)}(\heli)\}$ in the colour space spanned by
the colour basis elements $\{\calC_i\}$,
\bea
\calM_0 = \sum_i \calA^{(i)}_0(\heli)\, \calC_i\,.
\label{eq:treecolvect}
\eea
For a process with $n$ external
gluons and $m$ external $q\bar q$ pairs, each element of the basis has the
general colour structure
\bea
\calC_i \equiv \left(\calC^{a_{\sigma_1}\dots
a_{\sigma_n}}_i\right)^{\barj{\beta_1}\dots
\barj{\beta_m}}_{i_{\alpha_1}\dots i_{\alpha_m}}\,,
\label{eq:colbasis}
\eea
where the particle labels $\alpha_k$, $\beta_k$, $\sigma_k$,
and the corresponding colour indices $i_{\alpha_k}$,
$\barj{\beta_k}$, $a_{\sigma_k}$,
are attributed according to the 
labelling scheme defined in \refta{tab:colornumbering}.

\paragraph{Trace basis}
In \OL the colour basis is chosen as a so-called trace basis, where each
basis element \refeq{eq:colbasis}
is a product of chains of fundamental generators and traces
thereof. More precisely, each basis element is a product of building blocks
of type
\bea
\label{eq:colbasisA}
L(\beta,\alpha)&=&
\delta^{\barj{\beta}}_{i_\alpha}\,,\\
\label{eq:colbasisB}
L(k, \dots, l, \beta,\alpha)&=&
\left(T^{a_k}\cdots T^{a_l}\right)^{\barj{\beta}}_{i_\alpha}\,,\\
\label{eq:colbasisC}
L(k, \dots, l)&=&
\Tr\left(T^{a_k}\cdots T^{a_l}\right)\,.
\eea
As indicated on the 
lhs of the above equations,  each building block is uniquely identified through a sequence of 
integer particle labels. 
Sequences terminating with gluon labels and
antiquark--quark labels correspond, respectively,
to traces~\refeq{eq:colbasisC} and
chains~\refeq{eq:colbasisA}--\refeq{eq:colbasisB}.
Products of chains and traces are represented as
\bea
L(x_1,\dots x_k, 0, y_1,\dots)&=&
L(x_1,\dots x_k) L(y_1,\dots)\,,
\label{eq:tracebasisproducts}
\eea
\ie the individual sequences are concatenated 
using zeros as separators. 
With this notation each element of the colour basis can be 
encoded as an array of integers. For instance,  
for $q\bar q \to \gamma q  \bar q {Z} g  g g$
(see \refta{tab:colornumbering}) we have
\bea
L\left(8,2,5,0,7,9,0,4,1\right) 
\,=\,
\left(T^{a_8}\right)^{\barj{2}}_{i_5}\,
\mathrm{Tr}(T^{a_7}T^{a_9})\,
\delta^{\barj{4}}_{i_1}\,.
\label{eq:tracebasisrepr}
\eea

The explicit colour basis 
for a given scattering process can be accessed through the
interface functions 
\texttt{tree\_colbasis\_dim} and
\texttt{tree\_colbasis}. The former yields the number of 
elements of the basis, as well as the number of 
helicity configurations, while \texttt{tree\_colbasis}
returns the basis vectors in a format corresponding to
\refeq{eq:colbasisA}--\refeq{eq:tracebasisproducts}.
The complex-valued 
colour vector $\{\calA_0^{(i)}(\heli)\}$ in \refeq{eq:treecolvect} 
can be obtained through the function \texttt{evaluate\_tree\_colvect}.
Using $\{\calA_0^{(i)}(\heli)\}$ it is possible to calculate the
LO probability density~\refeq{eq:Wtree} as
\bea
\calW_{00} &=&
\frac{1}{\nhcs}\sum_{\heli}\sum_{i,j} \left[\calA_0^{(i)}(\heli)\right]^*\,\calK_{ij}\,
\calA_0^{(j)}(\heli)\,,
\label{eq:bornsqcolint}
\eea
where $\calK_{ij}$ is the colour-interference matrix defined
in~\refeq{eq:colintB}.

\paragraph{Colour-flow basis}
For the purpose of parton-shower matching in leading-colour approximation,
it is more convenient to use the 
colour-flow representation~\cite{Kanaki:2000ey,Maltoni:2002mq}, 
where gluon fields are handled as $3\times 3$ matrices
$\left(\mathcal{A}_\mu\right)_i^{\bar j} =
\frac{1}{\sqrt{2}}\mathcal{A}^a_\mu \left(T^a\right)_i^{\bar j}$,
and the colour structures of tree amplitudes with 
$m$ external quark--antiquark pairs and $n$
external gluons take the form
\begin{align}
  \calC \equiv \calC^{\barj{\beta_1}\dots\barj{\beta_{N}}}_{
    i_{\alpha_1}\dots i_{\alpha_{N}}},
\end{align}
with $N=m+n$. The elements of the colour-flow basis have the form
\begin{align}
  \calC_i^{\mathrm{flow}} = \delta^{\barj{\beta_1}}_{i_{\tilde\alpha_1}}\dots
  \delta^{\barj{\beta_N}}_{i_{{\tilde\alpha_N}}}\,,
  \label{eq:flowbasis}
\end{align}
where $\alpha_k\to \tilde \alpha_k=\pi(\alpha_k)$ is a permutation of
the quark particle labels, which encodes the 
colour connections 
between antiquarks ($\beta_k$)
and quarks ($\tilde\alpha_k$) in \refeq{eq:flowbasis}.

A basis element of the form~\refeq{eq:flowbasis}
is represented by an array of
$\npart$ integer pairs defined as
\begin{align}
  (\alpha_k,0) &~\text{for an incoming quark (outgoing anti-quark)
    with particle label $\alpha_k$,}\nonumber\\
  (0,\tilde \alpha_k)&~\text{for an incoming anti-quark (outgoing quark)
    with particle label $\beta_k$,}\nonumber\\
  (\alpha_k,\tilde\alpha_k)&~\text{for a gluon with particle label $\alpha_k$,}
    \nonumber\\
  (0,0)&~\text{for an uncoloured particle.}
  \label{eq:colourflowrepr}
\end{align}
The pairs are ordered according to the sequence of scattering particles
as registered by the user.
Each non-zero index will appear twice, indicating which particles are
colour connected.

In leading-colour approximation, the  trace and colour-flow
bases are related through the identities
\begin{align}
  \left(T^{a_1}T^{a_2}\cdots T^{a_{M-1}}T^{a_M}\right)^{\barj{\beta}}_{i_\alpha}
  &~=2^{-M/2}\,
  \delta^{\barj{\beta}}_{i_{a_1}}
  \delta^{\barj{a_1}}_{i_{a_2}} \dots
  \delta^{\barj{a_{M-1}}}_{i_{a_M}} \dots
  \delta^{\barj{a_M}}_{i_\alpha}
+ \text{sub-leading colour},
\nonumber\\
  \Tr\left(T^{a_1}T^{a_2}\cdots T^{a_{M-1}}T^{a_M}\right)
  &~=2^{-M/2}\,
  \delta^{\barj{a_M}}_{i_{a_1}}
  \delta^{\barj{a_1}}_{i_{a_2}} \dots
  \delta^{\barj{a_{M-1}}}_{i_{a_M}}
+ \text{sub-leading colour},\,
  \label{eq:colourflowrepl}
\end{align}
which imply a one-to-one correspondence between the elements of the two
bases, \ie 
\bea
\calC_i = \calC_i^{\mathrm{flow}}+ \text{sub-leading colour}\,.
\label{eq:traceflowcorrresp}
\eea

\paragraph{Squared colour vector}
In leading-colour approximation, the colour-correlation matrices in the 
trace and colour-flow basis are equivalent to each other and proportional to
the identity matrix,
\bea
\calK_{ij}&=&\sum_{\col}\,\calC_i^\dagger\,\calC_j
\,=\,
\sum_{\col}\,  \left(\calC_i^{\mathrm{flow}}\right)^\dagger\,\calC_j^{\mathrm{flow}}
+ \text{sub-leading colour}\nonumber\\
&=& \delta_{ij}\,{2^{-n}}\,N_{\mathrm{c}}^{n+m}
+ \text{sub-leading colour}\,,
\label{eq:tracecolintmat}
\eea
where $n$ and $m$ are defined as above.
Thus the LO probability density \refeq{eq:bornsqcolint} can be written as
\bea
\calW_{00} &=& \frac{N_{\mathrm{c}}^{n+m}}{2^n\nhcs}
\sum_i \big|\calA^{(i)}_{0}\big|^2 + \text{sub-leading colour}\,,
\label{eq:LCbornsqA}
\eea
with%
\footnote{Note that~\refeq{eq:bornsqcolvect} is computed
in the trace basis excluding off-diagonal $\calK_{ij}$  terms but including 
any other sub-leading-colour contributions.}
\bea
\big|\calA^{(i)}_{0}\big|^2
&=&
\sum_{\heli}\left[\calA_0^{(i)}(\heli)\right]^*\,
\calA_0^{(i)}(\heli)\,.
\label{eq:bornsqcolvect}
\eea
This squared colour vector can be evaluated through the interface function 
\texttt{evaluate\_tree\_colvect2}. Since each component of \refeq{eq:bornsqcolvect}
is associated with a given colour flow according to
\refeq{eq:traceflowcorrresp}, in the context of parton-shower matching the ratio
\bea
p^{(i)}=
\frac{\big|\calA^{(i)}_{0}\big|^2}{\sum_i\big|\calA^{(i)}_{0}\big|^2}
\label{eq:colflowprob}
\eea
can be used as the probability with which
the shower starts from the colour-flow configuration
$\calC_i^{\mathrm{flow}}$.

The explicit form of the colour-flow basis for a given process can be accessed
through the interface function \texttt{tree\_colourflow}, which 
returns an array of basis elements $\{\calC_i^{\mathrm{flow}}\}$
in a format corresponding to \refeq{eq:colourflowrepr}. 

The interface functions described in this section are supported under
\texttt{amptype=1,11}. So far they are implemented in a way that 
guarantees consistent results only for leading-QCD Born quantities, \ie terms  of 
order $\alphaS^p\alpha^q$ with maximal power $p$, which involve a single Born 
term of order $g_\rS^p e^q$.

\subsection{Reduction methods and stability system}\label{sec:stabsystem}

As discussed in \refse{sec:numstab},
tree--loop interferences and squared loop amplitudes
are computed using different methods for the
reduction to scalar integrals and the treatment of related
instabilities.

For all types of amplitudes, \OL{} chooses default settings for the stability
system that require adjustments only in very rare cases.

\paragraph{On-the-fly stability system} For tree--loop interferences, with
the only exception of the Higgs Effective Field Theory,
the reduction to scalar integrals is based on the on-the-fly method and the
stability system described in~\refse{se:OL2stab}.
Each processed object carries a cumulative
instability estimator\footnote{This estimate is based on the analytic form
of all presently known spurious singularities.  So far it was found to be
quite reliable.  However, it may have to be improved if new types of
instabilities are encountered.} that is propagated through the algorithm and
updated when necessary.
If the estimated instability exceeds a threshold value, the object at hand
and all subsequent operations connected to it are processed through the qp
channel.
The stability threshold is controlled by the interface parameter
\texttt{hp\_loopacc}, which plays the role of 
target numerical accuracy for the whole Born--loop interference $\calW_{01}$. 
Its default value is 8 and corresponds to 
$\delta \calW_{01}/\calW_{01}\sim 10^{-8}$.

In order to find an optimal balance between CPU performance 
and numerical accuracy, certain aspects of the
stability system can be activated or deactivated 
using the parameter \texttt{hp\_mode}.
Setting \texttt{hp\_mode=1} (default) enables all stability improvements 
described in \refse{se:OL2stab} and is recommended for
NLO calculations with hard kinematics.
Setting \texttt{hp\_mode=2} activates 
qp also for additional types of rank-two Gram-determinant instabilities 
that occur exclusively in IR regions.
This mode is supported only for QCD corrections and is 
recommended for real--virtual NNLO calculations.
Finally,  \texttt{hp\_mode=0} deactivates
the usage of qp through the hybrid-precision system, 
while keeping all stability improvements of analytic type in dp.

\paragraph{Stability rescue system}
For tree--loop interferences in the Higgs Effective Field Theory, the
reduction to scalar integrals is based on external libraries.
The primary reduction
library \texttt{redlib1} (default: \Coli{}-\Collier) is used to evaluate all points
in dp.
The fraction \texttt{stability\_triggerratio} (default: 0.2, meaning 20\,\%)
of the points with the largest $K$-factor is re-evaluated with the secondary
reduction library \texttt{redlib2} (default: \DD{}-\Collier).
If the relative deviation of the two results exceeds
\texttt{stability\_unstable} (default: 0.01, meaning 1\,\%), the point is
re-evaluated in qp with \Cuttools{} including a qp scaling test to estimate
the resulting accuracy.
If the estimated relative accuracy $\delta \calW_{01}/\calW_{01}$
in qp is less than \texttt{stability\_kill}
(default: 1, meaning 100\,\%), the result is set to zero, otherwise the
smaller of the scaled and unscaled qp results 
is returned. 
The accuracy argument of the matrix element routines
(e.g.\ \texttt{evaluate\_loop}) returns the relative deviation of the \Coli{}-\Collier
and \DD{}-\Collier results or, if qp was triggered, of the scaled and unscaled qp result.
In case of a single dp evaluation, the accuracy argument is set to $-1$.

Also squared loop amplitudes are reduced to scalar integrals 
using external libraries. To asses related instabilities,
for all phase-space points the reduction is carried out twice, 
using \texttt{redlib1} and
\texttt{redlib2}.
The option \texttt{stability\_kill2} (default: 10) sets the relative deviation
of the two results beyond which the result is set to zero.
Due to the double evaluation of all points, an accuracy estimate is always
returned by the matrix element routine \texttt{evaluate\_loop2}.

Setting \texttt{redlib1} and \texttt{redlib2}, as well as various other
options to control the stability system, is only possible in the so-called
``expert mode''. Further details can be obtained from the authors upon
request.

\section{Technical benchmarks}
\label{se:benchmarks}
In this section we present speed and stability benchmarks 
obtained with \OLtwo and compare them with the performance of \OLone.

\subsection{CPU performance}

\begin{table}
  \begin{center}
\renewcommand{\arraystretch}{1.3}
    \begin{tabular}[t]{|c|ccc|cc|cc|}\hline
& \multicolumn{3}{|c|}{$t^{\mathrm{def}}_{\mathrm{OL2}}$\,[ms]}
& \multicolumn{2}{|c|}{$t^{\mathrm{11digits}}_{\mathrm{OL2}}/t^{\mathrm{def}}_{\mathrm{OL2}}$}
&
\multicolumn{2}{|c|}{$t^{\mathrm{preset2}}_{\mathrm{OL1}}\,(t^{\mathrm{no\,stab}}_{\mathrm{OL1}})/t^{\mathrm{def}}_{\mathrm{OL2}}$}
\\\hline
process 
& QCD & EW & $\frac{\EW}{\QCD}$ & QCD & EW  & QCD & EW 
\\\hline
$gg\to t\bar t$      & 0.80 & 1.17 & 1.46 & 1.01 & 1.01 & 1.82\,(1.67) & 2.22\,(2.02) \\
$gg\to t\bar tg$           & 21.4 & 24.0  & 1.12 & 1.04 & 1.07 & 1.68\,(1.56) & 2.16\,(2.10)\\
$gg\to t\bar tgg$          & 600  & 582 & 0.97 & 1.15 & 1.22 & 2.18\,(2.17) &
2.64\,(2.59)\\
$gg\to t\bar tggg$      & 21145 & 16928 & 0.80 & 1.09 & 1.14 & 2.59\,(2.55) & 3.06\,(3.06) \\
\hline
$u\bar u\to t\bar t$       & 0.23 & 0.43 & 1.87 & 1.0  & 1.02 & 1.22\,(0.93) & 1.65\,(1.37)\\
$u\bar u\to t\bar t g$     & 3.1  & 8.0  & 2.58 & 1.06 & 1.08 & 1.28\,(1.19) & 1.36\,(1.28)\\
$u\bar u\to t\bar t gg$    & 73   & 176  & 2.41 & 1.16 & 1.19 & 1.45\,(1.45) & 1.64\,(1.63)\\
$u\bar u\to t \bar tggg$   & 2085 & 4862 & 2.33 & 1.26 & 1.28 & 1.88\,(1.88) & 2.05\,(2.04) \\
\hline
$b\bar b \to t\bar t$      & 0.22 & 0.92 & 4.18 & 1.01 & 1.01 & 1.78\,(1.53) & 2.01\,(1.73) \\
$b\bar b \to t\bar t g$      & 3.53 & 18.1 & 5.13 & 1.04 & 1.07 & 2.04\,(1.90) & 1.92\,(1.84) \\
$b\bar b \to t\bar t gg$     & 95  & 415 & 4.37 & 1.18 & 1.23 & 2.15\,(2.05) & 2.49\,(2.40) \\
\hline
$d\bar u\to W^- g$         & 0.33 & 0.71  & 2.15 & 1.03 & 1.03 & 0.96\,(0.79) & 1.45\,(1.17)\\
$d\bar u\to W^- gg$        & 5.6  & 12.9  & 2.30 & 1.05 & 1.10 & 0.99\,(0.92) & 1.14\,(1.05)\\
$d\bar u\to W^- ggg$       & 134  & 269 & 2.01 & 1.16 & 1.22 & 1.33\,(1.28) & 1.44\,(1.44)\\
$d\bar u\to W^- gggg$      & 3760  & 7442 & 1.98 & 1.14 & 1.18 & 1.41\,(1.41) & 1.69\,(1.68) \\
\hline
$d\bar u\to e^-\bar\nu_e$  &0.024 & 0.23 &  9.58 & 1.02 & 1.02 & 1.60\,(0.92) & 1.98\,(1.37)\\
$d\bar u\to e^-\bar\nu_eg$ &0.29  & 1.40 & 4.83 & 1.04 & 1.11 & 1.00\,(0.81) & 1.31\,(1.09)\\
$d \bar u \to e^- \bar\nu_e g g$
                           & 4.0  & 13.3 & 3.33 & 1.13 & 1.27 & 0.80\,(0.75) & 1.11\,(1.11)\\
\hline
$u \bar u \to W^+ W^-$    & 0.19 & 3.34 & 17.6 & 1.00  & 1.00  & 1.47\,(1.19) & 1.42\,(1.36)\\
$u \bar u \to W^+ W^- g$  & 6.7  & 25.7 & 3.84 & 1.16 & 1.06 & 1.31\,(1.24) & 1.46\,(1.40)\\
$u \bar u \to W^+ W^- gg$ & 154  & 379  & 2.46 & 1.19 & 1.15 & 1.63\,(1.60) & 2.03\,(2.01)\\
$u \bar u \to W^+ W^- g g g$
                          &3660   & 8606 & 2.35 & 1.17 & 1.15 & 2.18\,(2.18) & 2.44\,(2.44) \\
\hline
$d \bar d \to e^- \bar \nu_e \mu^+ \nu_\mu$ &
                           0.19   & 9.02 & 47.5 & 1.02 & 1.68 & 0.80\,(0.58) & 1.67\,(1.34) \\
$d \bar d \to e^- \bar \nu_e \mu^+ \nu_\mu g$  &
                           5.6    & 42.2 & 7.54 & 1.23 & 1.85 & 0.57\,(0.51) & 1.36\,(1.15) \\
\hline
    \end{tabular}
  \end{center}
  \caption{Runtimes for the calculation of the NLO QCD and NLO EW virtual
corrections (with respect to the leading QCD Born order) for various
partonic processes at the LHC.  Timings are given per phase-space point,
including colour and helicity sums, and averaged
over a 
sample of random points generated with \Rambo \cite{Kleiss:1985gy} 
at $\sqrt{s}=1$\,TeV without cuts.
The measurements have been carried out on a single Intel
i7-4790K @ 4.00GHz core using gfortran~7.4.0.
The reference \OLtwo timings ($t^{\mathrm{def}}_{\mathrm{OL2}}$) correspond to the
on-the-fly approach with default stability settings, while
$t^{\mathrm{11\,digits}}_{\mathrm{OL2}}$ illustrates the CPU overhead 
caused by augmenting the hybrid-precision target accuracy from 8 to 11 digits.  
Default \OLone timings ($t^{\mathrm{preset2}}_{\mathrm{OL1}}$) correspond to
the recommended stability setting (\texttt{preset}=2), where
tensor reduction is done with \Coli-\Collier and compared against 
\DD-\Collier for 20\% of the points with the largest $K$-factor;
differences beyond one percent between \Coli-\Collier and \DD-\Collier
trigger qp re-evaluations with \Cuttools+\OneLoop and a further
stability test via qp-rescaling.
For comparison, also \OLone
timings with disabled stability system ($t^{\mathrm{no\, stab}}_{\mathrm{OL1}}$) are
shown within parentheses.
}
\label{tab:perfnew}
\end{table}

The speed at which one-loop matrix elements are evaluated plays a
key role for the feasibility and efficiency of non-trivial
NLO Monte Carlo simulations.
In \refta{tab:perfnew} we present CPU timings for the
calculation of one-loop QCD and EW corrections for several processes of
interest at the LHC. Specifically, we consider the production of 
single $W$ bosons, $W^+W^-$ pairs and $t\bar t$ pairs in association with a variable
number of additional gluons and quarks.
For $W$ production we consider 
final states with on-shell bosons and, alternatively, off-shell $\ell\nu$ decay
products.  

The observed timings  are roughly proportional to the number of 
one-loop Feynman diagrams, which ranges from $\ord(10)$ for the
simplest $2\to 2$ processes to $\ord(10^5)$ for the
most complex $2\to 5$ processes.
Absolute timings correspond to \OLtwo with default settings,
\ie with all stability improvements in dp plus the 
hybrid-precision system with a target accuracy of 8 digits.
Augmenting the target accuracy to 11 digits causes a CPU overhead of 1\% to
50\%, depending on the process, while we have checked that switching off
hybrid precision (\texttt{hp\_mode}=0) 
yields only a speed-up of order one percent.

Comparing QCD to EW corrections, for processes without leptonic weak-boson
decays we observe timings of the same order.  More precisely, the QCD (EW)
corrections
tend to be comparatively more expensive in the presence of more 
external gluons (weak bosons).  %
In contrast, in processes with
off-shell weak bosons decaying into leptons EW corrections are drastically more
expensive than QCD corrections. This is due to the fact that, for each off-shell $W/Z$
decay to
leptons, at NLO~EW the maximum number of loop propagators 
increases by one, while at NLO QCD it remains unchanged.
Due to Yukawa interactions, also the presence of massive quarks
tends to increase the CPU cost of EW corrections.

Timings of \OLtwo are compared against \OLone with recommended stability
settings (\texttt{preset}=2, \texttt{preset} is deprecated in \OLtwo) and, alternatively,
with the stability rescue system switched off (``no stab'') in \OLone.  The
difference reflects the cost of
stability checks in \OLone, which is significantly  higher
 than in  \OLtwo.  Note that this cost depends very
strongly on the kinematics of the considered phase-space sample, and 
the values reported in \refta{tab:perfnew} should be understood as a lower bound.

Apart from few exceptions, \OLtwo is similarly fast or
significantly faster than \OLone.  In particular, for the most complex and
time consuming processes the new on-the-fly approach yields speed-up factors
between two and three.

\subsection{Numerical stability}

As discussed in \refse{sec:numstab}, the stability of 
one-loop amplitudes in exceptional phase-space regions is of crucial
importance for challenging multi-particle and multi-scale NLO calculations,
as well as for NNLO applications.  In the following we present \OLtwo stability benchmarks
for NLO~QCD and NLO~EW virtual corrections.
The level of numerical stability is quantified by comparing output in double
(dp) or hybrid (hp)
precision ($\calW^{\ssst{dp/hp}}_{01}$) against
quadruple-precision (qp) benchmarks ($\calW^{\ssst{qp}}_{01}$). The latter
are obtained using \OLtwo in combination with the \OneLoop library for scalar integrals.
More precisely, we define the numerical instability of a certain result
$\calW^{X}_{01}$ as
%
\bea
\label{eq:relativedeviation} 
\calA_X &=& \log_{10}
\left|\frac{\calW^X_{01}-\calW^{\ssst{qp}}_{01}}{\calW^{\ssst{qp}}_{01}}\right|\,,
\eea
which corresponds, up to a minus sign, to the number of 
stable digits. For the case of qp benchmark results ($X=\mathrm{qp}$)
the accuracy estimate \refeq{eq:relativedeviation}
corresponds to the result of a so-called rescaling
test, see Section \ref{se:OL1stab}(iii).

\begin{figure}
\centering
\begin{subfigure}{0.85\textwidth}
\includegraphics[width=\textwidth]{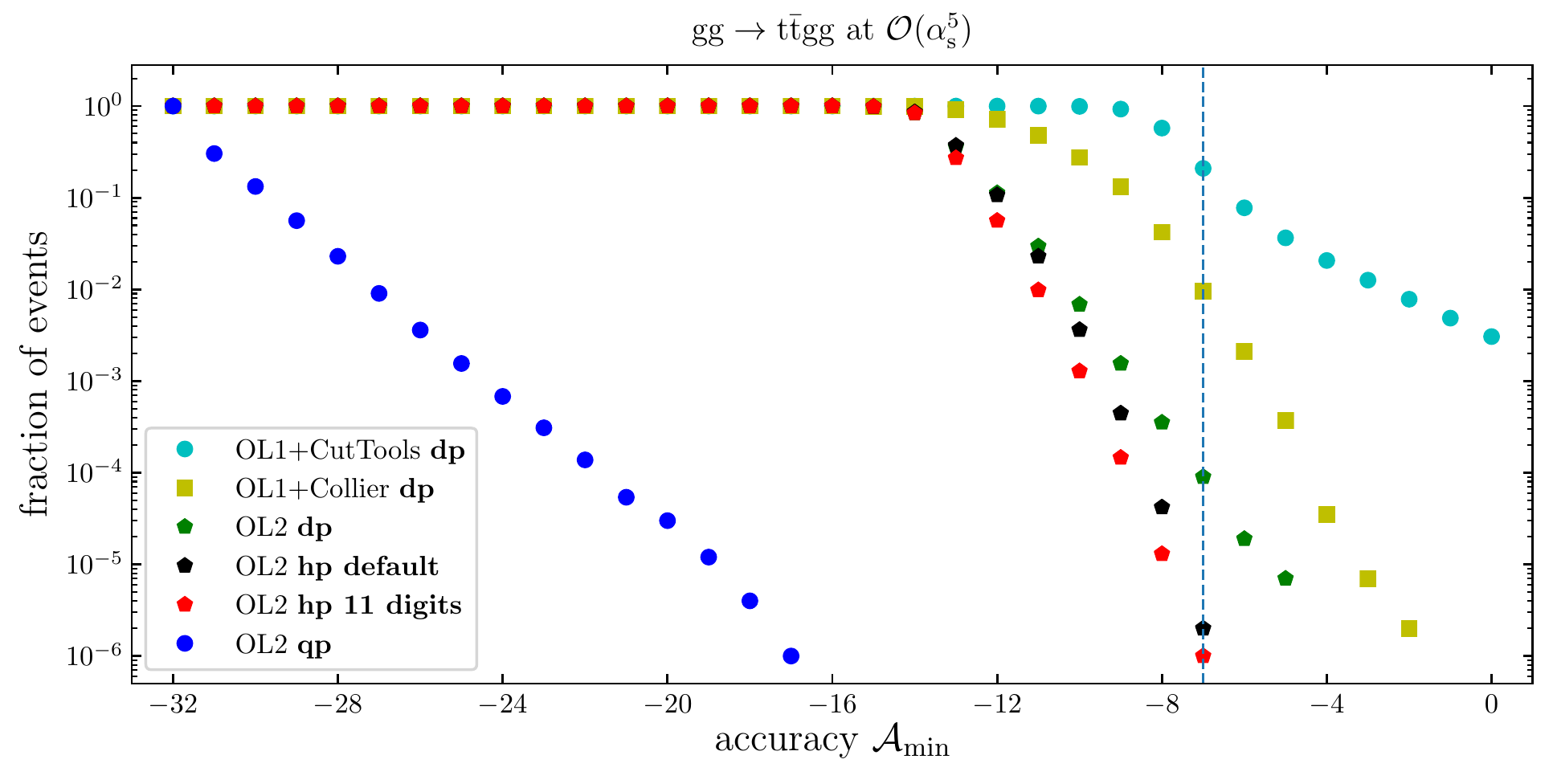}
\end{subfigure}
\vspace{8mm}

\begin{subfigure}{0.85\textwidth}
\includegraphics[width=\textwidth]{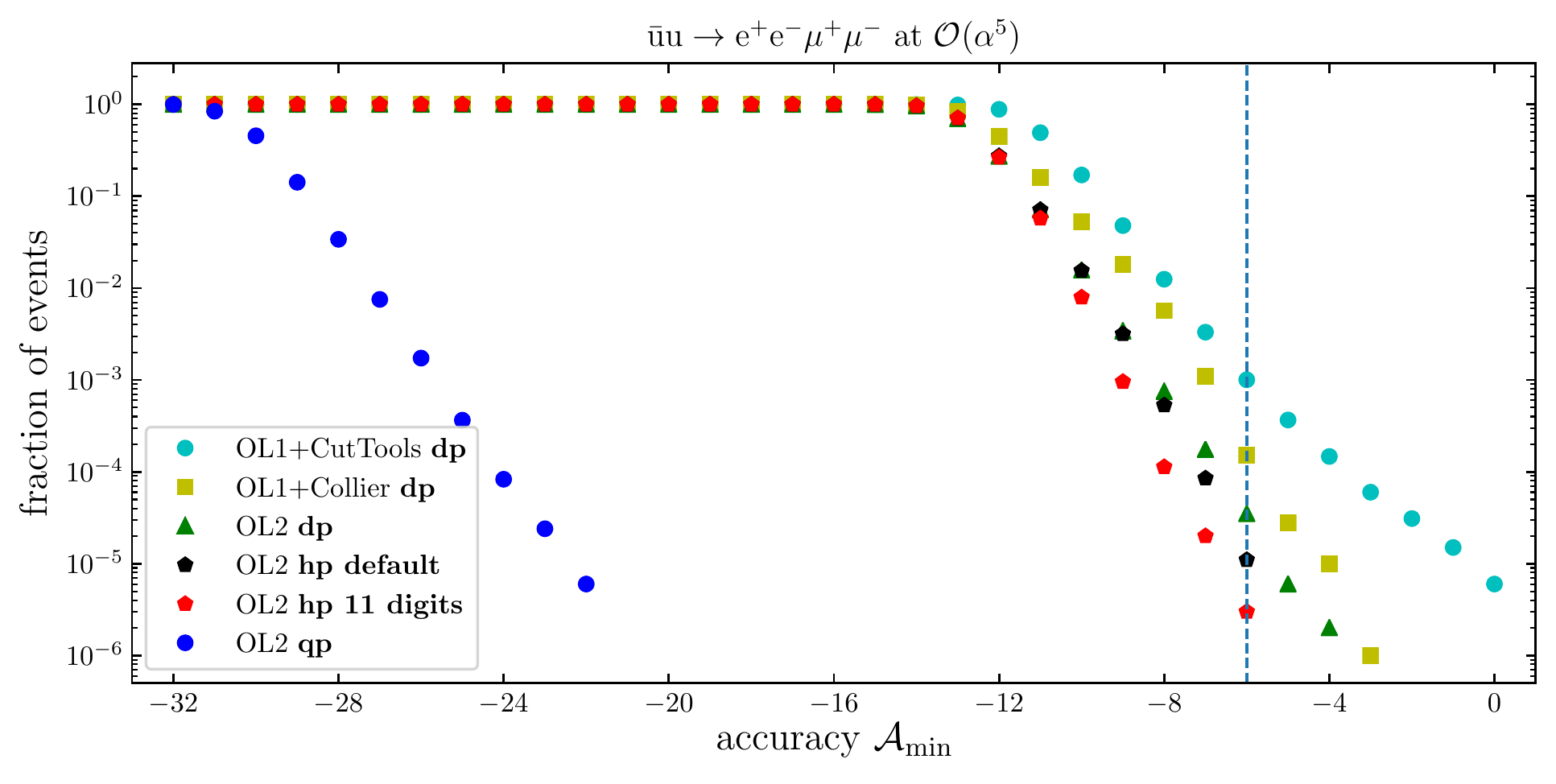}
\end{subfigure}
\caption{Probability of finding an instability
$\calA>\calA_{\mathrm{min}}$ as a function of $\calA_{\mathrm{min}}$ in a
sample of $10^6$ events for 
$gg\to t\bar t gg$ at NLO QCD (upper plot) 
and  $\bar u u \to e^+e^-\mu^+\mu^-$ at NLO EW (lower plot).
The stability of quad-precision benchmarks (blue) is compared to different 
variants of the \OLtwo on-the-fly reduction
(green, black, red) and to the \OLone algorithm interfaced with
\Collier (yellow) or \Cuttools (turquoise).
For \OLtwo, besides default stability settings  (black) we show the effect
of increasing the hybrid-precision target from 8 to 11 digits
($\texttt{hp\_loopacc=11}$, red), or disabling the hybrid precision system
($\texttt{hp\_mode=0}$, green).
The \OLone curves correspond to the level of stability 
that is obtained in dp without full re-evaluations of unstable points in qp.
}
\label{stabilityhard}
\end{figure}

The numerical stability of \OLtwo in the hard regions 
is illustrated in \reffi{stabilityhard} for two non-trivial 
$2\to 4$ processes at NLO QCD and NLO EW.
The plots correspond to $10^6$ homogeneously distributed \Rambo points
at $\sqrt{s}=1$\,TeV with $p_{i,\mathrm{T}} >
50$\,GeV and  $\Delta R_{ij} > 0.5$ for all massless final-state
particles.
As demonstrated by the reference qp curve, running \OLtwo in pure qp makes it
possible to produce one-loop results with up to 32 stable digits.  Such
high-precision qp benchmarks can be obtained as a by-product of the
hybrid-precision system and allow one to quantify the level of stability
with better than 16-digit resolution in the full phase space.
The results of \OLone with \Cuttools in dp illustrate the impact of
Gram-determinant instabilities, which result in a probability of one percent
of finding less than two stable digits in $gg\to t\bar t gg$.%
\footnote{In the tail of the \Cuttools curve (not shown) numerical
instabilities can reach and largely exceed $\ord(10^{10})$.}
Using \Collier reduces this probability by 
3--4 orders of magnitudes, while \OLtwo with 
one-the-fly reduction and hp-system leads to a further dramatic 
suppression of instabilities by four orders of magnitude, which corresponds to 
five extra stable digits.
The effect of hybrid-precision alone corresponds to about 
two digits or, equivalently, a factor 100 suppression of the tail. 
The EW corrections to $\bar u u \to e^+e^-\mu^+\mu^-$ feature a
qualitatively similar behaviour but a generally lower level of instability,
which is most likely a consequence of the lower tensor rank.\\

\begin{figure}
\centering
\begin{subfigure}{0.85\textwidth}
\includegraphics[width=\textwidth]{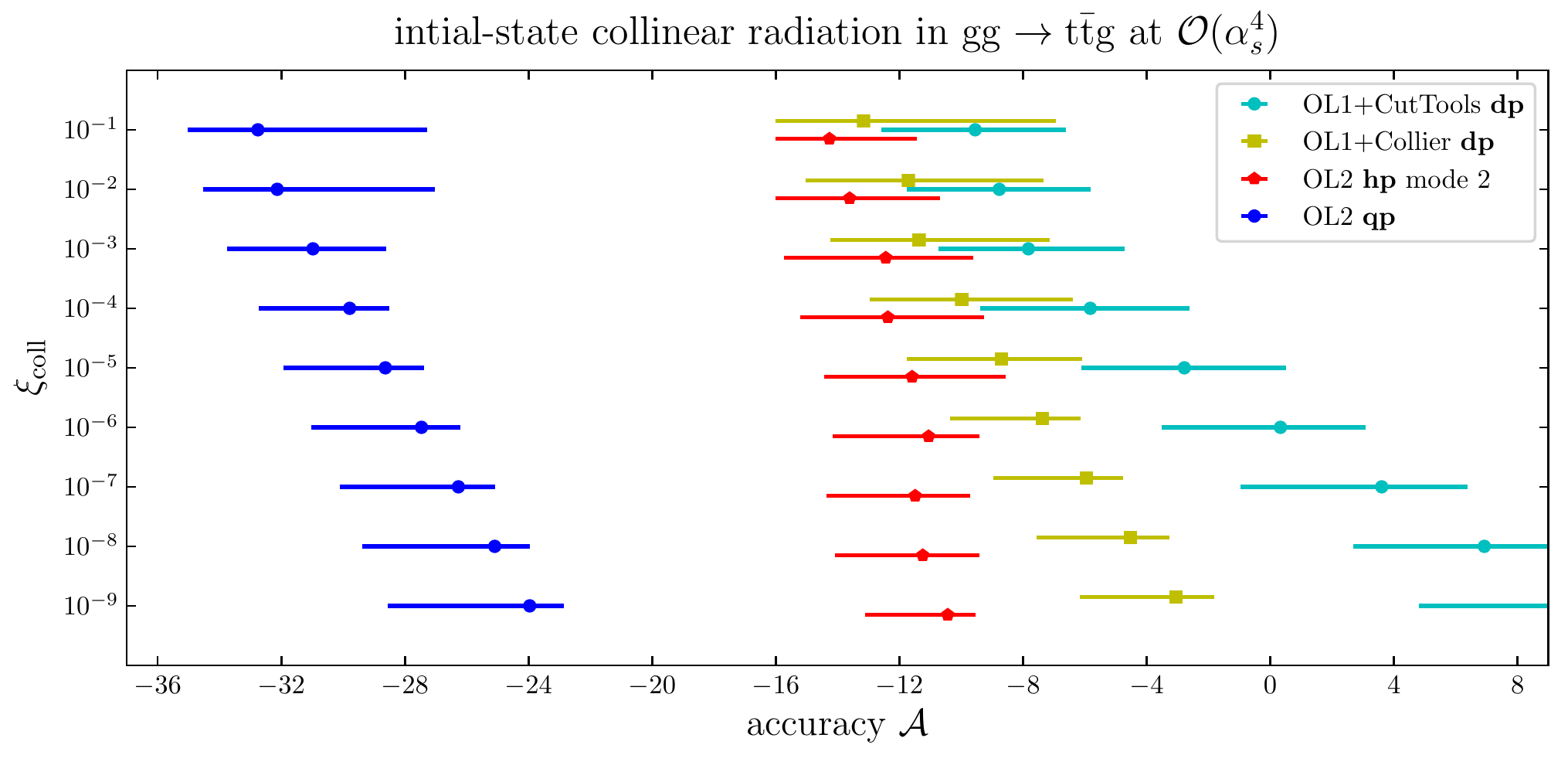}
\end{subfigure}
\vspace{8mm}

\begin{subfigure}{0.85\textwidth}
\includegraphics[width=\textwidth]{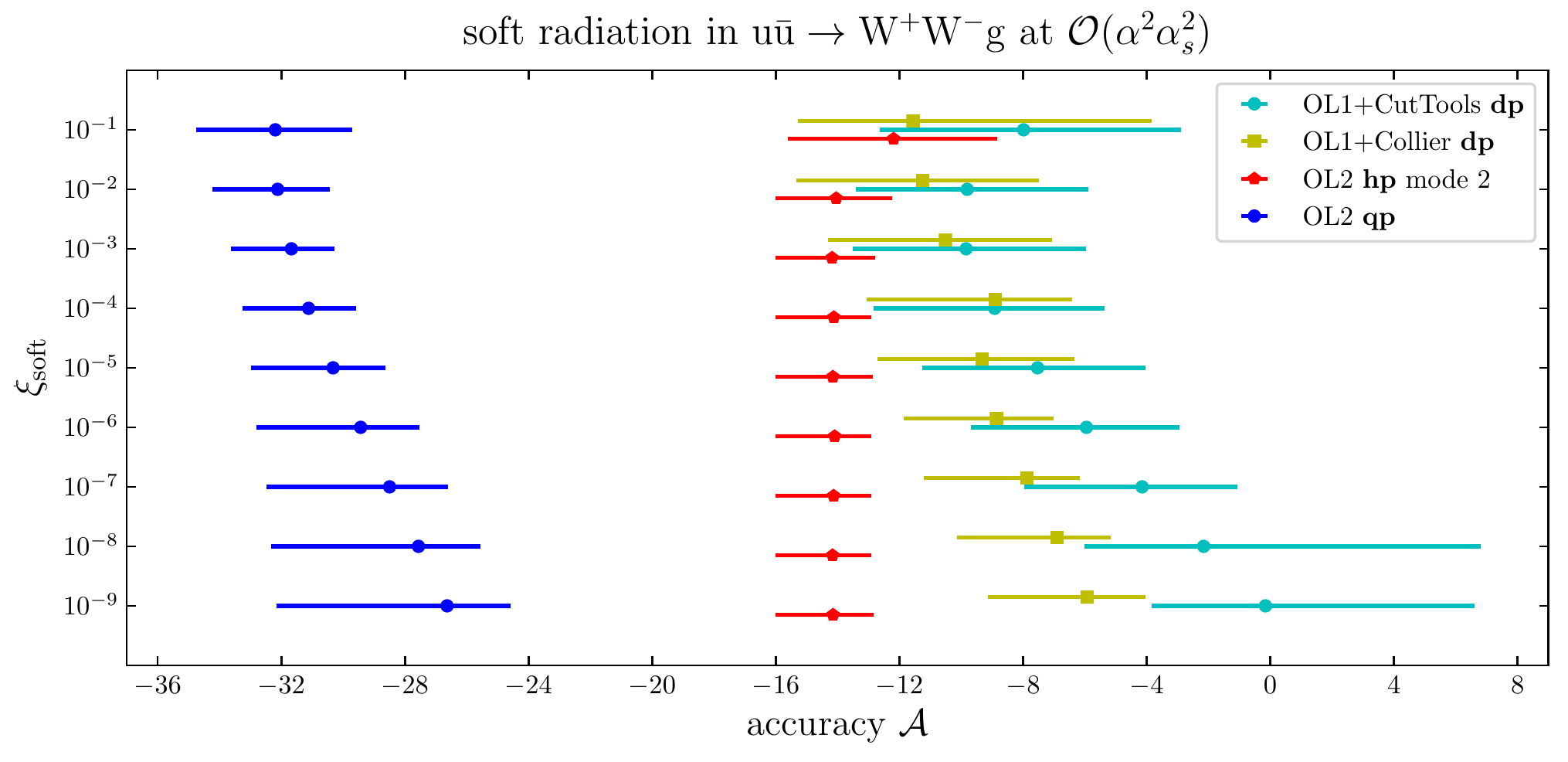}
\end{subfigure}
\caption{
Relative numerical accuracy 
$\calA$ for $gg\to t\bar t g$ (upper plot) 
and  $u \bar u \to W^+W^-g$ (lower plot) at NLO QCD
versus the degree 
of collinear ($\xi_{\mathrm{coll}}$) or soft singularity ($\xi_{\mathrm{soft}}$)
as defined in \refeq{eq:softcolldeg}.
For each value of $\xi_{\mathrm{coll/soft}}$
the numerical accuracy 
is estimated with a sample of $10^4$ randomly distributed 
underlying $2\to 2$ hard events.
The plotted central points and variation bands correspond, respectively,
to the average and $99.9\%$ confidence interval of $\calA$.
Quad-precision benchmarks (blue) are compared to 
\OLtwo with additional hybrid-precision improvements 
for IR regions (\texttt{hp\_mode=2}, red)
and also to \OLone with 
\Collier (yellow) or \Cuttools (turquoise) in dp.
}
\label{stabilityir}
\end{figure}

Example stability benchmarks relevant for $2\to 2$ calculations at NNLO
are shown in \reffi{stabilityir} for the case of the real-virtual 
QCD corrections to  $t\bar t$ and $W^+W^-$ hadron production. 
The instability $\calA$ is estimated using a sequence of 
$gg\to t\bar t g$ and $u\bar u \to W^+W^-g$ samples with increasing degree
of softness and collinearity, defined as
\bea
  \xi_\mathrm{soft} &=& \frac{E_j}{Q},\qquad
  \xi_\mathrm{coll} \,=\, \theta_{ij}^2\,.
\label{eq:softcolldeg}
\eea
Here $Q$ denotes the center-of-mass energy,
$E_j$ is the energy of the soft particle, 
and $\theta_{ij}$ is the angle 
of a certain collinear branching.
The parameters $\xi_\mathrm{soft/coll}$ are 
defined in such a way that the denominators of soft and collinear enhanced propagators
scale like $(p_i+p_j)^2 \propto \xi_\mathrm{soft/coll}\,Q^2$.
In practice, starting from a sample of $10^4$ hard $2\to 2$ events
with $Q=1$\,TeV, we have supplemented each event by an additional soft or collinear emission with 
$\xi_{\mathrm{soft/coll}}= 10^{-1}, 10^{-2},\dots,10^{-9}$.

In \reffi{stabilityir} the average level of instability and its spread 
are plotted versus $\xi_{\mathrm{coll}}$ in $gg\to t\bar t$  and  $\xi_{\mathrm{soft}}$ in $u\bar u \to  W^+W^-g$.
The stability of qp benchmarks is again very high in the 
whole phase space. In the deep IR regions numerical instabilities grow at a
speed that depends on the process, the  type of region
(soft/collinear), and the employed method.
For initial-state collinear radiation in $gg\to t\bar t g$, \Cuttools 
loses three digits per order of magnitude 
in $\xi_{\mathrm{coll}}$, resulting in huge average instabilities
of $\ord(10^{10})$ in the deep unresolved regime.
Using the \Collier library in dp we observe a more favourable scaling,
with losses of only one digit per order of magnitude in 
$\xi_{\mathrm{coll}}$, 
and an average of three stable digits in the tail.
Thanks to the hybrid-precision system, 
the level of stability of \OLtwo 
is even much higher. It stays always 
above 10 digits and 
is roughly independent of $\xi_{\mathrm{coll}}$.
For soft radiation in $u\bar u \to W^+W^-g$, 
apart from the fact that numerical instabilities are generally milder,
the various tools behave in a qualitatively similar way.\\

Similar tests of the \OLtwo stability system as the ones presented here have been carried out for various
$2\to 3,4, 5$ hard processes  and $2\to 3$ processes with an
unresolved parton, finding similar stability curves as shown here, and not a
single fully unstable result, i.e.\ one with zero correct digits.
A more comprehensive study on numerical instabilities will be presented in a
follow-up paper~\cite{OL2_stability}.

\section{Summary and conclusions}

We have presented \OLtwo, the latest version of the \OpenLoops tree 
and one-loop amplitude provider based on the open-loop recursion.
This new version introduces two significant novelties highly relevant for state-of-the art
precision simulations at high-energy colliders. First, the original algorithm has been extended to 
provide one-loop amplitudes in the full SM, \ie including, besides QCD
corrections,
also EW corrections from gauge, Higgs and Yukawa interactions.
The inclusion of EW corrections becomes mandatory for the control of
cross sections at the percent level, and even more
importantly in the tails of distributions at energies well above the EW scale.
Second, the original algorithm
has been extended to include the recently proposed on-the-fly reduction
method, which 
supersedes the usage of external reduction libraries for the calculation of tree--loop
interferences.
In this approach, loop amplitudes are constructed in a way that 
avoids high tensorial rank at all stages of the calculation, thereby 
preserving and often ameliorating (by up to a factor of three) 
the excellent CPU performance of \OLone.
The on-the-fly reduction algorithm has opened the door to a series of
new techniques that have reduced the level of numerical 
instabilities in exceptional phase-space regions
by up to four orders of magnitude.
These speed and stability improvements are especially
significant for challenging 
multi-leg NLO calculations and for real-virtual contributions in NNLO computations. 

In this paper we have presented the algorithms implemented in \OLtwo for the calculation
of squared tree, tree--loop interference and squared loop amplitudes. This entails 
a summary of the on-the-fly reduction method~\cite{Buccioni:2017yxi} 
and its stability system, which automatically identifies and cures 
numerical instabilities in exceptional phase-space regions.
This is achieved by means of Gram-determinant expansions and other analytic methods
in combination with a hybrid
double-quadruple precision system. The latter 
ensures an unprecedented level of numerical stability, while making use of
quadruple precision only for very small parts of the amplitude construction.
Details of these stability improvements and 
hybrid precision system will be presented in an upcoming publication~\cite{OL2_stability}.

In the context of the extension to calculations in the full SM, we 
presented a systematic discussion of the bookkeeping of QCD--EW interferences
and sub-leading one-loop contributions, which are relevant for processes 
with multiple final-state jets. We also detailed the input parameter schemes and
one-loop $\ord(\alphaS)$ and $\ord(\alpha)$ renormalisation as implemented in \OLtwo.
Here we emphasised crucial details in the implementation of the complex-mass scheme
for the description of off-shell unstable particles.
The flexible implementation of the complex-mass scheme in \OLtwo is
applicable to processes with both on-shell and off-shell unstable particles at NLO.
We also introduced a special treatment of processes with external photons,
handling photons of on-shell and off-shell type in different ways,
which is inherently required by the cancellation of 
fermion-mass singularities associated with the 
photon propagator and with collinear splitting processes.

While this manuscript as a whole provides detailed documentation of the algorithms implemented
in \OLtwo, Section~\ref{sec:program} together with Appendix~\ref{app:native}  
can be used as a manual, both in order to use \OLtwo as a standalone program
or to interface it to any Monte Carlo 
framework. Calculations at NLO and beyond  require, besides squared amplitude
information, 
also spin and colour correlators for the construction of infrared subtraction terms. 
To this end we documented
the available correlators and conventions available in \OLtwo, which comprise tree-tree and loop-loop
correlators as well as tree-loop correlators. The former are necessary for the construction of NLO
subtraction terms for standard and loop-induced processes. The latter are necessary in NNLO 
subtraction schemes. Furthermore, conventions and interfaces for the extraction of full
tree amplitude vectors in colour space are given. These are necessary ingredients for
parton shower matching at NLO.

The new functionalities of \OLtwo and their future 
improvement will open the door to a wide range of
new precision calculations in the High-Luminosity era of the LHC.

\appendix


\definecolor{lightgray}{gray}{0.95}

\lstdefinestyle{overallstyle}{
basicstyle=\small,
backgroundcolor=\color{lightgray}
}

\lstdefinestyle{cstyle}{
style=overallstyle,
language=C,
title=C/\cpp
}

\lstdefinestyle{fortstyle}{
style=overallstyle,
language=Fortran,
title=Fortran,
}

\section{Native \Fortran and C/C++ interfaces}
\label{app:native}

\OpenLoops can easily be integrated into Monte Carlo tools via its native interfaces in \Fortran and C or via the BLHA interface
\cite{Binoth:2010xt,Alioli:2013nda}.
The C interface can of course be used from \cpp as well.
We recommend to use the native interface, because it is easier to use, provides more functionality and does not require exchanging files between the tools. 
%
%
In this Appendix we 
%
present the various functionalities of the native \OpenLoops
interface. In doing so we will always refer to the names of the relevant \Fortran
interface functions.  The corresponding C functions are named in the same
way with an extra \texttt{ol\_} prefix.
In Appendix~\ref{app:native:generalities} we detail necessary modules to be loaded (\Fortran) and required header files
(C/\cpp) together with conventions for the format of phase-space points for the evaluation of scattering amplitudes.
In Appendix~\ref{app:native:parameters} the setting of
parameters is discussed and in Appendix~\ref{app:native:registration} the registration of processes.
In Sections~\ref{app:native:amplitudes}-\ref{app:native:colour} we detail the various
interfaces for the evaluation of squared scattering amplitudes, amplitude correlators and 
amplitude colour vectors.
Finally, in Appendix~\ref{app:native:examples} we give a basic example for the usage of the native \OL interface in \Fortran and C.

The implementation of the BLHA interface and the usage of \OL together with \Sherpa and \Powheg are discussed
in Appendix~\ref{app:otherinterfaces}.

\subsection{Generalities}
\label{app:native:generalities}

\paragraph{Fortran} 

In order to use the native \Fortran interface, the module \texttt{openloops} must be included with 
\begin{lstlisting}[style=overallstyle,language=Fortran]
use openloops
\end{lstlisting}
The module files are located in the directory \texttt{lib\_src/openloops/mod}, which should be added to the include path of the \Fortran compiler.

\noindent
Floating point numbers used in the interface are in double precision, denoted here by the kind type \texttt{dp} which can be obtained as follows:
\begin{lstlisting}[style=overallstyle,language=Fortran]
integer, parameter :: dp = selected_real_kind(15)
\end{lstlisting}
Phase space points \texttt{p\_ex} are passed as two-dimensional arrays declared as
\begin{lstlisting}[style=overallstyle,language=Fortran]
real(dp) :: p_ex(0:3,N)
\end{lstlisting}

Here and in the following $N$ stands for the number of incoming plus
outgoing external
particles of the considered process.  External particles are numbered from
$1$ to $N$ and are interpreted as incoming or outgoing according to the process
registration. See \refeq{eq:processdef} and below.
The entries \texttt{p\_ex(i,K)} correspond to the
energy (\texttt{i}=0) and the three physical momentum components (\texttt{i}=1,2,3) 
of particle $K$ in GeV units. 

\paragraph{C} 

The C interface is declared in the the header file \texttt{include/openloops.h} and can be included in C and \cpp code. Phase space points \texttt{pp} are passed as one-dimensional arrays with $5N$ components, where every fifth component is the mass of the corresponding external particle (BLHA
convention),
i.e.\ phase-space points in the C interface are declared as
\begin{lstlisting}[style=overallstyle,language=C]
double pp[5*N];
\end{lstlisting}
The fifth component is currently not used within \OpenLoops.

\subsection{Parameter setting}
\label{app:native:parameters}

In order to set the \OL parameter with name \texttt{key} to the value \texttt{val}, call
\begin{lstlisting}[style=fortstyle]
subroutine set_parameter(key, val, err)
  character(*), intent(in) :: key
  TYPE, intent(in) :: val
  integer, intent(out), optional :: err
\end{lstlisting}
where \texttt{TYPE} is \texttt{integer}, \texttt{real(dp)} or \texttt{character(*)} depending on the type of the parameter. It is possible to set parameters of \texttt{integer} or \texttt{real(dp)} type by passing the value in string representation. The error code \texttt{err} will be zero on success.

In C, the function to set a parameter depends on the parameter type:
%
\begin{lstlisting}[style=cstyle]
void ol_setparameter_int(const char *key, int val);
void ol_setparameter_double(const char *key, double val);
void ol_setparameter_string(const char *key, const char *val);
\end{lstlisting}
\texttt{ol\_setparameter\_string()} may be used to set integer or double precision values given in string representation. The functions do not return an error code, but it may be retrieved by calling
%
\begin{lstlisting}[style=cstyle]
int  ol_get_error();
\end{lstlisting}
right after setting a parameter. A return value of 0 means that no error occured in the preceeding call.

With the default settings, the program will terminate in case of an error. This can be changed by adjusting the warning level using the function
%
\begin{lstlisting}[style=fortstyle]
subroutine set_init_error_fatal(level)
  integer, intent(in) :: level
\end{lstlisting}
\begin{lstlisting}[style=cstyle]
void ol_set_init_error_fatal(int level)
\end{lstlisting}
where \texttt{level=0} means that errors are silently ignored, \texttt{level=1} means that a warning message is
printed,
and \texttt{level=2} (default) means that the program will be terminated on error.

\noindent
The current value of a parameter can be retrieved by calling
%
\begin{lstlisting}[style=fortstyle]
subroutine get_parameter(key, val, err)
  character(*), intent(in) :: key
  TYPE, intent(out) :: val
  integer, intent(out), optional :: err
\end{lstlisting}
\begin{lstlisting}[style=cstyle]
void ol_getparameter_int(const char *key, int *val);
void ol_getparameter_double(const char *key, double *val);
\end{lstlisting}
Retrieving parameter values is only supported for integer and double precision parameter types.

\noindent A list of all parameters can be written to a file
\begin{lstlisting}[style=fortstyle]
subroutine printparameter(file)
  character(*), intent(in) :: file
\end{lstlisting}
\begin{lstlisting}[style=cstyle]
void ol_printparameter(const char *file);
\end{lstlisting}
For an empty file name, \ie file='''', the output is written to stdout.

\subsection{Process registration}
\label{app:native:registration}

As detailed in Section~\ref{se:processselection} before evaluation a process has to be 
registered. This proceeds via
%
\begin{lstlisting}[style=fortstyle]
function register_process(process, amptype)
  integer :: register_process
  TYPE, intent(in) :: process
  integer, intent(in) :: amptype
\end{lstlisting}

which takes the \texttt{process} as a string in the format 
``${\rm PID}_{i,1} \dots {\rm PID}_{i,n} \,\texttt{->}\,  {\rm PID}_{f,1} \dots {\rm PID}_{f,m}$''
for a $n\to m$ process,
where the various particle identifiers (PID) are enetered in either of the two
particle labelling schemes specified in \refta{particleid}.
Alternatively, $2\to N-2$ processes can be registered by entering
\texttt{process} as an array of integers of length $N$, 
where the first two entries are interpreted
as initial-state particles. 
Additionally the amplitude type \texttt{amptype} has to be passed as argument. For the possible values of \texttt{amptype} see Tab.~\ref{amptypes}. The function \texttt{register\_process} returns the process ID to be used in the routines to evaluate matrix elements, where it is denoted as \texttt{id}.

In the corresponding C interface for process registration
\begin{lstlisting}[style=cstyle]
int ol_register_process(const char *process, int amptype);
\end{lstlisting}
the  \texttt{process} can only be passed as a string. Again, the process ID is returned.

\noindent
When all processes are registered the following function must be called before calculating matrix elements.
%
\begin{lstlisting}[style=fortstyle]
subroutine start()
\end{lstlisting}
\begin{lstlisting}[style=cstyle]
void ol_start();
\end{lstlisting}
When the calculation is finished, \ie no more matrix elements will be calculated, the following function should be called.
%
\begin{lstlisting}[style=fortstyle]
subroutine finish()
\end{lstlisting}
\begin{lstlisting}[style=cstyle]
void ol_finish();
\end{lstlisting}
While these calls are not strictly necessary, if log files are used, the files may not be updated at the end of the run and therefore lack information. Additionally, dynamically allocated memory will be deallocated upon the \texttt{finish} call. 

\subsection{Scattering amplitudes}
\label{app:native:amplitudes}

The following interface functions evaluate
the scattering probability densities~\refeq{eq:Wtree}--\refeq{eq:Wloop2}
and their building blocks described in \refse{sec:amplitudecalls}.
The required inputs are the integer identifier 
\texttt{id} of the desired process
and the phase-space point \texttt{p\_ex} (\Fortran) / \texttt{pp} (\cpp), as defined in \ref{app:native:generalities}.

\paragraph{Tree-level amplitudes}

The function \texttt{evaluate\_tree} evaluates the 
tree--tree probability density \refeq{eq:Wtree} returning 
\texttt{m2l0}=$\calW_{\tree}$ as output.
%
%
\begin{lstlisting}[style=fortstyle]
subroutine evaluate_tree(id, p_ex, m2l0)
  integer, intent(in) :: id
  real(dp), intent(in) :: p_ex(4,N)
  real(dp), intent(out) :: m2l0
\end{lstlisting}
\begin{lstlisting}[style=cstyle]
void ol_evaluate_tree(int id, const double *pp, double *m2l0);
\end{lstlisting}

\paragraph{One-loop NLO amplitudes}

The function \texttt{evaluate\_loop} evaluates the UV renormalised Born--one-loop  interference \eqref{eq:Wloop}
returning \texttt{m2l0}=$\calW_{\tree}$ and \texttt{m2l1}=$\{\calW_{\onel}^{(0)}, \calW_{\onel}^{(1)},\calW_{\onel}^{(2)}\}$  as output. The three values in \texttt{m2l1} represent  the finite part, 
and the coefficients of the IR single and  double poles\footnote{For performance reasons, by default the (negative) IR poles of the $\mathbf{I}$-operator, Eq.~\eqref{eq:ioperatorB},
are returned as IR poles in \texttt{m2l1}.  The true poles of the virtual amplitudes can be
obtained by setting the parameter \texttt{truepoles=1}. Alternatively setting \texttt{truepoles=2}
sums the virtual amplitude including its true poles and the $\mathbf{I}$-operator including its finite part 
and poles, which allows for easy pole cancellation checks. See more details
in \refse{sec:amplitudecalls}.}
of the Born--one-loop  interference, as defined in
Eq.~\eqref{eq:laurentseries}.
Together with the one-loop amplitude an accuracy estimate is returned (depending on the employed stability system) as \texttt{acc} with \texttt{acc}=-1 in case no stability estimate is available. 
When available, \texttt{acc} quantifies the relative accuracy $\delta
\calW_{\onel}^{(0)}/\calW_{\onel}^{(0)}$, and
\texttt{acc}=$10^{-a}$ corresponds to an estimated accuracy of $a$ decimal digits.
%
\begin{lstlisting}[style=fortstyle]
subroutine evaluate_loop(id, p_ex, m2l0, m2l1, acc)
  integer, intent(in) :: id
  real(dp), intent(in) :: p_ex(4,N)
  real(dp), intent(out) :: m2l0
  real(dp), intent(out) :: m2l1(0:2)
  real(dp), intent(out) :: acc
\end{lstlisting}
\begin{lstlisting}[style=cstyle]
void ol_evaluate_loop(int id, const double *pp,
                      double *m2l0, double *m2l1, double *acc);
\end{lstlisting}

As documented in \refse{sec:amplitudecalls}, various technical parameters
permit to activate and deactivated the different building blocks of one-loop
amplitudes and to change the normalisation convention for UV and IR poles.

\paragraph{Bare d=4 amplitudes}

The function \texttt{evaluate\_loopbare} evaluates  the unrenormalised Born--one-loop  interference  without UV and $R_2$ 
counterterm contributions (i.e.~with $d=4$ loop numerator) 
as defined in \eqref{eq:loopingredients}, returning \texttt{m2l0}=$\calW_{\tree}$,  
\texttt{m2l1bare}=$\{\calW_{\onel,\fourdim}^{(0)},\calW_{\onel,\fourdim}^{(1)},\calW_{\onel,\fourdim}^{(2)} \}$  and an accuracy estimate (see above) \texttt{acc}  as output. 
The three values in \texttt{m2l1bare} represent the finite part and the coefficients of the (UV and IR) single and the double poles.\footnote{ For performance reasons, by default the (negative) IR poles of the $\mathbf{I}$-operator
and UV counterterm are returned as poles in \texttt{m2l1bare}.  The true poles of the bare virtual amplitudes can be
obtained by setting the parameter \texttt{truepoles=1}. }
%
\begin{lstlisting}[style=fortstyle]
subroutine evaluate_loopbare(id, p_ex, m2l0, m2l1bare, acc)
  integer, intent(in) :: id
  real(dp), intent(in) :: p_ex(4,N)
  real(dp), intent(out) :: m2l0
  real(dp), intent(out) :: m2l1bare(0:2)
  real(dp), intent(out) :: acc
\end{lstlisting}
\begin{lstlisting}[style=cstyle]
void ol_evaluate_loopbare(int id, const double *pp,
                      double *m2l0, double *m2l1bare, double *acc);
\end{lstlisting}

\paragraph{UV counterterms}

The function \texttt{evaluate\_loopct} evaluates the UV counterterm matrix element, as defined in \eqref{eq:loopingredients} returning \texttt{m2l0}=$\calW_{\tree}$  and \texttt{m2ct}=$\{\calW_{\onel,\ct}^{(0)},\calW_{\onel,\ct}^{(1)},\calW_{\onel,\ct}^{(2)} \}$ as output.
  The three values in \texttt{m2ct} represent the finite part and  the coefficients of the (UV) single and double poles, where the latter is always zero.

\begin{lstlisting}[style=fortstyle]
subroutine evaluate_loopct(id, p_ex, m2l0, m2ct)
  integer, intent(in) :: id
  real(dp), intent(in) :: p_ex(4,N)
  real(dp), intent(out) :: m2l0
  real(dp), intent(out) :: m2ct(0:2)
\end{lstlisting}
\begin{lstlisting}[style=cstyle]
void ol_evaluate_loopct(int id, const double *pp, double *m2l0, double *m2ct);
\end{lstlisting}

For performance reasons we also provide the function \texttt{evaluate\_ct}, which evaluates only the finite
part of the UV counterterm, defined in \eqref{eq:loopingredients}, returning \texttt{m2ct0}=$\calW_{\onel,\ct}^{(0)}$
and \texttt{m2l0}=$\calW_{\tree}$ as output.

%
\begin{lstlisting}[style=fortstyle]
subroutine evaluate_ct(id, p_ex, m2l0, m2ct0)
  integer, intent(in) :: id
  real(dp), intent(in) :: p_ex(4,N)
  real(dp), intent(out) :: m2l0
  real(dp), intent(out) :: m2ct0
\end{lstlisting}
\begin{lstlisting}[style=cstyle]
void ol_evaluate_ct(int id, const double *pp, double *m2l0, double *m2ct0);
\end{lstlisting}

\paragraph{$R_2$ counterterms}

The function \texttt{evaluate\_r2} evaluates the $R_2$ counterterm matrix element defined in  \eqref{eq:loopingredients},
returning \texttt{m2r2}=$\calW_{\onel,R_2}$ and \texttt{m2l0}=$\calW_{\tree}$.
%
\begin{lstlisting}[style=fortstyle]
subroutine evaluate_r2(id, p_ex, m2l0, m2r2)
  integer, intent(in) :: id
  real(dp), intent(in) :: p_ex(4,N)
  real(dp), intent(out) :: m2l0
  real(dp), intent(out) :: m2r2
\end{lstlisting}
\begin{lstlisting}[style=cstyle]
void ol_evaluate_r2(int id, const double *pp, double *m2l0, double *m2ct);
\end{lstlisting}

\paragraph{Pole residues}

The function \texttt{evaluate\_poles} evaluates the residues of the UV and IR poles of all ingredients to a Born--one-loop
interference defined in \eqref{eq:loopingredients} including also the \textbf{I}-operator.
As output it returns  
\texttt{m2l0}=$\calW_{\tree}$, \texttt{m2bare}=$\{\calW_{\onel,\fourdim}^{(1,\uv)},\calW_{\onel,\fourdim}^{(1,\ir)},\calW_{\onel,\fourdim}^{(2,\ir)} \}$, \texttt{m2ct}=$\{\calW_{\onel,\ct}^{(1,\uv)},\calW_{\onel,\ct}^{(1,\ir)},\calW_{\onel,\ct}^{(2,\ir)} \}$, \texttt{m2ir}=$\{\calW_{\treeiop}^{(1,\uv)},\calW_{\treeiop}^{(1,\ir)},\calW_{\treeiop}^{(2,\ir)} \}$
and \texttt{m2sum}=\texttt{m2bare}+\texttt{m2ct}+\texttt{m2ir}.
The three values in \texttt{m2bare}, \texttt{m2ct}, \texttt{m2ir}, \texttt{m2sum} correspond respectively to the residues of the $1/\epsuv$, $1/\epsir$ and $1/\epsir^2$ poles.
For automated pole cancellation checks the output of this routine can automatically be printed to the screen upon amplitude registration when the parameter \texttt{check\_poles}=1 is set.

%
\begin{lstlisting}[style=fortstyle]
subroutine evaluate_poles(id, psp, m2l0, m2bare, m2ct, m2ir, m2sum)
  integer, intent(in) :: id
  real(dp), intent(in) :: p_ex(4,N)
  real(dp), intent(out) :: m2l0
  real(dp), intent(out) :: m2bare(0:2)
  real(dp), intent(out) :: m2ct(0:2)
  real(dp), intent(out) :: m2ir(0:2)
  real(dp), intent(out) :: m2sum(0:2)
\end{lstlisting}
\begin{lstlisting}[style=cstyle]
void ol_evaluate_poles(int id, const double *pp,
                      double *m2l0, double *m2bare, double *m2ct, 
                      double *m2ir, double *m2sum);
\end{lstlisting}

\paragraph{Squared one-loop amplitudes}

The function \texttt{evaluate\_loop2} evaluates the squared one-loop matrix element \refeq{eq:Wloop2} returning
\texttt{m2l2}$=\calW_{\onelsq}$ and a relative accuracy estimate
\texttt{acc}=$\delta\calW_{\onelsq}/\calW_{\onelsq}$ (depending on the stability settings) as output.
%
\begin{lstlisting}[style=fortstyle]
subroutine evaluate_loop2(id, p_ex, m2l2, acc)
  integer, intent(in) :: id
  real(dp), intent(in) :: p_ex(4,N)
  real(dp), intent(out) :: m2l2
  real(dp), intent(out) :: acc
\end{lstlisting}
\begin{lstlisting}[style=cstyle]
void ol_evaluate_loop2(int id, const double *pp, double *m2l2, double *acc);
\end{lstlisting}

\subsection{$\mathbf{I}$-operator}
\label{app:native:iop}

\paragraph{Tree--tree $\mathbf{I}$-operator insertions}

The function \texttt{evaluate\_iop} evaluates the $\mathbf{I}$-operator insertion into a squared Born amplitude, as defined in \eqref{eq:ioperator}, returning \texttt{m2l0}=$\calW_{\tree}$
and \texttt{m2ir}=$\{\calW_{\treeiop}^{(0)},$\linebreak$\calW_{\treeiop}^{(1)},\calW_{\treeiop}^{(2)}\}$.
The three values in \texttt{m2ir} represent the finite part and the coefficients of the (IR) single and double poles. 
%
\begin{lstlisting}[style=fortstyle]
subroutine evaluate_iop(id, p_ex, m2l0, m2ir)
  integer, intent(in) :: id
  real(dp), intent(in) :: p_ex(4,N)
  real(dp), intent(out) :: m2l0
  real(dp), intent(out) :: m2ir(0:2)
\end{lstlisting}
\begin{lstlisting}[style=cstyle]
void ol_evaluate_iop(int id, const double *pp, double *m2l0, double *m2ir);
\end{lstlisting}

\paragraph{Loop-loop $\mathbf{I}$-operator insertions}

The function \texttt{evaluate\_loop2iop} evaluates the $\mathbf{I}$-operator insertion into a squared one-loop amplitude
as defined in \eqref{eq:ioperator}, returning \texttt{m2l2}$=\calW_{\onelsq}$ and \texttt{m2l2ir}$=\{\calW_{\onelsqiop}^{(0)},\calW_{\onelsqiop}^{(1)},\calW_{\onelsqiop}^{(2)}\}$. The three values in \texttt{m2l2ir} represent the finite part and the coefficients of the (IR) single and double poles in a Laurent series similar to \eqref{eq:laurentseries}.
%
\begin{lstlisting}[style=fortstyle]
subroutine evaluate_loop2iop(id, p_ex, m2l2, m2l2ir)
  integer, intent(in) :: id
  real(dp), intent(in) :: p_ex(4,N)
  real(dp), intent(out) :: m2l2
  real(dp), intent(out) :: m2l2ir(0:2)
\end{lstlisting}
\begin{lstlisting}[style=cstyle]
void ol_evaluate_loop2iop(int id, const double *pp,
                          double *m2l2, double *m2l2ir);
\end{lstlisting}

\subsection{Colour and charge correlators}

\paragraph{Tree--tree colour correlators}

The function \texttt{evaluate\_ccmatrix} returns the full  matrix of  colour-correlated squared tree amplitudes
as defined in \eqref{eq:treecolcordef}, returning \texttt{m2l0}=$\calW_{\tree}$ and a two-dimensional array \texttt{m2ccmatrix(i,j)}=$\colcorr{p,q}{ij}{00}{\ssLO\, \ssQCD} $ (\Fortran) or a one-dimensional array \texttt{m2ccmatrix}[(i-1)*N+j-1]=$\colcorr{p,q}{ij}{00}{\ssLO\, \ssQCD} $  (C). \texttt{m2ewcc} is reserved for the associated charge-correlated born amplitude, but is currently not in use.

%
\begin{lstlisting}[style=fortstyle]
subroutine evaluate_ccmatrix(id, p_ex, m2l0, m2ccmatrix, m2ewcc)
  integer, intent(in) :: id
  real(dp), intent(in) :: p_ex(4,N)
  real(dp), intent(out) :: m2l0
  real(dp), intent(out) :: m2ccmatrix(N,N)
  real(dp), intent(out) :: m2ewcc
\end{lstlisting}

\begin{lstlisting}[style=cstyle]
void ol_evaluate_ccmatrix(int id, const double *pp,
                          double *m2l0, double *m2ccmatrix, double *m2ewcc);
\end{lstlisting}

Alternatively the function \texttt{evaluate\_cc} evaluates only the $N(N-1)/2$ independent colour-correlated squared tree amplitudes \eqref{eq:treecolcordef} in the BLHA convention, returning \texttt{m2l0}=$\calW_{\tree}$ and \linebreak  \texttt{m2cc(i+(j-1)(j-2)/2)}=$\colcorr{p,q}{ij}{00}{\ssLO\, \ssQCD} $ (\Fortran) rsp.\  \texttt{m2cc[i+(j-1)(j-2)/2-1]} =$\colcorr{p,q}{ij}{00}{\ssLO\, \ssQCD}$ (C)  with $1 \leq i < j \leq N$. 

%
\begin{lstlisting}[style=fortstyle]
subroutine evaluate_cc(id, p_ex, m2l0, m2cc, m2ewcc)
  integer, intent(in) :: id
  real(dp), intent(in) :: p_ex(4,N)
  real(dp), intent(out) :: m2l0
  real(dp), intent(out) :: m2cc(N*(N-1)/2)
  real(dp), intent(out) :: m2ewcc
\end{lstlisting}
\begin{lstlisting}[style=cstyle]
void ol_evaluate_cc(int id, const double *pp,
                    double *m2l0, double *m2cc, double *m2ewcc);
\end{lstlisting}

\paragraph{Tree--tree charge correlators}

The function \texttt{evaluate\_ccewmatrix} returns the full  matrix of  charge-correlated squared tree amplitudes,  as defined in \eqref{eq:treechargedef}, returning \texttt{m2l0}=$\calW_{\tree}$ and a two-dimensional array \texttt{m2ccewmatrix(i,j)}=$\colcorr{p,q}{ij}{00}{\ssLO\, \ssQED} $ (\Fortran) or a one-dimensional array\linebreak  \texttt{m2ccewmatrix}[(i-1)*N+j-1]=$\colcorr{p,q}{ij}{00}{\ssLO\, \ssQED} $  (C).

%
\begin{lstlisting}[style=fortstyle]
subroutine evaluate_ccewmatrix(id, p_ex, m2l0, m2ccewmatrix)
  integer, intent(in) :: id
  real(dp), intent(in) :: p_ex(4,N)
  real(dp), intent(out) :: m2l0
  real(dp), intent(out) :: m2ccewmatrix(N,N)
\end{lstlisting}

\begin{lstlisting}[style=cstyle]
void ol_evaluate_ccewmatrix(int id, const double *pp,
                          double *m2l0, double *m2ccewmatrix);
\end{lstlisting}

\paragraph{Loop--loop colour correlators}

The function \texttt{evaluate\_ccmatrix2} returns the full  matrix of  colour-correlated squared loop amplitudes 
as defined in \eqref{eq:treecolcordef}, returning \texttt{m2l2}=$\calW_{\onelsq}$
as a two-dimensional array  \texttt{m2ccmatrix(i,j)}=$\colcorr{p,q}{ij}{11}{\ssLO\, \ssQCD} $ (\Fortran) or as a one-dimensional array  \texttt{m2ccmatrix}[(i-1)*N+j-1]=
$\colcorr{p,q}{ij}{11}{\ssLO\, \ssQCD} $ (C). 
\texttt{m2ewcc} is reserved for the associated charge-correlated loop-squared amplitude, but is currently not in use.

\begin{lstlisting}[style=fortstyle]
subroutine evaluate_ccmatrix2(id, p_ex, m2l2, m2ccmatrix, m2ewcc)
  integer, intent(in) :: id
  real(dp), intent(in) :: p_ex(4,N)
  real(dp), intent(out) :: m2l2
  real(dp), intent(out) :: m2ccmatrix(N,N)
  real(dp), intent(out) :: m2ewcc
\end{lstlisting}
\begin{lstlisting}[style=cstyle]
void ol_evaluate_ccmatrix2(int id, const double *pp,
                           double *m2l2, double *m2ccmatrix, double *m2ewcc);
\end{lstlisting}

Similarly as for the colour-correlated Born correlators (see above), the function \texttt{evaluate\_cc2}
evaluates only the  independent colour-correlated loop-squared amplitudes in the BLHA convention  returning \texttt{m2l2}=$\calW_{\onelsq}$ and  \texttt{m2cc(i+(j-1)(j-2)/2)}=$\colcorr{p,q}{ij}{11}{\ssLO\, \ssQCD} $ (\Fortran) rsp.\ \texttt{m2cc[i+(j-1)(j-2)/2-1]} =$\colcorr{p,q}{ij}{11}{\ssLO\, \ssQCD}$ (C)  with $1 \leq i < j \leq N$. 

%
\begin{lstlisting}[style=fortstyle]
subroutine evaluate_cc2(id, p_ex, m2l2, m2cc, m2ewcc)
  integer, intent(in) :: id
  real(dp), intent(in) :: p_ex(4,N)
  real(dp), intent(out) :: m2l0
  real(dp), intent(out) :: m2cc(N*(N-1)/2)
  real(dp), intent(out) :: m2ewcc
\end{lstlisting}
\begin{lstlisting}[style=cstyle]
void ol_evaluate_cc2(int id, const double *pp,
                     double *m2l2, double *m2cc, double *m2ewcc);
\end{lstlisting}

\paragraph{Loop--Loop charge correlators}

The function \texttt{evaluate\_ccewmatrix2} computes  the full  matrix of  charge-correlated squared loop amplitudes  
as defined in \eqref{eq:treechargedef}. As output it returns \texttt{m2l2}=$\calW_{\onelsq}$ and a two-dimensional array\linebreak \texttt{m2ccewmatrix(i,j)}=$\colcorr{p,q}{ij}{11}{\ssLO\, \ssQED} $ (\Fortran) or a one-dimensional array  \texttt{m2ccewmatrix}[(i-1)*N+j-1]=$\colcorr{p,q}{ij}{11}{\ssLO\, \ssQED}$~(C).

%
\begin{lstlisting}[style=fortstyle]
subroutine evaluate_ccewmatrix2(id, p_ex, m2l2, m2ccewmatrix)
  integer, intent(in) :: id
  real(dp), intent(in) :: p_ex(4,N)
  real(dp), intent(out) :: m2l2
  real(dp), intent(out) :: m2ccewmatrix(N,N)
\end{lstlisting}

\begin{lstlisting}[style=cstyle]
void ol_evaluate_ccewmatrix2(int id, const double *pp,
                             double *m2l2, double *m2ccewmatrix);
\end{lstlisting}

\paragraph{Tree--loop colour correlators}

The function \texttt{evaluate\_loopccmatrix2} returns the full  matrix of the finite parts of the colour-correlated Born--loop interferences,  as defined in \eqref{eq:treeloopcolcordef}, returning \texttt{m2l0}=$\calW_{\tree}$,  \texttt{m2l1}=$\{\calW_{\onel}^{(0)}, \calW_{\onel}^{(1)},\calW_{\onel}^{(2)}\}$  and a two-dimensional array  \texttt{m2ccmatrix(i,j)}=$\colcorr{P,Q}{ij}{01}{\ssNLO\, \ssQCD} $ (\Fortran) or as a one-dimensional array  \texttt{m2ccmatrix}[(i-1)*N+j-1]=
$\colcorr{P,Q}{ij}{01}{01, \ssNLO\, \ssQCD} $ (C). 
\texttt{m2ewcc} is reserved for the associated charge-correlated Born--loop interference, but is currently not in use.

\begin{lstlisting}[style=fortstyle]
subroutine evaluate_loopccmatrix(id, p_ex, m2l0, m2l1, m2ccmatrix, m2ewcc)
  integer, intent(in) :: id
  real(dp), intent(in) :: p_ex(4,N)
  real(dp), intent(out) :: m2l2
  real(dp), intent(out) :: m2l1(0:2)
  real(dp), intent(out) :: m2ccmatrix(N,N)
  real(dp), intent(out) :: m2ewcc
\end{lstlisting}
\begin{lstlisting}[style=cstyle]
void ol_evaluate_loopccmatrix2(int id, const double *pp, double *m2l0,
                               double *m2l1, double *m2ccmatrix, 
                               double *m2ewcc);
\end{lstlisting}

Similarly as for the colour-correlated Born correlators (see above), the function \texttt{evaluate\_loopcc}
evaluates only  the  independent colour-correlated Born--loop interference
amplitudes (finite parts only) in the BLHA convention returning
\texttt{m2l0}=$\calW_{\tree}$,  \texttt{m2l1}=$\{\calW_{\onel}^{(0)}, \calW_{\onel}^{(1)},\calW_{\onel}^{(2)}\}$
 and  \texttt{m2cc(i+(j-1)(j-2)/2)}=\linebreak$\mathcal{C}^{(ij)}_{01, \ssNLO\, \ssQCD} $ (\Fortran) rsp.\ \texttt{m2cc[i+(j-1)(j-2)/2-1]} =$\mathcal{C}^{(ij)}_{01, \ssNLO\, \ssQCD}$ (C)  with $1 \leq i < j \leq N$. 

%
\begin{lstlisting}[style=fortstyle]
subroutine evaluate_loopcc(id, p_ex, m2l2, m2cc, m2ewcc)
  integer, intent(in) :: id
  real(dp), intent(in) :: p_ex(4,N)
  real(dp), intent(out) :: m2l0
  real(dp), intent(out) :: m2l1(0:2)
  real(dp), intent(out) :: m2cc(N*(N-1)/2)
  real(dp), intent(out) :: m2ewcc
\end{lstlisting}
\begin{lstlisting}[style=cstyle]
void ol_evaluate_loopcc2(int id, const double *pp, double *m2l0,
                    double *m2l1, double *m2cc, double *m2ewcc);
\end{lstlisting}

\subsection{Spin correlators}

\paragraph{Tree--tree spin correlators}

The function \texttt{evaluate\_sc} evaluates the colour-spin-correlated squared tree amplitudes 
\eqref{eq:treecolspincorscalar} for a given gluon/photon \texttt{emitter} $j$ and polarisation vector $\texttt{polvect}=\kperp$ 
fulfilling $\kperp\cdot p_j=0$. It returns
\texttt{m2sc(k)}=$\spincorrB{p,q}{jk}{LL,\ssLO}(\kperp)$ (Fortran), rsp.\
\texttt{m2sc[k-1]}=$\spincorrB{p,q}{jk}{LL,\ssLO}(\kperp)$ (C) with $1\le k
\le N$.

%
\begin{lstlisting}[style=fortstyle]
subroutine evaluate_sc(id, p_ex, emitter, polvect, m2sc)
  integer, intent(in) :: id
  real(dp), intent(in) :: p_ex
  integer, intent(in) :: emitter
  real(dp), intent(in) :: polvect(4)
  real(dp), intent(out) :: m2sc(N)
\end{lstlisting}
\begin{lstlisting}[style=cstyle]
void ol_evaluate_sc(int id, const double *pp,
                    int emitter, double *polvect, double *m2sc);
\end{lstlisting}

The function \texttt{evaluate\_sctensor} evaluates the  colour-spin-correlated squared tree tensor \eqref{eq:treecolspincortensor} for an \texttt{emitter} $j$ returning \texttt{m2l0}=$\calW_{\tree}$ and as a N$\times$4$\times$4 array \texttt{m2munu(k,mu,nu)}=$\spincorr{p,q}{jk}{\mu\nu}{00,\ssLO}$
(\Fortran), rsp.\ a vector of length $(16N)$,
\texttt{m2munu[(k-1)*N+(mu-1)*4+(nu-1)]}=$\spincorr{p,q}{jk}{\mu\nu}{00,\ssLO}$
(C), with $1\le k \le N$ and $1 \le \texttt{mu,nu} \le 4$.

%
\begin{lstlisting}[style=fortstyle]
subroutine evaluate_sctensor(id, p_ex, emitter, m2l0, m2munu)
    integer, intent(in) :: id
    real(dp), intent(in) :: p_ex
    integer, intent(in) :: emitter
    real(dp), intent(out) :: m2l0
    real(dp), intent(out) :: m2munu(N,4,4)
\end{lstlisting}
\begin{lstlisting}[style=cstyle]
void ol_evaluate_sctensor(int id, const double *pp,
                    int emitter, double *m2l0, double *m2munu);
\end{lstlisting}

The function \texttt{evaluate\_stensor} evaluates the  spin-correlated squared tree tensor \eqref{eq:treespincortensor}
(\Powheg convention) for an \texttt{emitter} $j$ returning \texttt{m2l0}=$\calW_{\tree}$ and as a 4$\times$4 array \texttt{m2munu(mu,nu)}\linebreak=$\spincorr{p,q}{j}{\mu\nu}{00,\ssLO}$
(\Fortran), rsp.\ a vector of length $16$,
\texttt{m2munu[(mu-1)*4+(nu-1)]}=$\spincorr{p,q}{j}{\mu\nu}{00,\ssLO}$ (C),
with $1 \le \texttt{mu,nu} \le 4$.

%
\begin{lstlisting}[style=fortstyle]
subroutine evaluate_stensor(id, p_ex, emitter, m2l0, m2munu)
    integer, intent(in) :: id
    real(dp), intent(in) :: p_ex
    integer, intent(in) :: emitter
    real(dp), intent(out) :: m2l0
    real(dp), intent(out) :: m2munu(4,4)
\end{lstlisting}
\begin{lstlisting}[style=cstyle]
void ol_evaluate_stensor(int id, const double *pp,
                         int emitter, double *m2l0, double *m2munu);
\end{lstlisting}

\paragraph{Loop--loop spin correlators}

The function \texttt{evaluate\_sc2} evaluates the colour-spin-correlated loop-squared amplitudes 
\eqref{eq:treecolspincorscalar} for a given gluon/photon \texttt{emitter} $j$ and polarisation vector $\texttt{polvect}=\kperp$ 
fulfilling $\kperp\cdot p_j=0$. It returns an array of length $N$,
\texttt{m2sc(k)}=$\spincorrB{p,q}{jk}{11,\ssLO}(\kperp)$ (Fortran), rsp.\
\texttt{m2sc[k-1]}=$\spincorrB{p,q}{jk}{11,\ssLO}(\kperp)$ (C) with $1 \le k \le N$.

%
\begin{lstlisting}[style=fortstyle]
subroutine evaluate_sc2(id, p_ex, emitter, polvect, m2sc)
  integer, intent(in) :: id
  real(dp), intent(in) :: p_ex
  integer, intent(in) :: emitter
  real(dp), intent(in) :: polvect(4)
  real(dp), intent(out) :: m2sc(N)
\end{lstlisting}
\begin{lstlisting}[style=cstyle]
void ol_evaluate_sc2(int id, const double *pp,
                     int emitter, double *polvect, double *m2sc);
\end{lstlisting}

The function \texttt{evaluate\_sctensor2} evaluates  the colour-spin-correlated loop-squared tensor \eqref{eq:treespincortensor}
(\Powheg convention) for an \texttt{emitter} $j$ returning \texttt{m2l2}=$\calW_{\onelsq}$ and as a N$\times$4$\times$4 array \linebreak \texttt{m2munu(k,mu,nu)}=$\spincorr{p,q}{jk}{\mu\nu}{11,\ssLO}$
(\Fortran), rsp.\ a vector of length $16N$ \linebreak
\texttt{m2munu[(k-1)*N+(mu-1)*4+(nu-1)]}=$\spincorr{p,q}{jk}{\mu\nu}{11,\ssLO}$
(C) with $1 \le k \le N$ and $1 \le \texttt{mu,nu} \le 4$.

%
\begin{lstlisting}[style=fortstyle]
subroutine evaluate_sctensor2(id, p_ex, emitter, m2l2, m2munu)
    integer, intent(in) :: id
    real(dp), intent(in) :: p_ex
    integer, intent(in) :: emitter
    real(dp), intent(out) :: m2l2
    real(dp), intent(out) :: m2munu(N,4,4)
\end{lstlisting}
\begin{lstlisting}[style=cstyle]
void ol_evaluate_sctensor2(int id, const double *pp,
                           int emitter, double *m2l2, double *m2munu);
\end{lstlisting}

Alternatively the function \texttt{evaluate\_stensor2} evaluates  the spin-correlated loop-squared tensor \eqref{eq:treecolspincortensor}
(\Powheg convention) for an \texttt{emitter} $j$ returning \texttt{m2l2}=$\calW_{\onelsq}$ and as a 4$\times$4 array \linebreak \texttt{m2munu(mu,nu)}=$\spincorr{p,q}{j}{\mu\nu}{11,\ssLO}$
(\Fortran), rsp.\ a vector of length $16$,
\texttt{m2munu[(mu-1)*4+(nu-1)]}=\linebreak$\spincorr{p,q}{j}{\mu\nu}{11,\ssLO}$
(C) with $1 \le  \texttt{mu,nu} \le 4$.

%
\begin{lstlisting}[style=fortstyle]
subroutine evaluate_stensor2(id, p_ex, emitter, m2l2, m2munu)
    integer, intent(in) :: id
    real(dp), intent(in) :: p_ex
    integer, intent(in) :: emitter
    real(dp), intent(out) :: m2l2
    real(dp), intent(out) :: m2munu(4,4)
\end{lstlisting}
\begin{lstlisting}[style=cstyle]
void ol_evaluate_stensor2(int id, const double *pp,
                          int emitter, double *m2l2, double *m2munu);
\end{lstlisting}

\paragraph{Tree--loop spin correlators}

The function \texttt{evaluate\_loopsc} evaluates the colour-spin-correlated Born--loop interference (finite part) 
\eqref{eq:treeloopcolspincortensor} for a given gluon/photon \texttt{emitter} $j$ and polarisation vector $\texttt{polvect}=\kperp$ 
fulfilling $\kperp\cdot p_j=0$. It returns an array of length $N$,
\texttt{m2sc(k)}=$\spincorrB{P,Q}{jk}{01,\ssNLO}(\kperp)$ (Fortran), rsp.\
\texttt{m2sc[k-1]}=$\spincorrB{P,Q}{jk}{01,\ssNLO}(\kperp)$ (C) with $1 \le k \le N$.

%
\begin{lstlisting}[style=fortstyle]
subroutine evaluate_loopsc(id, p_ex, emitter, polvect, m2sc)
  integer, intent(in) :: id
  real(dp), intent(in) :: p_ex
  integer, intent(in) :: emitter
  real(dp), intent(in) :: polvect(4)
  real(dp), intent(out) :: m2sc(N)
\end{lstlisting}
\begin{lstlisting}[style=cstyle]
void ol_evaluate_loopsc(int id, const double *pp,
                        int emitter, double *polvect, double *m2sc);
\end{lstlisting}

The function \texttt{evaluate\_loopsctensor} evaluates  the colour-spin-correlated  Born--loop interference tensor (finite part) \eqref{eq:treeloopspincortensorB}
(\Powheg convention) for an \texttt{emitter} $j$ returning
\texttt{m2l0}=$\calW_{\tree}$,  \texttt{m2l1}=\linebreak $\{\calW_{\onel}^{(0)}, \calW_{\onel}^{(1)},\calW_{\onel}^{(2)}\}$
and a N$\times$4$\times$4 array \texttt{m2munu(k,mu,nu)}=$\spincorr{P,Q}{jk}{\mu\nu}{01,\ssNLO}$
(\Fortran), rsp.\ a vector of length $16N$,
\texttt{m2munu[(k-1)*N+(mu-1)*4+(nu-1)]}=$\spincorr{P,Q}{j}{\mu\nu}{11,\ssNLO}$
(C) with $1 \le k \le N$ and $1 \le \texttt{mu,nu} \le 4$.

%
\begin{lstlisting}[style=fortstyle]
subroutine evaluate_loopsctensor(id, p_ex, emitter, m2l0, m2l1, m2munu)
    integer, intent(in) :: id
    real(dp), intent(in) :: p_ex
    integer, intent(in) :: emitter
    real(dp), intent(out) :: m2l0
    real(dp), intent(out) :: m2l1(0:2)
    real(dp), intent(out) :: m2munu(N,4,4)
\end{lstlisting}
\begin{lstlisting}[style=cstyle]
void ol_evaluate_loopsctensor(int id, const double *pp,
                     int emitter, double *m2l0, double *m2l1, double *m2munu);
\end{lstlisting}

Alternatively the function \texttt{evaluate\_loopstensor} evaluates  the spin-correlated  Born--loop interference tensor (finite part) \eqref{eq:treeloopcolspincortensorB}
(\Powheg convention) for an \texttt{emitter} $j$ returning
\texttt{m2l0}=$\calW_{\tree}$,  \texttt{m2l1}=$\{\calW_{\onel}^{(0)}, \calW_{\onel}^{(1)},\calW_{\onel}^{(2)}\}$
and a 4$\times$4 array \texttt{m2munu(mu,nu)}=$\spincorr{P,Q}{j}{\mu\nu}{01,\ssNLO}$
(\Fortran), rsp.\ a vector of length $16$, \texttt{m2munu[(mu-1)*4+(nu-1)]}=\linebreak=$\spincorr{P,Q}{j}{\mu\nu}{11,\ssNLO}$ (C)
with $1 \le \texttt{mu,nu} \le 4$.

%
\begin{lstlisting}[style=fortstyle]
subroutine evaluate_loopstensor(id, p_ex, emitter, m2l0, m2l1, m2munu)
    integer, intent(in) :: id
    real(dp), intent(in) :: p_ex
    integer, intent(in) :: emitter
    real(dp), intent(out) :: m2l0
    real(dp), intent(out) :: m2l1(0:2)
    real(dp), intent(out) :: m2munu(4,4)
\end{lstlisting}
\begin{lstlisting}[style=cstyle]
void ol_evaluate_loopstensor(int id, const double *pp,
                    int emitter, double *m2l0, double *m2l1, double *m2munu);
\end{lstlisting}

\subsection{Colour basis and tree amplitudes in colour space}
\label{app:native:colour}

Besides calculating squared and colour-summed matrix elements, \OL also provides tree-level amplitudes with full colour information, see Section~\ref{sec:colourbasis}, 
required for the matching of parton showers to matrix elements. 
In the following we describes how to retrieve the colour basis used for a process and the amplitude as a vector in the colour space which is spanned by these basis elements.

\paragraph{Dimension of colour basis and number of helicities}

The colour basis elements are encoded as integer arrays and must be retrieved once for each process. First one must obtain the following information:
\begin{itemize}
  \item \texttt{ncolb}: the number of basis elements,
  \item \texttt{colelemsz}: the size of the longest basis element,
  \item \texttt{nheltot}: the total number of helicity configurations (including vanishing configurations).
\end{itemize}
These are returned by the function \texttt{tree\_colbasis\_dim} for a given process.

\begin{lstlisting}[style=fortstyle]
subroutine tree_colbasis_dim(id, ncolb, colelemsz, nheltot)
  integer, intent(in) :: id
  integer, intent(out) :: ncolb, colelemsz, nheltot
\end{lstlisting}
\begin{lstlisting}[style=cstyle]
void ol_tree_colbasis_dim(int id, int *ncolb, int *colelemsz, int *nheltot);
\end{lstlisting}

\paragraph{Trace basis}
The function \texttt{tree\_colbasis} returns the actual colour basis as a trace basis  in a format corresponding to
\refeq{eq:colbasisA}--\refeq{eq:tracebasisproducts},  encoded 
as a two-dimensional integer array of the size
\texttt{basis(colelemsz,ncolb)} (Fortran) rsp.\ \texttt{basis[ncolb][colelemsz]} (C).  
Trailing zeros should be ignored.
The two-dimensional array \texttt{needed} indicates if a certain colour interference contributes to the squared amplitude or not. If \texttt{needed[i][j]=1}, the interference of basis elements \texttt{i} and \texttt{j} contributes, if \texttt{needed[i][j]=0} it does not.

\begin{lstlisting}[style=fortstyle]
subroutine tree_colbasis(id, basis, needed)
  integer, intent(in) :: id
  integer, intent(out) :: basis(colelemsz,ncolb), needed(ncolb,ncolb)
\end{lstlisting}
\begin{lstlisting}[style=cstyle]
void ol_tree_colbasis(int id, int *basis, int *needed);
\end{lstlisting}

\paragraph{Colour-flow basis}

Alternatively the function \texttt{tree\_colourflow} returns the basis in colour flow representation, as defined in Eq.~\eqref{eq:colourflowrepr}. The format of the basis is \texttt{flowbasis(2,N,ncolb)} (Fortran) rsp.\ \texttt{flowbasis[ncolb][N][2]} (C),  defining \texttt{ncolb}  colour flows.

%
\begin{lstlisting}[style=fortstyle]
subroutine tree_colourflow(id, flowbasis)
  integer, intent(in) :: id
  integer, intent(out) :: flowbasis(2,N,ncolb)
\end{lstlisting}
\begin{lstlisting}[style=cstyle]
void ol_tree_colourflow(int id, int *flowbasis);
\end{lstlisting}

\paragraph{Tree amplitudes in colour space}
Now, the function \texttt{evaluate\_tree\_colvect} returns the (complex) tree-level amplitude \texttt{amp}=$\{\calA_0^{(i)}(\heli)\}$, defined in \eqref{eq:treecolvect}, as a vector in the colour space spanned by the colour basis elements for each of the \texttt{nhelnonv} non-vanishing helicity configurations, which may be smaller than the total number of helicity configurations \texttt{nheltot} returned by\linebreak  \texttt{tree\_colbasis\_dim()}. 
In \Fortran \texttt{amp(:,h)} for \texttt{h=1..nhelnonv} is an array of \texttt{ncolb} complex numbers such that the element \texttt{amp(i,h)} corresponds to the colour basis element \texttt{basis(:,i)}. In C \texttt{amp[h]} for \texttt{h=0..nhelnonv-1} is an array of \texttt{2*ncolb} real numbers such that the elements \texttt{amp[h][2*i]} and \texttt{amp[h][2*i+1]} are the real and imaginary parts of the amplitude which corresponds to the colour basis element \texttt{basis[i]}.

Note that colour and helicity average factors and symmetry factors 
must still be applied when the squared amplitude is built from these results.
See \refeq{eq:bornsqcolint} and
\refeq{eq:LCbornsqA}--\refeq{eq:bornsqcolvect}.
%
\begin{lstlisting}[style=fortstyle]
subroutine evaluate_tree_colvect(id, p_ex, amp, nhelnonv)
  integer, intent(in) :: id
  real(dp), intent(in) :: p_ex
  complex(dp), intent(out) :: amp(ncolb,nheltot)
  integer, intent(out) :: nhelnonv
\end{lstlisting}

\begin{lstlisting}[style=cstyle]
void ol_evaluate_tree_colvect(int id, const double *pp,
                              double *amp, int *nhelnonv);
\end{lstlisting}

\paragraph{Squared tree amplitudes in colour space}
Finally, the function \texttt{evaluate\_tree\_colvect2} evaluates the squared amplitudes for the colour basis elements, \ie the diagonal elements of the colour interference matrix \eqref{eq:bornsqcolvect}, returning a vector of \texttt{ncolb} elements as \texttt{m2arr}(i)=$\big|\calA^{(i)}_{0}\big|^2$ (\Fortran), rsp.\ \texttt{m2arr}[i-1]=$\big|\calA^{(i)}_{0}\big|^2$ (C).
This is meant to calculate the probability with which a matched parton shower should start from the corresponding colour flow. Note that the results are only correct to leading colour approximation and may contain (or even be purely) sub-leading colour contributions.

%
\begin{lstlisting}[style=fortstyle]
subroutine evaluate_tree_colvect2(id, psp, m2arr)
  integer, intent(in) :: id
  real(dp), intent(in) :: psp
  real(dp), intent(out) :: m2arr(ncolb)
\end{lstlisting}
\begin{lstlisting}[style=cstyle]
void ol_evaluate_tree_colvect2(int id, const double *pp, double *m2arr);
\end{lstlisting}
%

\subsection{Basic examples}
\label{app:native:examples}

Here we give a basic example, both for \Fortran and C, which illustrates the usage of the native \OL interface.
In these examples the process $d\bar{d}\to Zu\bar{u}$ is registered via \texttt{order\_ew=1}, \ie the leading tree-level order corresponds to $\ord(\alphaS^2\alpha)$ and the one-loop order corresponds to the $\ord(\alphaS^3\alpha)$ NLO QCD corrections. Similar examples are shipped with the \OL installation as \texttt{./examples/OL\_fortran.f90} and \texttt{./examples/OL\_cpp.cpp} respectively.
%
\begin{lstlisting}[style=fortstyle]
program main
  use openloops
  implicit none
  integer :: id
  real(selected_real_kind(15)) :: muren = 100, alpha_s = 0.1, sqrts=1000
  real(selected_real_kind(15)) :: p_ex(0:3,5), m2_tree, m2_loop(0:2), acc

  call setparameter_int("order_ew", 1)
  id = register_process("1 -1 -> 23 2 -2", 11);
  ! or id = register_process([1,-1,23,2,-2], 11)
  ! register more processes as needed
  call start();
  ! calculate matrix elements, e.g.
  if (id > 0) then
    ! generate a random phase-space point with Rambo
    call phase_space_point(id, sqrts, p_ex)

    ! set strong coupling
    call set_parameter("alpha_s", alpha_s)
    ! set renormalisation scale
    call set_parameter("muren", muren)

    ! evaluate tree matrix element and print result
    call evaluate_tree(id, p_ex, m2_tree)
    print *, "evaluate_tree"
    print *, "Tree:       ", m2_tree

    ! evaluate loop matrix element and print result
    call evaluate_loop(id, p_ex, m2_tree, m2_loop(0:2), acc)
    print *, "evaluate_loop"
    print *, "Tree:       ", m2_tree
    print *, "Loop ep^0:  ", m2_loop(0)
    print *, "Loop ep^-1: ", m2_loop(1)
    print *, "Loop ep^-2: ", m2_loop(2)
    print *, "accuracy:   ", acc
  end if

  call finish();
end program main
\end{lstlisting}

\begin{lstlisting}[style=cstyle]
#include "openloops.h"

int main() {
  double sqrts = 1000., muren = 100., mZ = 91.2, alphas = 0.1;
  double m2_tree, m2_loop[3], acc;

  ol_setparameter_int("order_ew", 1);
  int id = ol_register_process("1 -1 -> 23 2 -2", 11);
  /* register more processes as needed */
  ol_start();
  /* calculate matrix elements, e.g. */
  if (id > 0) {
    /* Set parameter: strong coupling */
    ol_setparameter_double("alpha_s", alphas);
    /* Set parameter: renormalisation scale */
    ol_setparameter_double("muren", muren);

    /*  generate a random phase-space point with Rambo */
    double pp[5*ol_n_external(id)];
    ol_phase_space_point(id, sqrts, pp);

    /* evaluate tree matrix element and print result */
    ol_evaluate_tree(id, pp, &m2_tree);
    std::cout << "ol_evaluate_tree" << std::endl;
    std::cout << "Tree:       " << m2_tree << std::endl;

    /* evaluate loop matrix element and print result */
    ol_evaluate_loop(id, pp, &m2_tree, m2_loop, &acc);
    std::cout << "ol_evaluate_loop" << std::endl;
    std::cout << "Tree:       " << m2_tree << std::endl;
    std::cout << "Loop ep^0:  " << m2_loop[0] << std::endl;
    std::cout << "Loop ep^-1: " << m2_loop[1] << std::endl;
    std::cout << "Loop ep^-2: " << m2_loop[2] << std::endl;
    std::cout << "Accuracy:   " << acc << std::endl;
  }

  ol_finish();
  return 0;
}
\end{lstlisting}

\section{Other interfaces}
\label{app:otherinterfaces}

\OpenLoops{} has been integrated in a number of Monte Carlo frameworks. In particular \OpenLoops{} can be used in conjunction with
\Sherpa{}~\cite{Gleisberg:2008ta,Bothmann:2019yzt}, \Munich/\Matrix{}~\cite{Grazzini:2017mhc}, \Herwig{}~\cite{Bellm:2015jjp},  \Powheg{}~\cite{Alioli:2010xd}, \Whizard{}~\cite{Kilian:2007gr} and \Geneva~\cite{Alioli:2012fc}.
In Appendix~\ref{app:blha} we detail the BLHA interface within \OL, and in Appendices~\ref{app:sherpa} and \ref{app:powheg} 
the usage of \OL within \Sherpa and \Powheg respectively. Finally in Appendix~\ref{app:python} we briefly introduce the
OpenLoops {\sc Python} command line tool.

\subsection{BLHA interface}
\label{app:blha}

\OpenLoops offers an interface in the Binoth-Les-Houches-Accord in both versions BLHA1~\cite{Binoth:2010xt} and BLHA2~\cite{Alioli:2013nda}.
In order to use the \Fortran BLHA interface, the module \texttt{openloops\_blha} must be included with 
\begin{lstlisting}[style=fortstyle]
use openloops_blha
\end{lstlisting}
The module files are located in the directory \texttt{lib\_src/openloops/mod}, which should be added to the include path of the \Fortran compiler. In a C/\cpp program the \texttt{openloops.h} header has to be included.
In the following we list the scope of the BLHA interface within a \cpp program. Usage within a \Fortran program proceeds analogous.

Within a \cpp program an BLHA contract file is read by OpenLoops via

\begin{lstlisting}[style=cstyle]
OLP_Start(char *contract_file_name, int *error);
\end{lstlisting}

The answer file is either written to the same file or in a file specified in the contract file via

\begin{verbatim}
Extra AnswerFile ole_answer_file_name
\end{verbatim}

Parameters are either set via the contract file or directly via the procedure

\begin{lstlisting}[style=cstyle]
OLP_SetParameter(char *name, double *real_value, double *imag_value,
                 int *error);
\end{lstlisting}

Furthermore a list of the actual parameter settings can be written to a file \texttt{filename} via

\begin{lstlisting}[style=cstyle]
OLP_PrintParameter(char *filename);
\end{lstlisting}

At runtime the tree and loop amplitudes for a phase-space point of N external particles with momenta \texttt{pp}, as specified in the BLHA1/BLHA2 standards, are obtained via

\begin{lstlisting}[style=cstyle]
OLP_EvalSubProcess(int *id, const double *pp, double *muren,
                   double *alphaS, double *result);
\end{lstlisting}

Here, \texttt{id} is the ID of the corresponding subprocess (specified in the answer file), \texttt{muren} the renormalisation scale and \texttt{alphaS} the strong coupling constant. The result is written into the array \texttt{result}, where \texttt{result[3]} gives the tree amplitude and \texttt{result[2]} the finite part, \texttt{result[1]} the single pole and \texttt{result[0]} the double pole of the one-loop amplitude $\calW_{\onel}$.

A corresponding routine of the BLHA2 standard is also implemented:

\begin{lstlisting}[style=cstyle]
OLP_EvalSubProcess2(int *id, double *pp, double *mu, double *result,
                    double *acc);
\end{lstlisting}

Here, additionally an accuracy measure of the corresponding amplitude is returned as \texttt{acc}.
When not available \texttt{acc}=-1 is returned.
For further details see the specification of the BLHA1~\cite{Binoth:2010xt} and BLHA2~\cite{Alioli:2013nda} standards.
An example illustrating the usage of the BLHA interface with \OpenLoops is shipped as \texttt{./examples/OL\_blha.cpp}.

%

\subsection{Sherpa}
\label{app:sherpa}

OpenLoops can be used as a plug-in of \Sherpa 2.1.0 or later. Within upcoming releases of \Sherpa also the EW
subtraction~\cite{Schonherr:2017qcj} will become publicly available. For the installation of \Sherpa and the usage of 
\SherpaOpenLoops please also refer to the \Sherpa documentation available at

\begin{verbatim}
https://sherpa.hepforge.org.
\end{verbatim}

In order to use \OpenLoops together with \Sherpa the \SherpaOpenLoops interface has to be compiled together with \Sherpa passing the \texttt{-{}-enable-openloops} option together with the \OL installation path to the Sherpa configure script.
The \OpenLoops installation path can be modified at runtime by setting (in the Sherpa run card or command line):

\begin{verbatim}
OL_PREFIX=PATH_TO_OPENLOOPS
\end{verbatim}

In order to run \Sherpa in combination with \OpenLoops it is sufficient to add 
to the \Sherpa run card the statement

\begin{verbatim}
ME_SIGNAL_GENERATOR Comix Amegic OpenLoops;
\end{verbatim}

which includes \OpenLoops in the list of available matrix element generators,
and to set in the processes section of the \Sherpa run card the flag

\begin{verbatim}
Loop_Generator OpenLoops;
\end{verbatim}

Sherpa will now automatically use the one-loop matrix elements from \OpenLoops when for example a parton-shower matched simulation is requested via (in the processes section of the run card)

\begin{verbatim}
NLO_QCD_Mode MC@NLO;
\end{verbatim}

For details on these modes and many other options we refer to the \Sherpa documentation.

An example run card illustrating the use of \SherpaOpenLoops can be found within the installation of \Sherpa in the file

\begin{verbatim}
PATH_TO_SHERPA/AddOns/OpenLoops/example/Run.dat
\end{verbatim}

Additional examples of \SherpaOpenLoops run cards can be found in the \Sherpa manual.

In general Sherpa automatically handles all the necessary parameter initialisation of
\OpenLoops. However, user-defined parameters can be passed from the \Sherpa run card 
(or command line) to \OpenLoops via

{\small
\begin{verbatim}
OL_PARAMETERS FIRST_PARAM_NAME FIRST_PARAM_VAL SECOND_PARAM_NAME SECOND_PARAM_VAL ...;
\end{verbatim}

\subsection{POWHEG-BOX}
\label{app:powheg}

Internally the {\sc Powheg-Box+OpenLoops} framework automatically compiles, loads and manages all required 
\OpenLoops amplitude libraries. The interface provides the subroutines \texttt{openloops\_born}, \linebreak
 \texttt{openloops\_real}, and \texttt{openloops\_virtual} with interfaces identical to the corresponding 
\Powheg routines \texttt{setborn}, \texttt{setreal}, and \texttt{setvirtual} including colour- and 
spin-correlated tree-level amplitudes in the format required by the \Powheg. 
Additionally, the interface provides the routines \texttt{openloops\_init}, 
\texttt{openloops\_borncolour} and \texttt{openloops\_realcolour}. 
The former synchronises all parameters between \OpenLoops and the \Powheg and should be called at the 
end of the init processes subroutine of the \Powheg. The latter two provide  colour information required 
for parton-shower matching, \ie they return a colour-flow of the squared Born and real matrix elements 
in leading-colour approximation, on a probabilistic basis. Further details are given in 
Appendix A.3 of \cite{Jezo:2016ujg}.

\subsection{Python}
\label{app:python}

\OL provides a \Python module \texttt{openloops.py} in the directory
\texttt{pyol/tools} that wraps a subset of the functionality of the native interface.
Its main application is to provide a simple command line tool to evaluate matrix elements.
The documentation of the command line tool can be obtained via
\begin{verbatim}
./openloops run --help
\end{verbatim}

For example the following command evaluates the tree and one-loop amplitudes for $n=10$ random phase-space points with a center-of-mass energy
$\sqrt{\hat s}=500$~GeV for the process $u \bar u \to Z g g$ using $M_Z=91$~GeV and prints the result to the screen:
 \begin{verbatim}
./openloops run "u u~ > Z g g" order_ew=1 mass\(23\)=91 -e 500 -n 10
\end{verbatim}
The random phase-space points are generated with \Rambo~\cite{Kleiss:1985gy}.

\section{List of input parameters}
\label{app:input_parameters}

In Tabs.~\ref{app:tab1}-\ref{app:tab3} we list all input parameters and switches 
available in \OpenLoops. Within the general purpose Monte Carlo frameworks (e.g.\ \Sherpa, \Powheg 
and \Herwig) these parameters are synchronised automatically.

In Tab.~\ref{app:tab1} input parameters relevant for the process registration are listed, 
in Tab.~\ref{app:tab2} model input parameters are listed and in Tab.~\ref{app:tab1} input 
parameters relevant for the stability system are summarised.

\begin{table}[h]
  \begin{center}
    \begin{tabular}[t]{ccl}

     \multicolumn{3}{c}{process registration}  \\ \hline\hline
      parameter  & type/options & description \\ \hline\hline
         \texttt{order\_ew}  &     int, default=-1 & requested fixed (Born \& one-loop) power of the  \\ 
                    &     &   electromagnetic coupling at the squared-amplitude level \\ \hline
         \texttt{order\_qcd}  &     int, default=-1 & requested fixed (Born \& one-loop) power of the  \\ 
                    &     &   strong coupling at the squared-amplitude level \\ \hline
         \texttt{loop\_order\_ew}  &    int, default=-1 & requested one-loop power of the electromagnetic coupling 	 \\ 
                     &     &    constant at the squared-amplitude level (any Born)\\ \hline
                             \texttt{loop\_order\_qcd}  &    int, default=-1 & requested one-loop power of the strong coupling 	 \\ 
                                                 &     &    constant at the squared-amplitude level (any Born) \\ \hline
        \texttt{nf}  &    int, default=6 &  number of active quark flavours \\ \hline
               \multirow{2}{*}{\texttt{ckmorder}}    &        0 (default)&  diagonal CKM matrix \\
       &1 &  non-diagonal CKM matrix \\ \hline 
                \texttt{model}  &    str, default=''sm'' &  model selection. Available models: ``sm'', ``heft''  \\ \hline
         \texttt{install\_path}  &    str, default='''' & set installation path of process libraries if different from\\ 
                  &   &  \OL default installation \\ \hline
         \texttt{approx}  &    str, default='''' &  approximation \\ \hline
                  \texttt{allowed\_libs}  &    str, default='''' &  whitespace separated list of allowed libraries \\ \hline
       \texttt{check\_poles} & int, default=0 & 1: print pole cancellation checks upon amplitude registration  \\ \hline           
       \hline 
    \end{tabular}
  \end{center}
  \caption{Available input parameters and switches in \OL relevant for the process registration. Possible input types include int:\ integer or str:\ string. For details see Section~\ref{se:processselection}.}
  \label{app:tab1}
\end{table}

\begin{table}[h]
  \begin{center}
    \begin{tabular}[t]{ccl}

     \multicolumn{3}{c}{model input parameters}  \\ \hline\hline
      parameter  & type/options & description \\ \hline\hline
             \texttt{muren}  &    dp${}_+$ &  renormalisation scale $\mur$ \\ \hline
              \texttt{mureg}  &    dp${}_+$ &  dimensional regularisation scale $\mudim$ \\ \hline
               \texttt{alphas}  &    dp${}_+$ & strong coupling constant $\alphaS$	 \\ \hline
               \texttt{nf\_alphasrun}  &     int, default=0 & minimum number of quark flavours that contribute
 \\
                                &     & to the running of $\alphaS$ 		 \\ \hline
               \multirow{3}{*}{\texttt{ew\_scheme}}    &        &  0: $\alpha(0)$-scheme for electromagnetic couplings, \\
       &int, default=1 &  1: $\GF$-scheme for electromagnetic coupling,  \\
       & & 2: $\alphamz$-scheme for electromagnetic coupling \\
       \hline
            \texttt{alpha\_qed\_0}  &    dp${}_+$ &  $\alpha(0)$: electromagnetic coupling constant \\ 
           &     &  in the Thomson limit \\ \hline
            \texttt{alpha\_qed\_mz}  &    dp${}_+$ &  $\alphamz$: electromagnetic coupling constant	at $M_Z$ \\ \hline
             \texttt{gmu}  &    dp${}_+$ &  $\GF$:  Fermi constant as input for electromagnetic  \\ 
                                 &         &  coupling constant in  $\GF$-scheme \\ \hline
             \texttt{mass(PID)}  &  dp${}_+$ &  mass of particle with given PID	 \\ \hline
              \texttt{width(PID)}  &  dp${}_+$ &  width of particle with given PID	 \\ \hline
              \texttt{lambdam(PID)}  &  dp${}_+$ &  $\msbar$ renormalisation scale for mass of particle PID	 \\ \hline
              \texttt{yuk(PID)}  &  dp${}_+$ &  Yukawa mass of particle with given PID \\ 
                       &   &  (only NLO QCD) \\ \hline
                \texttt{yukw(PID)}  &  dp${}_+$ &  imaginary part of Yukawa mass  \\
                &  &  of particle with given PID (only NLO QCD) \\ \hline
                              \texttt{lambday(PID)}  &  dp${}_+$ &  $\msbar$ renormalisation scale for Yukawa mass of particle PID	 \\ \hline
                \texttt{freeyuk\_on} & int, default=0 & Switch to allow for Yukawa masses (\texttt{yuk/yukw/lambday})\\ 
                  & &  independent of  masses. \\ \hline
                \texttt{VCKMXY}&dp& CKM matrix elements (real part),\\
                                          &dp& \texttt{XY=\{du, su, bu, dc, sc, bc, dt, st, bt\}}\\ \hline
                 \texttt{VCKMIXY}&dp& CKM matrix elements (imaginary part),\\
                                          &dp& \texttt{XY=\{du, su, bu, dc, sc, bc, dt, st, bt\}}\\ \hline
                 \texttt{kappa\_hhh}  &  dp &  coupling multiplier for trilinear Higgs coupling $\lambda_H^{(3)}$	 \\ \hline
                 \texttt{kappa\_hhhh}  &  dp &  coupling multiplier for quartic Higgs coupling $\lambda_H^{(4)}$	 \\ \hline
                 
                                 \texttt{complex\_mass\_scheme} & int, default=1 & 0: on-shell scheme, 1: mixed on-shell--complex-mass-scheme,  \\ 
                  & &  2: pure complex-mass-scheme \\ \hline
                \texttt{onshell\_photons\_lsz} &b, default=1&  switch for rescaling/shift of external on-shell photons\\
                & &  to  $\alpha(0)$-scheme. \\   \hline
                \texttt{offshell\_photons\_lsz} &b, default=1& switch for  rescaling/shift of external off-shell photons     \\
                & & including regularisation prescription   \\ \hline
                \texttt{all\_photons\_dimreg} &b, default=0 & switch to treat all photons in dimensional (1) instead \\
                & & of numerical (0) regularisation  \\
                
       \hline
    \end{tabular}
  \end{center}
  \caption{Available model input parameters and switches in \OL. Possible input types include dp:\ double, dp${}_+$:\ positive double, int:\ integer, and b:\ integer 0 or 1. For details see Sections~\ref{sec:ewschemes}-\ref{se:renormalisation}.}
    \label{app:tab2}
        \label{app:inputparameters}
\end{table}

\begin{table}[h]
  \begin{center}
 
          \begin{tabular}[t]{ccl}
     \multicolumn{3}{c}{stability system: general}  \\ \hline\hline
      parameter & options & decription \\ \hline\hline
             \texttt{psp\_tolerance}  &   dp${}_+$, default=$10^{-9}$ &  Tolerance for warnings triggered by phase-space consistency   \\
                         &   &   checks (momentum conservation and on-shell conditions)  \\
       \hline\hline

    \end{tabular}

  \vspace{1cm}

    \begin{tabular}[t]{ccl}
     \multicolumn{3}{c}{stability system: born--loop interferences}  \\ \hline\hline
      parameter & options & description \\ \hline\hline
\multirow{3}{*}{\texttt{hp\_mode}}    &        1 (default) &  hybrid precision mode for hard regions \\
       &2 &  hybrid precision mode for IR regions (restricted to NLO QCD) \\
       &0 & hybrid precision mode turned off \\
       \hline
             \texttt{hp\_loopacc}  &   dp${}_+$, default=8. &  target precision in number of correct digits  \\
       \hline\hline

    \end{tabular}

  \vspace{1cm}

    \begin{tabular}[t]{ccl}\\
      \multicolumn{3}{c}{stability system: HEFT and loop--loop interferences}\\ \hline\hline
      parameter & options & description \\ \hline\hline
      \texttt{stability\_triggerratio}
      & dp${}_+$, default=0.2
      & The fraction of points with the largest $K$-factor\\
      && to be re-evaluated with the secondary reduction\\
      && library\\ \hline
      \texttt{stability\_unstable}
      & dp${}_+$, default=0.01
      & Relative deviation of two Born-loop interference\\
      &&  results for the same point above which the \\
      && qp evaluation is triggered\\ \hline
      \texttt{stability\_kill}
      & dp${}_+$, default=1.
      & Accuracy below which an unstable point is discarded\\
      && after qp evaluation for Born-loop interferences.\\ \hline
      \texttt{stability\_kill2}
      & dp${}_+$, default=10.
      & Accuracy below which an unstable point is\\
      && discarded in loop-loop interferences\\ \hline
      \multirow{4}{*}{\texttt{stability\_log}}
      & 0 (default) & no stability logs are written\\
      & 1           &  stability logs written  on \texttt{finish()} call\\
      & 2           & stability logs written adaptively\\
      & 3           & stability logs written for every phase-space point\\ \hline
      \texttt{stability\_logdir}
      & str
      & set the (relative) path for the stability log files\\ \hline\hline
    \end{tabular}
    \end{center}
  \caption{Available input parameters and switches in \OL relevant for the
  stability system. Possible inputs include dp${}_+$:\ positive double, int:\ integer,
  str:\ string. For details see Section~\ref{sec:stabsystem}.}
  \label{app:tab3}
  \label{tab:stability}
\end{table}

\clearpage

\subsection*{Acknowledgements}
We are thankful to Andreas van Hameren for supporting OneLOop, and to Ansgar Denner and Stefan Dittmaier
for supporting \Collier{}. We are indebted to Stefan Kallweit for numerous bug reports and pre-release tests.
Also we would like to thank the \Sherpa and \Matrix collaborations for continuous collaboration and
discussions. We thank the ATLAS and CMS Monte Carlo groups for valuable feedback.
J.M.L. would like to thank the Theoretical Particle Physics group at Sussex University for 
the hospitality during the completion of this work.
F.B., J.-N.L., S.P., H.Z., M.Z. acknowledge support
from the Swiss National Science Foundation (SNF) under contract
BSCGI0-157722.
M.Z. acknowledges support by the Swiss National Science Foundation (Ambizione grant PZ00P2-179877).



\bibliographystyle{elsarticle-num}
\bibliography{OL2_literature}

\end{document}